\documentclass[prc,showpacs,preprint,preprintnumbers,amsmath,amssymb,floatfix]{revtex4}

\usepackage{graphicx} % Include figure files
\usepackage{dcolumn}  % Align table columns on decimal point
\usepackage{bm}       % bold math
\usepackage{epsfig}

\newcommand{\asy}{A}               % raw uncorrected asymmetry
\newcommand{\acorr}{A^{\rm corr}}  % asymmetry with additive corrections
\newcommand{\aexp}{A^{\rm exp}}    % asymmetry with polarization correction
\newcommand{\aphys}{A^{\rm phys}}  % asymmetry with all corrections
\newcommand{\apv}{A^{\rm PV}}      % theoretical asymmetry
\newcommand{\ath}{A^{\rm th}}      % analyzing power
\newcommand{\af}{A_{{F}}}
\newcommand{\pg}{P_{\gamma}}

\begin{document}

% \preprint{JLab-HAPPEX-v6.1}

% \title{The Hall A Proton Parity Experiment\\HAPPEX}
\title{Parity-violating electroweak asymmetry in $\vec{e}  p$ scattering
}

% Author list
% Both authors and institutions are in alphabetical order.
% Within REVTEX4, to get superscripts from multiple authors
% tied to one institution, while typing the institution
% name only once, is a bit awkward.  We use the ``edef''
% macro to define successive superscript numbers in the
% same order as the ``affiliations'' are listed.
% These superscripts (e.g. \calstate) are then connected
% to authors.  To add another insitution, put the
% appropriate ``edef'' line and ``affiliation'' line.

\newcounter{univ_counter}
\setcounter{univ_counter} {0}

\addtocounter{univ_counter} {1} 
\edef\calstate{$^{\arabic{univ_counter}}$ } 

\addtocounter{univ_counter} {1} 
\edef\pascal{$^{\arabic{univ_counter}}$ } 

\addtocounter{univ_counter} {1} 
\edef\ekentucky{$^{\arabic{univ_counter}}$ } 

\addtocounter{univ_counter} {1} 
\edef\fiu{$^{\arabic{univ_counter}}$ } 

\addtocounter{univ_counter} {1} 
\edef\fourier{$^{\arabic{univ_counter}}$ } 

\addtocounter{univ_counter} {1} 
\edef\georgia{$^{\arabic{univ_counter}}$ } 

\addtocounter{univ_counter} {1} 
\edef\hampton{$^{\arabic{univ_counter}}$ } 

\addtocounter{univ_counter} {1} 
\edef\harvard{$^{\arabic{univ_counter}}$ } 

\addtocounter{univ_counter} {1} 
\edef\infnbari{$^{\arabic{univ_counter}}$ } 

\addtocounter{univ_counter} {1} 
\edef\infnroma{$^{\arabic{univ_counter}}$ } 

\addtocounter{univ_counter} {1} 
\edef\infnsanita{$^{\arabic{univ_counter}}$ } 

\addtocounter{univ_counter} {1} 
\edef\tjnaf{$^{\arabic{univ_counter}}$ } 

\addtocounter{univ_counter} {1} 
\edef\kent{$^{\arabic{univ_counter}}$ } 

\addtocounter{univ_counter} {1} 
\edef\kentucky{$^{\arabic{univ_counter}}$ } 

\addtocounter{univ_counter} {1} 
\edef\kharkov{$^{\arabic{univ_counter}}$ } 

\addtocounter{univ_counter} {1} 
\edef\kyungpook{$^{\arabic{univ_counter}}$ } 

\addtocounter{univ_counter} {1} 
\edef\maryland{$^{\arabic{univ_counter}}$ } 

\addtocounter{univ_counter} {1} 
\edef\umass{$^{\arabic{univ_counter}}$ } 

\addtocounter{univ_counter} {1} 
\edef\mitlns{$^{\arabic{univ_counter}}$ } 

\addtocounter{univ_counter} {1} 
\edef\newhamp{$^{\arabic{univ_counter}}$ } 

\addtocounter{univ_counter} {1} 
\edef\nccu{$^{\arabic{univ_counter}}$ } 

\addtocounter{univ_counter} {1} 
\edef\norfolk{$^{\arabic{univ_counter}}$ } 

\addtocounter{univ_counter} {1} 
\edef\olddom{$^{\arabic{univ_counter}}$ } 

\addtocounter{univ_counter} {1} 
\edef\princeton{$^{\arabic{univ_counter}}$ } 

\addtocounter{univ_counter} {1} 
\edef\regina{$^{\arabic{univ_counter}}$ } 

\addtocounter{univ_counter} {1} 
\edef\rutgers{$^{\arabic{univ_counter}}$ } 

\addtocounter{univ_counter} {1} 
\edef\saclay{$^{\arabic{univ_counter}}$ } 

\addtocounter{univ_counter} {1} 
\edef\stonybrook{$^{\arabic{univ_counter}}$ } 

\addtocounter{univ_counter} {1} 
\edef\syracuse{$^{\arabic{univ_counter}}$ } 

\addtocounter{univ_counter} {1} 
\edef\temple{$^{\arabic{univ_counter}}$ } 

\addtocounter{univ_counter} {1} 
\edef\tohoku{$^{\arabic{univ_counter}}$ } 

\addtocounter{univ_counter} {1} 
\edef\triumf{$^{\arabic{univ_counter}}$ } 

\addtocounter{univ_counter} {1} 
\edef\virginia{$^{\arabic{univ_counter}}$ } 

\addtocounter{univ_counter} {1} 
\edef\wandm{$^{\arabic{univ_counter}}$ } 

\author{K.~A.~Aniol\calstate}
\author{D.~S.~Armstrong\wandm}
\author{T.~Averett\wandm}
\author{M.~Baylac\saclay$\!\!^,$\tjnaf}
\author{E.~Burtin\saclay} 
\author{J.~Calarco\newhamp} 
\author{G.~D.~Cates\princeton$\!\!^,$\virginia} 
\author{C.~Cavata\saclay} 
\author{Z.~Chai\mitlns} 
\author{C.~C.~Chang\maryland} 
% ...

\date{\today} 

\begin{abstract}
We have measured the parity-violating
electroweak asymmetry in the elastic 
scattering of polarized electrons from protons.
Significant contributions to this asymmetry could
arise from the contributions of strange 
form factors in the nucleon. 
The measured asymmetry is 
$A = -15.05 \pm 0.98 ({\rm stat}) \pm 0.56 ({\rm syst})$
ppm at the kinematic point 
$\langle \theta_{\rm lab} \rangle = 12.{3^\circ} $
and 
$\langle Q^2 \rangle = 0.477$ (GeV/c)$^2$.
Based on these data as well as data on electromagnetic form factors, 
we extract the linear combination of strange form factors 
$G^s_E + 0.392 G^s_M = 0.014 \pm 0.020 \pm 0.010$  
where the first error arises from this experiment and the second arises
from the electromagnetic form factor data.
This paper provides a full description of the special experimental
techniques employed for precisely measuring the small asymmetry, 
including the first use of a strained GaAs crystal
and a laser-Compton polarimeter
in a fixed target parity-violation experiment.

\end{abstract}

\keywords{happex, parity violation, strange form factors}
\pacs{13.60.Fz; 11.30.Er; 13.40.Gp; 14.20.Dh}
\maketitle

\section{Introduction}
\label{sec:paper_intro}

In recent years, the role of strange quarks in nucleon structure has 
been a topic of great interest.  Data from the
European Muon Collaboration (EMC) \cite{emc}
showed that valence quarks contribute
less than half of the proton spin and also
suggested that significant spin may be carried by
the strange quarks.  Based on these observations,
Kaplan and Manohar \cite{kaplan_manohar}
pointed out that strange quarks might also contribute
to the magnetic moment and charge radius of 
the proton, {\it i.e.} to the vector matrix elements.
It turns out that a practical way to measure
these strange vector matrix elements is by measuring
the electroweak asymmetry in polarized electron scattering
\cite{mckeown1,beise_mckeown,beck1}.

In the work presented here, we have measured the 
parity-violating asymmetry
$A = {( \sigma_R - \sigma_L )} / {(\sigma_R + \sigma_L)}$
where $\sigma_{R(L)}$ is the differential cross section 
for elastic scattering of right($R$) and left($L$) handed
longitudinally polarized electrons from protons.
The kinematics 
$\langle \theta_{\rm lab} \rangle = 12.{3^\circ} $
and 
$\langle Q^2 \rangle = 0.477$ (GeV/c)$^2$
correspond to the smallest angle and largest energy possible
with the available spectrometers.  Under reasonable assumptions for the
$Q^2$ dependence of the strange form factors, these kinematics
maximize the figure of merit for a first measurement.
Results were obtained in two separate 
data-taking runs, in 1998 and 1999 in Hall A at the 
Thomas Jefferson National 
Accelerator Facility (Jefferson Lab). 
The experimental
conditions were somewhat different in the
two runs, here referred to as the 
``1998 run'' and ``1999 run''. 
In the 1998 run we used a 100 $\mu$A beam 
with 38\% polarization produced from a 
bulk GaAs crystal.  In the 1999 run we
ran with a strained GaAs crystal with
polarization $P$=70\% and $I$=35 $\mu$A.  This gave
an improvement in $P^2 I$, providing a greater
effective rate of taking data, but also creating new challenges
in controlling systematic errors.
The 1999 run was subdivided
into two periods of several weeks each,
the primary difference being the availability of the 
Compton polarimeter, which provided an independent measurement of
the beam polarization, for the latter part. 

Brief reports of these results have 
been published~\cite{aniol1,aniol2}; the present 
paper presents the experimental technique, 
data analysis, and physics implications in 
much more detail.
Further details can be found in several 
dissertations~\cite{Gary,Wilson,Johan,Bill,Brian,Baris}.

This paper is organized as follows.  
In section \ref{sec:motivation}
we explain the motivation for this experiment.
Section \ref{sec:expt_method} covers the experimental method
used to measure such small asymmetries 
of order 10 parts-per-million (ppm) in electron scattering.
A crucial aspect of the measurement
is the control of systematic errors, 
as described in section \ref{sec:systematics}.
Section \ref{sec:rsh_asy} discusses the data analysis of 
the asymmetries, the sensitivities to
beam parameters, and the resulting
helicity correlated systematic corrections
due to the beam.
In section \ref{sec:phys_asy} the extracted physics asymmetry 
is presented with all corrections to
the data including the beam polarization, backgrounds,
$Q^2$ measurements, radiative corrections,
kinematics, and acceptance.
Section \ref{sec:interpretation} presents the
results and their interpretation, which requires
corrections for form factors.
Section \ref{sec:strange}
provides the physics interpretation in the 
context of models of nucleon strangeness.
Finally, \ref{sec:conclusions} draws the conclusions of this work.

\section{Motivation}
\label{sec:motivation}

Measurements of the contribution of strange quarks 
to nucleon structure provide a unique window on
the quark-antiquark sea and make an 
important impact on our understanding of the 
low-energy QCD structure of nucleons. 
Since the mass of the strange quark is 
comparable to the strong interaction scale
it is reasonable to expect that
strangeness $q \bar q$ pairs should make observable 
contributions to the properties of nucleons, 
for instance the mass, spin, 
momentum, and the electromagnetic form factors.
Indeed, charm production in deep inelastic 
neutrino scattering \cite{charm1}
has shown that strange quarks carry about 3\% 
of the momentum of the 
proton at $Q^2 = 2$ (GeV/c)$^2$.
Much of the interest in the 
strangeness content of the nucleon originates 
from the EMC experiment 
\cite{emc} and related recent experiments \cite{dis1,abe}
which studied the spin structure functions of the
proton and neutron in deep inelastic scattering.
These experiments have established that the Ellis-Jaffe sum rule
\cite{ellis_jaffe}
is violated and that relatively little of the proton's
spin is carried by the valence 
quarks.  The initial paper also suggested
that significant spin was carried by strange quarks.  More
recent work has indicated that this latter conclusion is difficult
to establish convincingly \cite{badeva98};
see also the recent reviews by 
Kumar and Souder\cite{kumar_souder},
Beck and McKeown \cite{beck_mckeown},
Beck and Holstein \cite{beck_holstein},
and Musolf {\em et al.} \cite{musolf_physrep94}.

In the aftermath of the EMC results, 
it was suggested\cite{kaplan_manohar} that strange quarks
might contribute to the vector matrix elements
of the nucleon.  Indeed, numerous calculations of strange
matrix elements have been computed in the context of
various models.  The theoretical approaches include
dispersion relations \cite{jaffe1,hammer1,hammer2,forkel1}, 
vector dominance models with $\omega-\phi$ mixing \cite{forkel2}, 
the chiral bag model \cite{chiral_bag1},
unquenched quark model \cite{geiger_isgur},
perturbative chiral quark model \cite{pquark},
light-cone diquark model \cite{ma1},
chiral quark model \cite{riska1,riska2}, 
Skyrme model \cite{skyrme_1,skyrme_2}, 
Nambu-Jona-Lasinio soliton model \cite{weigel_nlj},
meson-exchange models \cite{meissner1},
kaon loops \cite{koepf1,musolf1,ito1}, 
an SU(3) chiral quark-soliton model \cite{soliton},
heavy baryon chiral 
perturbation theory \cite{hemmert1,hemmert2},
quenched chiral perturbation theory \cite{lattice_2},
as well as 
lattice QCD calculations \cite{lattice_1,lattice_2}.
These calculations have elucidated the physics behind
strange matrix elements and have provided numerical
estimates of the size of possible effects that have
served for the design goals of our experiment.

Parity violating electron scattering is a practical method
to measure the strange vector matrix elements
\cite{mckeown1,beise_mckeown,beck1}.
Purely electromagnetic scattering at a given kinematics can 
measure only two linear combinations 
of the Sachs form factors:
\begin{equation} \label{eq:electr}
 G^{\gamma p}_{E,M} = 
\frac {2}{3} G^u_{E,M} - 
\frac {1}{3} G^d_{E,M} - 
\frac {1}{3} G^s_{E,M} \end{equation}
\begin{equation} G^{\gamma n}_{E,M} = 
\frac {2}{3} G^d_{E,M} - 
\frac {1}{3} G^u_{E,M} - 
\frac {1}{3} G^s_{E,M} \end{equation}
where $G^f_{E,M}$ is the electric $(E)$
or magnetic $(M)$ form factor for quark flavor $f$
in the proton. Here it is assumed that 
the quark flavors $u$, $d$, and $s$ contribute.
Charge symmetry between proton $p$ and neutron $n$
is also assumed, so that for the quark form factors
\begin{equation} \label{eq:isospin}
G^u_p = G^d_n
\ {\rm ;} \ \ \ 
G^d_p = G^u_n
\ {\rm ;} \ \ \ 
G^s_p = G^s_n \end{equation}
where now the subscripts $p$ and $n$ are 
for proton and neutron.

Additional information is needed to determine 
whether or not there is a contribution from the
strangeness form factors $G^s_{E,M}$.
This is provided by  parity violation in the scattering 
from protons, measuring a new pair of linear
combinations
{\setlength\arraycolsep{2pt}
\begin{eqnarray} \label{eq:weakff}
 G^{Z p}_{E,M} & = &
\left(  \frac{1}{4} - 
\frac{2}{3}\sin^2 \theta_W \right)
G^u_{E,M} \hskip 0.04in  + 
\nonumber \\ 
&  & 
\left( - \frac{1}{4} + \frac{1}{3}
\sin^2 \theta_W \right)
\times \left[ G^d_{E,M} + G^s_{E,M} \right]
\end{eqnarray}}
where $Z$ stands for the 
$Z^0$ boson of the neutral weak interaction.

Thus by measuring these neutral weak form factors, 
in conjunction with
the electromagnetic form factors, we can extract
the strange quark contribution.
The explicit dependence of the parity violating asymmetry
on the strangeness content is written as follows
in terms of the Sachs form factors introduced above, 
the neutral weak axial form factor $G^{Zp}_A$,
the Weinberg angle
$\theta_{W}$, Fermi constant $G_F$,
fine-structure constant $\alpha$,
and kinematic factors
$Q^2$, $\tau$, $\epsilon$, and $\epsilon^{\prime}$ 

\begin{widetext}
{\setlength\arraycolsep{2pt}
\begin{eqnarray} \label{eq:pvasy}
\apv & =  &
- \frac { G_F {| Q |}^2 }
{ 4 \pi \alpha \sqrt2}
\hskip 0.04in \times \hskip 0.04in 
\rho' \biggl[
(1 - 4\kappa'\sin^2 \theta_W) - 
\nonumber \\
&  &
\frac{ \epsilon G^{\gamma p}_E
( G^{\gamma n}_E +
G^s_E )
+ \tau G^{\gamma p}_M 
( G^{\gamma n}_M
+ G^s_M ) 
\hskip 0.04in - \hskip 0.04in
2 {\epsilon}^{\prime} \hskip 0.02in
 (1 - 4\sin^2 \theta_W) \hskip 0.02in
G^{\gamma p}_M
G^{Z p}_A }
 { \epsilon {(G^{\gamma p}_E)}^2 +
  \tau {(G^{\gamma p}_M)}^2 }
\biggr]
\end{eqnarray}}
\end{widetext}
The kinematic factors are
$Q^2 = -q^2_{\mu} > 0$,
the square of the four-vector momentum transfer,
$\tau = Q^2 / 4 M^2$ where $M$ is the proton mass,
$\epsilon = {[1+2(1+\tau)\tan^2(\theta/2)]}^{-1}$
where $\theta$ is the scattering angle,
and $\epsilon^{\prime} = \sqrt{\tau(1+\tau)(1-{\epsilon}^2)}$.
The parameters $\rho' = 0.9879$ and $\kappa' = 1.0029$ arise
from electroweak radiative corrections
\cite{barnet}.

Note that the asymmetry also contains a term with
the neutral weak axial form factor
$G^{Z p}_A$ which as
explained in \cite{sample_prl} 
can be estimated by combining
information from neutron beta decay \cite{barnet},
polarized deep inelastic
scattering \cite{abe}, and 
calculations of the axial radiative
correction \cite{axial_corr1,musolf_physrep94}; 
\hskip 0.05in it is suppressed in the HAPPEX
kinematics since $\epsilon^\prime \sim 0.08$
and $1 - 4 \sin^2 \theta_W \sim 0.08$, 
and contributes only a few percent.

\section{Experimental Method}
\label{sec:expt_method}

\subsection{Overview}
\label{sec:expt_overview}

The experiment measured the helicity-dependent
left-right asymmetry in the scattering of longitudinally
polarized 3.2 GeV electrons from a 15 cm long unpolarized liquid
hydrogen target. Since the anticipated asymmetry was of the order
of $10^{-5}$ or 10 parts per million (ppm), there were two
characteristics that dictated the overall experimental design.
First, the physical properties of the incident beam on target and
the experimental environment as a whole had to be identical for
the left- and right-handed beams to a very high degree so as to
minimize spurious asymmetries. Second, in order to accumulate the
required statistics at a high rate, the relative scattered flux
was measured by integrating the response of the detector rather
than by counting individual particles.

A GaAs photocathode was optically pumped by circularly polarized
laser light to produce polarized electrons, with the ability to
rapidly and randomly flip the sign of the electron beam
polarization. The asymmetry was extracted by generating the
incident electron beam as a pseudorandom time sequence of helicity
``windows" at 30 Hz and then measuring the fractional difference in
the integrated scattered flux over window pairs of opposite
helicity.

The elastically scattered electrons with $\theta_{\rm
lab}\sim 12.5^\circ$ were focused by two high-resolution
spectrometers (HRS) onto detectors consisting of lead-lucite sandwich calorimeters. The
\v{C}erenkov light from each detector was collected by a
photomultiplier tube, integrated over the duration of each
helicity window and digitized by analog to digital converters
(ADCs). The HRS pair has sufficient resolution to spatially
separate the elastic electrons from inelastic electrons at the
$\pi^0$ threshold. The amount of background was measured in
separate calibration runs using conventional drift chambers,
resulting in a small correction with negligible systematic errors.

The experiment was carefully designed to minimize the impact of
random as well as of helicity-correlated fluctuations of the
measured scattered flux. The electrical environment around the
ADCs in particular and the data acquisition and control system as
a whole were configured so that the observed fluctuations in the
integrated scattered flux were dominated by counting statistics.

Apart from random jitter, an important class of potential false
asymmetries might arise from helicity-correlated fluctuations in
the physical properties of the beam, such as intensity, energy and
trajectory. The helicity-correlated intensity asymmetry was
maintained to be less than 1 ppm by an active feedback loop. The
physical properties of the electron beam were monitored with high
precision by beam monitors. The sensitivity of the scattered flux
to fluctuations in the beam parameters was evaluated continuously
and accurately by modulating judiciously placed corrector coils
in the beam line leading to the hydrogen target. Separate data
runs under different conditions determined that target density
fluctuations were negligible for our kinematics.

\begin{figure}
\begin{center}
\includegraphics[width=3.5in]{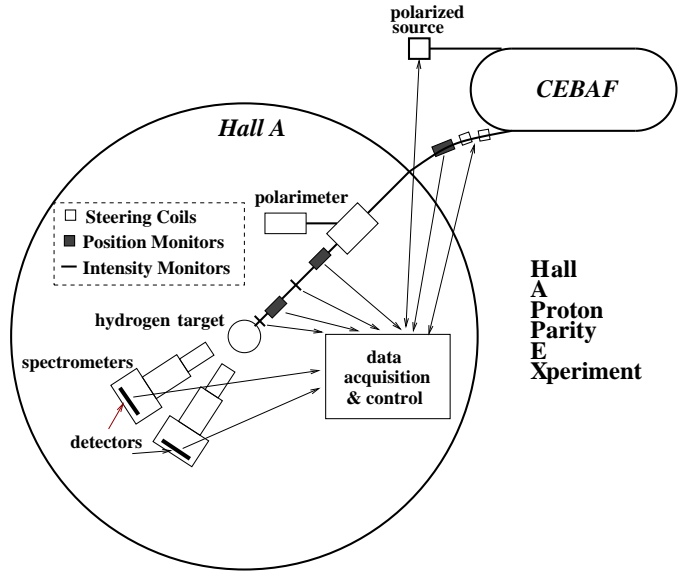}
\caption{Schematic Overview of the HAPPEX Experiment.}
\label{Fig1_KK_over} 
\end{center}
\end{figure}

The electron beam polarization was measured by three different
techniques at varying intervals: Mott scattering, M\o ller
scattering and Compton scattering. Figure~\ref{Fig1_KK_over} shows
a schematic diagram of the important components of the HAPPEX
experiment. In the following sections we elaborate on the above
considerations in detail.

\subsection{Polarized Electron Beam}
\label{sec:pol_beam}

\subsubsection{The Polarized Source and Laser Optics}\label{source_optics}
\label{sec:pol_source}

The longitudinally-polarized electron beam at Jefferson Lab is
produced by illuminating a GaAs photocathode with circularly
polarized laser light. For the 1998 run, a
``bulk" GaAs photocathode was used, which delivered a beam
intensity up to 100~$\mu$A with a polarization $\sim 38$\%. For
the 1999 run, a ``strained" GaAs
photocathode was used, which produced a beam intensity of $\sim
40~\mu$A with a polarization of $\sim 70$\%. This experiment was
the first to use a strained GaAs photocathode to measure a
parity-violating asymmetry in fixed-target electron scattering.

\begin{figure}
\begin{center}
\includegraphics[width=3.5in]{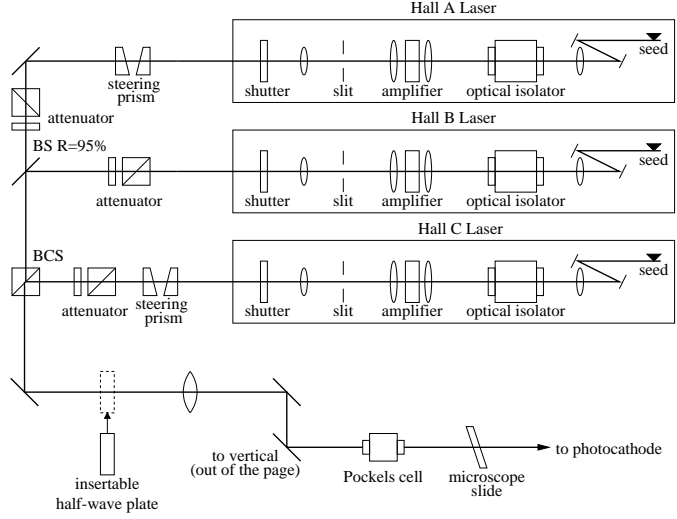}
\caption{Schematic diagram of the polarized source laser system
showing the seed laser, diode amplifier, and components
to steer, focus, and attenuate the beam and define 
its polarization.
BS: beam splitter.  BCS: beam combiner and
splitter.}
\label{Fig2_KK_laser} 
\end{center}
\end{figure}

The source laser system provided laser light with the 1497 MHz
microstructure of the JLab electron beam. A diagram of the source
laser system is shown in Fig.~\ref{Fig2_KK_laser}. There were three
lasers, which provided beams to the three different experimental
halls, allowing individual control of beam intensities. Each
laser system consisted of a gain-switched diode seed laser and a
single-pass diode optical amplifier. Each seed laser was driven
at 499 MHz, $120^\circ$ out of phase with the others. 
The seed laser light was focused into a
diode optical amplifier, whose respective drive current
controllers allowed precise control of the beam intensity into
each experimental hall.  

The Hall A laser light was guided through an attenuator
consisting of a remotely rotatable half-wave plate and a linear
polarizer, allowing a clean way to control the average beam
intensity without affecting the properties of the diode amplifier.
The three laser beams were then combined 
to produce the 1497 MHz pulse train.  
This beam was guided into a Pockels
cell, which is essentially a voltage controlled retardation
plate. The Pockels cell is configured to convert the linearly
polarized light to right- or left-circularly polarized light. The
polarity of the potential difference across the Pockels cell face
determines the handedness of the laser beam at the exit of the
cell.

Also shown in the figure are an insertable half-wave plate and a
microscope slide. The half-wave plate is aligned with its fast
axis at $45^\circ$ with respect to the linear polarization of the
laser beam, so that it rotates the incoming linear polarization by
$90^\circ$, which in turn switches the handedness of the circular
polarization exiting the Pockels cell. This was a powerful way of
reversing the sign of the experimental asymmetry with minimal
changes to the experiment. The microscope slide was used in
conjunction with the feedback scheme to control the
helicity-correlated intensity asymmetry. For the final phase of
running with the ``strained'' photocathode, an additional half-wave
plate was used downstream in order to control helicity-correlated
position fluctuations. These details will be discussed in
Sec.~\ref{sec:laser}.

\subsubsection{Helicity Control Electronics}
\label{sec:helicity_elect}

\begin{figure}
\begin{center}
\includegraphics[width=3.4in]{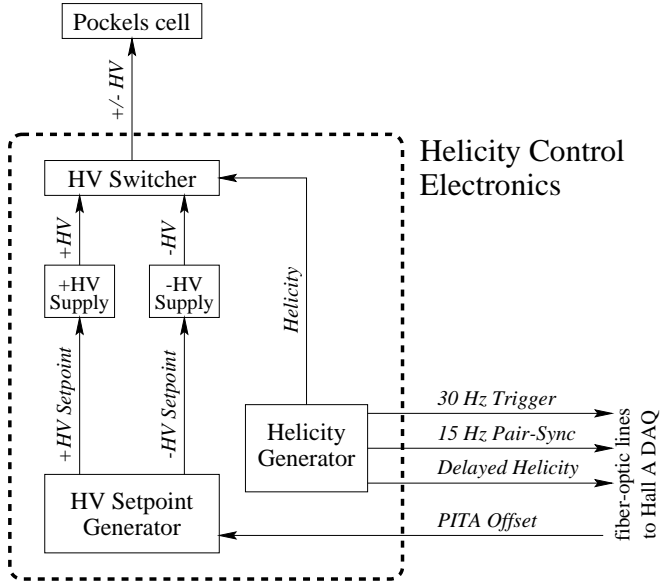}
\caption{Schematic diagram of the helicity control electronics.
The {\it Helicity} signal drives the high voltage on the Pockels cell.
The system is electrically isolated from the rest
of the lab (dashed box).}
\label{Fig3_KK_helicity} 
\end{center}
\end{figure}

A schematic diagram of the helicity control electronics is shown
in Fig.~\ref{Fig3_KK_helicity}. The high voltage (HV) switcher provided the
Pockels cell with positive or negative high voltage depending on
the state of a digital control signal. The programmable HV
supplies were set to correspond to $\pm\lambda/4$ retardation for
the Pockels cell, which was approximately $\pm 2.5$~kV. The net
effect of the system was that the helicity of the electron beam
depended on the state of the digital control signal, the {\it
Helicity} signal.

The {\it Helicity} signal was provided by the Helicity Generator,
a custom-built logic circuit which controlled the
helicity sequence and timing structure of the polarization of the
electron beam. As shown in the figure, the Helicity Generator also
produced three other control signals that provided principal
triggers to the data acquisition system.

The helicity of the beam was changed rapidly to minimize the
possibility that slow drifts might bias the measured asymmetry.
We chose to integrate over two
60 Hz cycles, setting the helicity every 33.33 ms. We denoted
each 33 ms period of constant helicity as a ``window". Sensitivity
to other, unforeseen frequencies was reduced by choosing the
helicity using a pseudo-random number generator sequence at 15
Hz. The helicity sequence was thus a train of ``window pairs": the
helicity of the first window was chosen pseudo-randomly, while
the second window was chosen to be the corresponding complement.

\begin{figure}
\begin{center}
\includegraphics[width=3.5in]{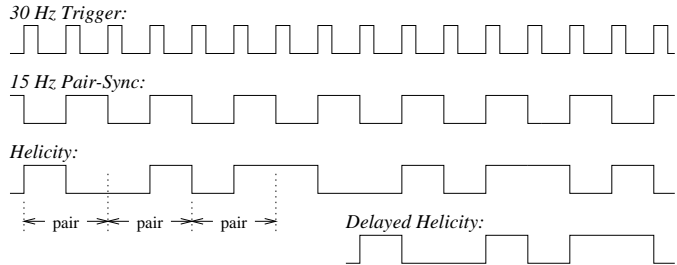}
\caption{Timing diagram of important control 
signals related to the beam helicity.}
\label{Fig4_KK_timing} 
\end{center}
\end{figure}

All signals to and from the Helicity Generator were routed via
fiberoptic cable, thus allowing complete ground isolation of the
helicity generator circuit from the rest of the experiment. This
was a powerful way to reduce the possibility of
helicity-correlated crosstalk and ground loops in the rest of the
experiment, which could lead to spurious asymmetries. As a
further precaution to suppress crosstalk, the true helicity of
each window was fed into an 8-bit shift register, and the helicity
that was transmitted to the data stream of the data acquisition
system arrived 8 windows later, breaking any correlation
with the helicity of the event. The timing
signals described above are depicted in
Fig.~\ref{Fig4_KK_timing}.

The system had one important input from the online analyzer of
the data acquisition system: a DC level that allowed for small
changes to the precise high voltages of the HV power supplies.
This signal, labeled as ``PITA offset" in
Fig.~\ref{Fig3_KK_helicity}, allowed for precise control of the
helicity-correlated intensity asymmetry of the electron beam, and
will be described in detail 
in section \ref{sec:pita_feedback}. In order to
preserve the ground isolation, the DC level was transmitted as a
frequency over fiberoptic cable and then converted to an analog
signal by a frequency-to-voltage converter.

\subsection{Beam Fluctuations}
\label{sec:beam_fluc}

The detected scattered flux $D$ in each spectrometer,
and the beam current $I$,
were measured independently for every window.  From these 
we obtained the normalized flux $d_i \equiv D_i/I_i$ and 
the cross section asymmetry $(\asy_d)_i$ for the
$i$th window pair. The raw asymmetry was then obtained by
appropriate averaging of $N$ measurements:
\begin{eqnarray} \label{eq:asyraw}
(\asy_d)_i & \equiv &
{\left(\frac{d^+-d^-}
{d^++d^-}\right)}_i 
\equiv {\left(\frac{\Delta d}
{2d}\right)}_i 
\nonumber \\
\quad \delta(\asy_d) & = &
\sigma(\asy_d)/\sqrt{N}. \label{eqn:pulsepair}
\end{eqnarray}
where $+$ and $-$ denote the two helicity states in a pair.
One goal of the experimental design was that $\sigma(\asy_d)$ should be
dominated by the counting statistics in the scattered flux,
greatly minimizing potential problems in the averaging procedure.
As will be seen in Section \ref{sec:rsh_rawasy}, this goal was met.  
This is a result of the extraordinary characteristics of the electron beam
and the associated beam instrumentation, which we discuss in this
section. 

The RMS noise in the asymmetry $\sigma(\asy_d)$ was 
found to be $3.8\times 10^{-3}$ at
a beam current of approximately 100$~\mu$A, which implied that
roughly 70,000 electrons were recorded in the detectors during
each beam window for a total rate of 2 MHz which was the
expected rate and consistent with the rate extrapolated
from lower currents.
Since the experimental cross section is a function of the
physical parameters of the beam, fluctuations in these
parameters may contribute significantly to $\sigma(\asy_d)$. All
electronic signals in the experiments are designed so that
electronic noise is small compared to $\sigma(\asy_d)$.

There are two key parameters for each experimentally measured
quantity $M$, such as detector rate, beam intensity {\em etc.} The first
is $\sigma(\Delta M)$, the size of the relative window pair-to-window pair
fluctuations in $\Delta M\equiv M_{+} - M_{-}$, which is affected by
real fluctuations in the electron flux. The second is
$\delta(\Delta M)$, the relative accuracy with which the window
pair differences in $M$ can be measured compared to the true
value, which is dominated by instrumentation noise.

If $\sigma(\Delta M)$ is large enough, it might mean that there
are non-statistical contributions to $\sigma(\asy_d)$ so that the
latter is no longer dominated by counting statistics. In this
case, it is crucial that $\delta(\Delta M)\ll\sigma(\Delta M)$ so
that window pair to window pair corrections for the fluctuations
in $\Delta M$ can be made.

\subsubsection{Random Fluctuations}
\label{sec:random_fluc}

As stated in \ref{sec:beam_fluc}, we desire that
$\sigma(\asy_d)$ be dominated by counting statistics.
An example of possible non-statistical contributions
is window-to-window relative beam intensity
fluctuations, $\sigma(A(I)) \equiv \sigma(\Delta I/2I)$,
which were observed to vary between $2\times 10^{-4}$ and
$2\times 10^{-3}$, depending on the quality of the laser and the
beam tune. This is remarkable and a unique feature of the beam at
Jefferson lab, since $\sigma(\asy_I)<\sigma(\asy_d)$. Nevertheless, the
detector-intensity correlation can be exploited to remove the
dependence of beam charge fluctuations on the measured asymmetry:
\begin{equation}
(\asy_d)_i \simeq
\left(\frac{\Delta D}{2D} - \frac{\Delta I}{2I}\right)_i \equiv
(\asy_D - \asy_I)_i.
\end{equation}
(This is equation \ref{eq:asyraw}
to first order.)

Similarly, $\sigma(\asy_d)$ might be affected by random beam
fluctuations in energy, position and angle. The corrections can be
parameterized as follows:
\begin{equation}
(\acorr_d)_i = \left(\frac{\Delta D}{2D} - \frac{\Delta I}{2I}\right)_i
-\sum_j{\left( {\alpha_j(\Delta X_j)_i}\right)}.
\end{equation}
Here, $X_j$ are beam parameters such as energy, position and
angle and $\alpha_j \equiv \partial D/\partial X_j$ are
coefficients that depend on the kinematics of the specific
reaction being studied, as well as the detailed spectrometer and
detector geometry of the experiment.

By judicious choices of beam position monitoring devices (BPMs)
and their respective locations, several measurements of beam
position can be made from which the average relative energy,
position, and angle of approach of each ensemble of electrons in a
helicity window on target can be inferred. One can then write
\begin{equation} 
(\acorr_d)_i = \left(\frac{\Delta D}{2D} - \frac{\Delta I}{2I}\right)_i
-\sum_j{\left({\beta_j(\Delta M_j)_i}\right)}.
\end{equation}
Here $M_i$ are a set of 5 BPMs that span the parameter space of
energy, position, and angle on target, and $\beta_i \equiv \partial
D/\partial M_i$. It is worth noting that this approach of making
corrections window by window automatically accounts for occasional
random instabilities in the accelerator (such as klystron
failures) that are characteristic of normal running conditions.

During HAPPEX running, we found that $\sigma(\Delta M_j)$ varied
between 1 and 10 $\mu$m and $\sigma (\asy_E)$ was typically less
than $10^{-5}$. These fluctuations were small enough that their
impact on $\sigma (\asy_d)$ was negligible. Indeed, we believe that a
significant contribution to the fluctuations in each monitor
difference $\Delta M$ was the intrinsic measurement precision
$\delta(\Delta M_i)$. We elaborate on this in section
~\ref{sec:beam_mon},
where we discuss the monitoring instrumentation.

Another important consideration is the accuracy with which the
coefficients $\beta_i$ are measured. As mentioned earlier, these
coefficients were evaluated using beam modulation, and will be
discussed in Sect.~\ref{bmod}.

\subsubsection{Beam Monitoring}
\label{sec:beam_mon}

The above discussion regarding measurement accuracy and its impact
on $\sigma(\asy_d)$ is particularly relevant in the monitoring of the
electron beam properties such as beam intensity, trajectory and
energy.

\begin{figure}[tb]
\begin{center}
\includegraphics[width=3.6in]{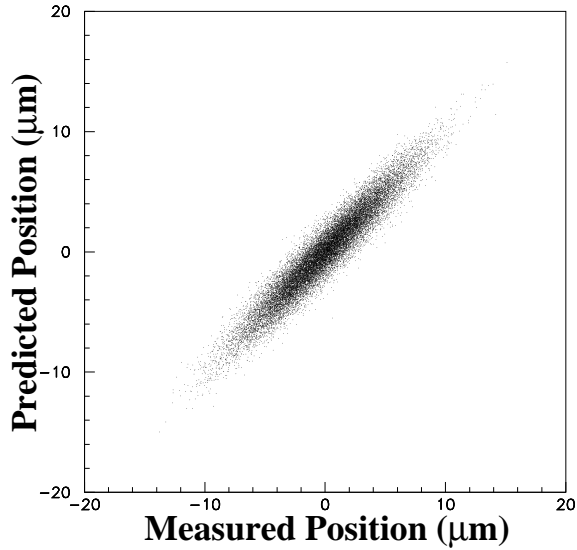}
\caption{Window-to-window beam jitter as measured
by a BPM is 
plotted along the $x$ axis. On the $y$ axis is plotted the beam
position as predicted by nearby BPMs. The residuals are smaller
than 1 $\mu$m.}
\label{fig5:jlabcorr}
\end{center}
\end{figure}

At Jefferson Lab, the beam position is measured by ``stripline"
monitors~\cite{stripline}, each of which consists of a set of four plates 
placed
symmetrically around the beam pipe. The plates act as antennae
that provide a signal (modulated by the microwave structure of the
electron beam) proportional to the beam position as well as
intensity. Figure~\ref{fig5:jlabcorr} shows the correlation between
the measured position at a BPM near the target compared with the
predicted position using neighboring BPMs for a beam current of
100 $\mu$A ($2\times 10^{13}$ electrons per window). A precision
for $\delta(\Delta X_i)$ close to 1 $\mu$m was obtained for the
average beam position for a beam window containing $2\times
10^{13}$ electrons.

To measure the beam intensity, microwave cavity BCMs have been
developed at Jefferson Lab~\cite{A-NIM}. The precision $\delta(\asy_I)$ that has
been achieved for a 30 ms beam window at 100 $\mu$A is $4\times
10^{-5}$. This superior resolution is a 
result of good radiofrequency (rf)
instrumentation as well as a high resolution 16-bit ADC, which
will be discussed in section \ref{sec:daq}.

The absolute calibration of the beam current 
was performed with a parametric
current transformer, the `Unser monitor'~\cite{Unser}.
Although the absolute calibration was not important
for HAPPEX, the Unser monitor was useful to establish
the pedestals and understand the linearity of the
cavity current monitors.

\subsubsection{Systematic Fluctuations}
\label{sec:syst_fluc}

Assuming that $\sigma(\asy_d)$ has negligible contributions from
window-to-window beam fluctuations and instrumentation noise,
there is still the possibility that there are 
helicity-correlated systematic effects at the sub-ppm
level. If one considers the cumulative corrected asymmetry
$\acorr_d$ over many window pairs, one can write
\begin{eqnarray}
\acorr_d & \equiv & \langle (\acorr_d)_i\rangle = 
\nonumber \\
&  & 
\left\langle\left(\frac{\Delta D}{2D}\right)_i\right\rangle -
\left\langle\left(\frac{\Delta I}{2I}\right)_i\right\rangle
- \sum_j {\beta_j\left\langle{(\Delta M_j)_i}\right\rangle} 
\nonumber \\
&=& \asy_D -  \asy_I - \sum_j \asy_{Mj}.
\end{eqnarray}

For most of the running conditions during data collection,
$\acorr_d\simeq \asy_D\simeq 10$~ppm, which meant that all
corrections were negligible. The cumulative average for $\asy_I$ was
maintained below 0.1 ppm. For $\asy_{Mj}$, the cumulative averages
were found to be below 0.1 ppm during the run with the ``bulk'' GaAs
photocathode. This resulted from the fact that the accelerator
damped out position fluctuations produced at the source by a large
factor (section \ref{sec:adiabatic_damp}). 
The averaged position differences 
on target were kept below 10 nm.

However, during data collection with ``strained'' GaAs, position
differences as large as several $\mu$m were observed in the
electron beam at a point in the accelerator where the beam energy is
5 MeV. Continuous adjustment of the circular polarization of the
laser beam was required to reduce the differences to about 0.5
$\mu$m. This resulted in observed position differences on target
ranging from 10 nm to 100 nm, which in turn resulted in $\asy_{Mj}$
in the range from 0.1 to 1 ppm.

The control of the asymmetry corrections within the aforementioned
constraints was one of the central challenges during data
collection. A variety of feedback techniques on the laser and
electron beam properties were employed in order to accomplish
this; these methods are discussed in Sec.~\ref{sec:laser}.

\subsection[Target]{Target}
\label{sec:target}

The Hall A cryogenic target system 
\cite{A-NIM} was used for this experiment. The
target system consists of three separate cryogenic target loops in 
an evacuated scattering chamber, along with subsystems for cooling,
temperature and pressure monitoring, target motion, gas-handling,
controls, and a solid and dummy target ladder.  Of the three cryogenic
loops (hydrogen, deuterium, and helium), only the hydrogen loop was
used in this experiment and will be described here.  The hydrogen loop
has two separate target cells, of 15 cm and 4 cm in length,
respectively; only the 15 cm cell was used here.

The liquid hydrogen loop was operated at a temperature of 19 K and a
pressure of $\sim 26$ psia, leading to a density of about 0.0723
g/cm$^3$. The Al-walled target cells were 6.48 cm in diameter, and
were oriented horizontally, along the beam direction. The upstream
window thickness was 0.071 mm, the downstream window thickness was
0.094 mm, and the side wall thickness was 0.18 mm. Also mounted on the
target ladder were solid thin targets of carbon, and aluminum dummy
target cells, for use in background and spectrometer studies.

The target was mounted in a cylindrical scattering chamber of 104 cm
diameter, centered on the pivot for the spectrometers. The scattering
chamber was maintained under a $10^{-6}$ torr vacuum. The
spectrometers view exit windows in the scattering chamber that were
made of 0.406 mm thick Al foil.

The target coolant, $^4$He gas at 15 K, was provided by the End
Station Refrigerator (ESR), with a flow rate controlled using
Joule-Thompson valves, which could be adjusted either locally or
remotely. At the beam currents used here (up to 100 $\mu $A) the beam
heating load was of order 600 W.  Including the heating from the
target circulation fans, and a small ($\sim$ 45 W) target heater, the
load could reach 1 kW, which could be adequately supplied by the
ESR. In addition to the 45 W target heater, used in a feedback system
in order to stabilize the target temperature, a high power heater (up
to 1 kW) was automatically switched on when the beam dropped out
suddenly.  This target has achieved a luminosity of $5 \times 10^{38}$
${\rm cm}^{-2} {\rm s}^{-1}$.

The target temperature was monitored continuously using 1)~radiation
hard semiconductor-based sensors, Lakeshore CERNOX \cite{gar_CERNOX},
2)~Allen-Bradley resistive sensors \cite{gar_ALLEN_BRADLEY}, and 
3)~vapor-pressure transducers.  The temperature control system was
computer controlled using a PID (proportion, integral and derivative)
feedback system. The control system was based on 
the EPICS \cite{gar_EPICS} system.

\begin{figure}
\begin{center}
\includegraphics[width=3.5in]{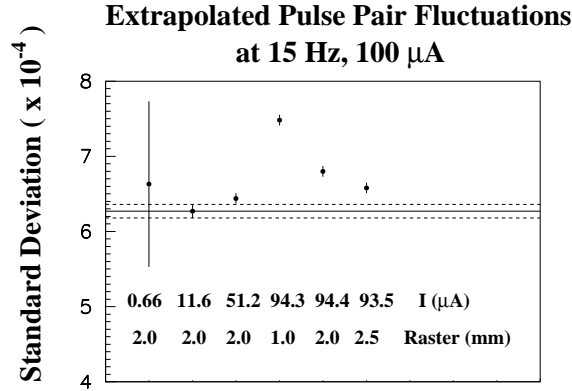}
\caption{Noise in pulse pair asymmetries
vs. beam current and raster size.
The width of asymmetries is extrapolated to
15 Hz, 100 $\mu$A to check if it is consistent
(within the dashed bars) with expectation.
A value above this indicates target 
density fluctuations that increase the noise.
For reasonably large raster patterns we saw
little noise at 94 $\mu$A.}
\label{fig6_gar_boil_kk}
\end{center}
\end{figure}

The normal electron beam spot size of about 50 $\mu$m is small enough
to potentially damage the target cells at full beam current, as well
as to cause local boiling in the target even at reduced currents. A
beam rastering system was used to distribute the heat load throughout
the target cell. The beam was rastered at 20 kHz by two sets of
steering magnets 23 m upstream of the target. These magnets deflected
the beam by up to $\pm 2.5$ mm in $x$ and $y$ at
the target.  For the 1998 run, a rectangular raster pattern
was used, while for the 1999 run a helical pattern was adopted, 
which
provided a more uniform distribution of heat load.  Local target
boiling would manifest itself as an increase in fluctuations in the
measured scattering rate, which would lead to an increase in the
standard deviation of the pulse-pair asymmetries in the data, above
that expected from counting statistics. Studies of the pulse-pair
asymmetries for various beam currents and raster sizes were performed,
at a lower $Q^2$ and thus at a higher scattering rate. Figure
\ref{fig6_gar_boil_kk} shows the standard deviation of the 
pulse-pair asymmetries, extrapolated to full current values, for
various beam currents and raster sizes. A significant increase over
pure counting statistics, indicating local boiling effects, was
observed only for the combination of a small raster (1.0 mm) size and
large beam current (94 $\mu$A).  During the experiment we used
larger raster sizes for which there was little boiling noise.

\subsection {High Resolution Spectrometers in JLab Hall A}
\label{sec:hrs}
\par

The Hall A high resolution spectrometers (HRS) at
Jefferson Lab consist of a pair
of identical spectrometers of QQDQ design,
together with detectors for detecting the 
scattered particles \cite{A-NIM}.
For HAPPEX, the spectrometer and their
standard detector package served the following 
purposes:
1)~to suppress background from inelastics
and low-energy secondaries;
2)~to study the backgrounds in separate runs 
at or near the HAPPEX kinematics;
3)~to measure the momentum transfer $Q^2$;
4)~to measure and monitor
the attenuation in the HAPPEX detector through the use of tracking;
and 
5)~to measure the detector amplitude weighting
factors for fine bins in $Q^2$ (section \ref{sec:qsq}).

\par 
The spectrometers are designed to have a
large acceptance with excellent
resolution and absolute accuracy
in the reconstructed
four--vectors of the events and, 
of less relevance for HAPPEX,
precise normalization of the cross section.
The momentum resolution is
necessary for HAPPEX 
to separate the elastically
scattered electrons from
inelastic background,
thus allowing the integrating technique.
To measure $Q^2$ with sufficient accuracy
requires good knowledge of the transfer matrix for
the spectrometer to reconstruct events at the
scattering point, as well as
good pointing accuracy for the location of the
spectrometers and precise measurements 
of beam position and angle.
The achieved properties of the HRS
are listed in Table ~\ref{bob_tableprops}.
The spectrometer detector package include
scintillators for triggering and vertical
drift chambers for reconstruction of 
particle trajectories, in addition to 
\v{C}erenkov and lead glass detectors 
for particle identification.  
The trigger is formed in programmable
CAMAC electronics and is configurable
to include various combinations of
the scintillator and other
detectors including the HAPPEX detector 
(see section ~\ref{sec:fp_det}).

\begin{table}
\centering
\caption{\bf Properties of the Hall A Spectrometers}
\begin{tabular}{|l|l|}  
Magnet Configuration  & QQDQ \\
Luminosity & $ 10^{38}   {\rm cm}^{-2} {\rm sec}^{-1}$ \\
Momentum Range (spectrometer 1) &  0.2 - 4.3 GeV/c \\
Momentum Range (spectrometer 2) &  0.2 - 3.2 GeV/c \\
Bend Angle  &  $45^{\circ}$ \\
Optical Length & 23.4 m \\
Dispersion &  12.4 cm/\% \\
Momentum Acceptance    &   $\pm$ 4.5\% \\
Momentum Resolution (FWHM)   &   2$\times 10^{-4}$\\
Solid Angle Acceptance  &   6 msr \\
Horizontal Angle Acceptance & $\pm$ 28 mrad \\
Vertical Angle Acceptance   & $\pm$ 60 mrad \\
Target Length Acceptance ($90^{\circ}$) & 10 cm \\
Transverse Position Resolution (FWHM)  & 1.5 mm \\
Missing Energy Resolution (FWHM) & 1.3 MeV \\
\end{tabular}
\label{bob_tableprops}
\end{table}

\subsection[Detector]{Focal Plane Detector}
\label{sec:fp_det}

\begin{figure}
{\includegraphics[width=2.5in,angle=-90]{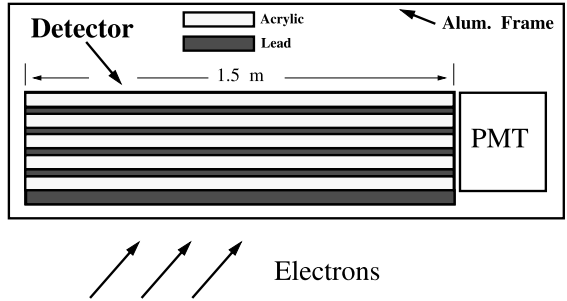}}
\caption{Schematic of the focal plane detector.  The
scattered electrons strike a lead-acrylic shower
counter whose light is collected by a PMT and 
integrated over a helicity period. }
\label{fig7_dsa_detector}
\end{figure}

  A total absorption shower counter was located in the focal plane of 
each spectrometer to detect the elastically
scattered electrons. These detectors were based on a layered
lead-acrylic geometry. \v{C}erenkov light in the shower propagates along
the acrylic and is detected at one end using a 
single photomultiplier tube (PMT); see Fig. ~\ref{fig7_dsa_detector}. 

  These simple focal plane detectors were chosen over, for example,
lead glass, because of their superior resistance to radiation damage.
The radiation dose expected per detector was approximately 40 Gy 
in a 30 day data-taking run, which would cause significant 
decrease in optical transmission for a lead glass detector. Acrylic is 
significantly less susceptible to radiation damage. The insensitivity 
of such a detector to low-energy backgrounds was also an important design
criterion. 

   The detectors were made up of 4 layers of 6.4 mm thick lead
sheets sandwiched between 5 layers of 1.27 cm thick acrylic
(Bicron BC-800 UVT Lucite). 
Each layer of acrylic was wrapped with a Teflon sheet, which
does not adhere to the surface, thereby preserving
internal reflection from the acrylic-air interface.
The incident electrons first
passed through a 1.9 cm Teflon spacer and 2 layers of 6.4 mm
lead sheets acting as a pre-radiator. The segmentation was 
chosen in order to provide a sufficiently good energy resolution
(15\% $\sigma$) with the use of commercially available thicknesses
of acrylic while maintaining mechanical simplicity. The detector
energy resolution affects the error on the physics asymmetry 
via 
\begin{eqnarray}
\delta A \sim \frac{1}{\sqrt{N}}~\sqrt{ 1 + \left( 
\frac{\Delta E}{\left< E \right>} \right) ^2 }
\end{eqnarray}
where
$N$ is the number of window pairs,
$\Delta E$ is the energy resolution of the detector, and 
$\left< E \right>$ is the average detected energy. The width (10 cm)
and length (150 cm) of the sandwich stack was chosen in order to contain
the entire image of the elastically scattered electrons in the focal
plane, as well as much of the radiative tail, and yet not detect events
from inelastic scattering. The width of the distribution of elastic
events on the focal plane was 3 cm, so edge effects were small. 

  The detector sandwich was viewed at one end by a single 12.7 cm
diameter Burle 8854 photomultiplier tube. A pair of blue LEDs was
mounted in the middle acrylic layer, at the opposite end from the PMT,
for use in study of detector linearity and attenuation. Tests
using the LEDs indicated that the non-linearity of the detector was less
than 1.5\% at typical operating voltages.

  Bench tests of the detectors using cosmic rays showed that the 
signal output was a strong function of the incident particle's 
position along the detector's length, due primarily to bulk absorption 
of light in the acrylic. While the Bicron BC-800 UVT Lucite
acrylic is transparent to wavelengths shorter 
than for ordinary acrylic, it
has a strong attenuation for wavelengths shorter than about 350 nm. 
Given that the PMT used has significant sensitivity down to 250 nm, 
and given the short wavelengths of typical \v{C}erenkov light, the bulk
attenuation in the acrylic led to a measured decrease in the light output
of 50\%$/{\rm m}$. To decrease this attenuation, a single sheet of Plexiglass
was installed directly in front of the PMT to filter out the UV
light. After installation of this filter the dependence of light
output on position along the detector was reduced to 9\%$/{\rm m}$,
at the cost of a reduction in the total gain, which was acceptable
for this experiment. 

\begin{figure}
{\epsfxsize=0.50\textwidth \epsfbox{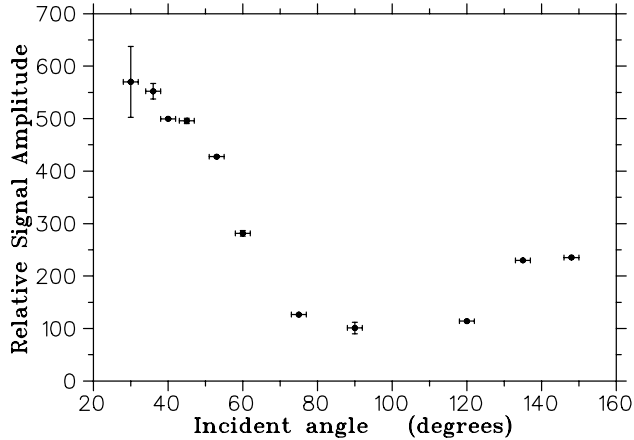}}
\caption{Focal plane detector response versus angle of incident particle
with respect to the long axis of 
the focal plane detector, measured using cosmic rays.
}
\label{fig8_dsa_detector_angle}
\end{figure}

   The detector, as expected, also exhibited a strong sensitivity
to the angle of the incident particles, with a maximum output when
the angle was such that part of the \v{C}erenkov cone pointed directly
at the PMT (see Fig.~\ref{fig8_dsa_detector_angle}). This angular
sensitivity was an advantage. Since the elastic electrons
arrive at the focal plane at well-known angles, the detector
orientation can then be adjusted to maximize the sensitivity to 
the elastically scattered events while minimizing the 
sensitivity to backgrounds that arrive at other angles. 

Due to the optics of the spectrometer, the incident angle of the 
elastically-scattered electrons varies with their position along the 
detector's length. Thus the crossing angle sensitivity leads 
to an additional variation of the detector's response with
position along the detector.  The total effect of variation along
the detector position was measured periodically
during data-taking and was $(17.3 \pm 0.5)\% /{\rm m}$. 
This value was stable during the run,
indicating no significant degradation of the optical
properties of the detector due to radiation damage. 

  The detector was mounted in a light-tight aluminum box with 1 cm
thick walls and was supported over the vertical drift chambers in a
frame that allowed adjustment of the horizontal location, as well as
the pitch, roll and yaw angle of the detector. The detector's strong
sensitivity to the incident angle of the incoming electron
necessitated the ability to orient the detector precisely.
All material used in the detector box and
support frame near the active region was chosen to be non-ferric in
order to reduce the possibility of false asymmetries due to M{\o}ller
scattering of electrons off magnetized material.
More information on the detectors 
is available in \cite{Gary}.

\subsection{Data acquisition and Custom ADCs}
\label{sec:daq}

Signals from the various detectors and monitors 
are integrated and digitized by custom-built 
VME integrating ADCs in a data acquisition system (DAQ) based on 
the CODA DAQ package \cite{CODA} 
triggered at nominally 30 Hz,
synchronized to the end of each helicity
window. In addition to these ADCs, the DAQ reads
scalers and input/output registers which count
various information such as helicity pulses.

\begin{figure*}
\includegraphics[width=6.5in]{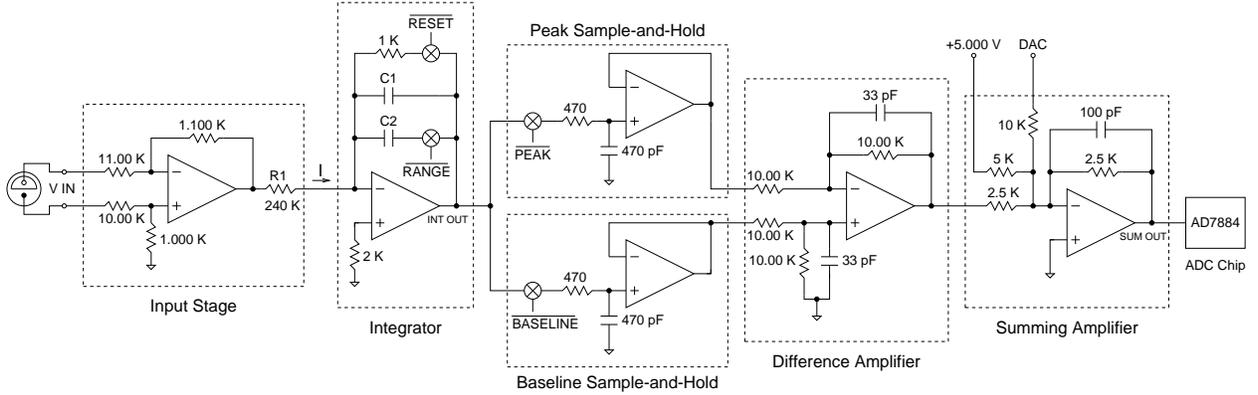}
\caption{\label{fig9_bob_adc}
Circuit diagram of one channel of the custom 16 bit
integrating ADC.}
\end{figure*}

The custom ADCs integrate the data over 
most of the 33 msec helicity pulse.  The first 0.5 msec of the pulse
is blanked off to remove instabilities due to the switching of
HV on the Pockels cell which controls the beam polarization.
The ADCs are designed to achieve high resolution (16 bits) with
low differential nonlinearity ($\le 0.1$\%).  Each ADC channel
(Fig. ~\ref{fig9_bob_adc}) consists of an input amplifier,
an integrating circuit, two sample-and-hold circuits,
a difference amplifier, a summing circuit, and a 16 bit
ADC chip. The input amplifier converts the input voltage
to a scaled current which is integrated in the next 
stage; for current signals such as PMTs this amplifier
stage is bypassed and the signal is integrated directly.  
The integrator output is sampled and held once 700 $\mu$sec
after the beginning of the helicity pulse, and again
32 msec later near the end of the pulse.  The
difference between these two is the integrated 
result.  The circuit components were chosen
to emphasize low noise
at the expense of speed.
Noise widths of 3 ADC channels FWHM
have been achieved.  

To achieve the nonlinearity specification,
a pseudorandom DAC voltage (``DAC noise'') is added
to the integrated result prior to digitization
by the ADC, then subtracted later in analysis.
DAC noise solves a problem of nonlinearity
that arises generally in the digitization of data
which leads to a systematic error in the asymmetry
that can be estimated as follows.
Consider a signal of average value $S$ (ADC channels)
and RMS width $\sigma$, and let the deviation from
ideal linear response be denoted $D$ which 
is typically the least count bit.
Denote the helicity correlated asymmetry
in the signal by $A$.
Then if $A S \ll \sigma$ 
the relative systematic error in the asymmetry
will be $dA/A \approx K D / \sigma$ with
$K \simeq 1$.
(For Gaussian signals $K = 2 / \sqrt{2 \pi}$.) 
Thus, the DAC noise smears the data 
over many ADC channels, which
reduces systematic errors from bit resolution.  
Since the noise is later subtracted it does 
not increase the statistical error.

The data acquisition software is 
based on the {\sc CODA} 1.4 package~\cite{CODA}. 
The trigger interrupt service routine in the VME controller assembles the
following data into an event record: ADC data, ADC flags, scaler data,
trigger controller data, VME flags, beam modulation data, and Pockels cell
high voltage offsets.  The ADC data include the digitized ADC outputs
and the value of the DAC noise that had been added
to the ADC signal.
The ADC flags govern
various options for each ADC board.  Data from the trigger controller
include a flag indicating the helicity of the first
window of the pair, and a flag
indicating whether the window is the first or the
second of a pair.  As described in section \ref{sec:helicity_elect},
the helicity flag is delayed at the polarized source 
and applies to the eighth window
preceding the one with which it is collected.
The VME flags govern
various options for the VME controller.  Beam modulation
data describe the state of the beam modulation system including the
object being modulated, the size of its offset, and flags indicating
whether the object's state was stable during the event.

The complete event record is then sent over
the network to
the data acquisition workstation, where the data files 
are written to disk
and are processed by an online analyzer.  

A separate process on the VME controller is able to handle requests
via a TCP/IP socket to change or report various system parameters,
including the ADC and VME flags, beam intensity feedback parameters,
and the Pockels cell high voltage offset, and to enable or disable
the beam modulation system.

The online analyzer verifies the integrity of the data, determines
where cuts due to beam off or computer dead time are required,
associates the delayed helicity information with its proper window,
groups windows into opposite-helicity pairs, subtracts DAC noise from
each ADC signal, computes $x$ and $y$ positions from the BPM data, and
packages the data into files in the PAW ntuple format for further analysis.

Another function of the online analyzer is to handle beam intensity
feedback.  Beam intensity asymmetries are averaged over a user-defined
interval, typically 2500 pairs, termed a ``minirun''.  At the end of
each minirun the change to the Pockels cell high voltage offset
required to null the observed intensity asymmetry is computed.  The analyzer
then issues a request for the VME controller to make the appropriate
change to the offset.

\subsection{Polarimetry}
\label{sec:pol_intro}
The experimental asymmetry $\aexp$ is related 
to the corrected asymmetry by 
\begin{equation}
\aexp=\acorr_d/P_e
\label{eq:a_exp}
\end{equation}
where $P_e$ is the beam polarization. 
Three beam polarimetry techniques were available at JLab for the
HAPPEX experiment: A Mott polarimeter in the injector, 
and both a M{\o}ller
and a Compton polarimeter in the experimental hall.

\subsubsection{Mott Polarimeter}
\label{sec:mpt_exp_meth}
A Mott polarimeter \cite{Price}
is located near the injector to the first linac, where the
electrons have reached 5 MeV in energy. Mott polarimetry is based on the 
scattering of polarized electrons from unpolarized high-Z nuclei. The
spin-orbit interaction of the electron's spin with the magnetic field it sees due to its
motion relative to the nucleus causes a differential cross section

\begin{equation}\label{mottcrosssection}
\sigma(\theta) = I(\theta)
\left[1 + S(\theta){\vec P_e}\cdot \widehat{n}\right]\;\;,
\end{equation}
where $S(\theta)$,  known as the Sherman function,  is the analyzing power of the polarimeter,
and 
$I(\theta)$ is the spin-averaged scattered intensity
\begin{equation}
I(\theta)= \frac{Z^2 e^4}{4m^2\beta^4 c^4 \sin^4(\theta/2)} \left[1 - \beta^2\sin^2(\theta/2)\right](1 - \beta^2) \;\;\; .
\end{equation}
 The unit vector ${\widehat n}$ is normal to the scattering plane, defined by
$
\widehat n = (\vec k \times \vec k')/\vert \vec k \times \vec k' \vert
$
where $\vec k$ and $\vec k'$ are the electron's momentum before and
after scattering, respectively. Thus $\sigma(\theta)$ depends on the electron beam polarization $P_e$.
Defining an asymmetry
\begin{equation}
A(\theta) = \frac{N_L - N_R}{N_L + N_R} \; ,
\end{equation}
where $N_L$ and $N_R$ are the number of electrons scattered to the left and
right, respectively, we have 
\begin{equation}
A(\theta) = P_e \; S(\theta) \;,
\end{equation}
and so knowledge of the Sherman function $S(\theta)$
allows $P_e$ to be extracted from the measured asymmetry. 

The 5 MeV Mott polarimeter employs a 0.1 $\mu$m gold foil target, and
four identical plastic scintillator total-energy detectors, located
symmetrically around the beam line at a scattering angle of
$172^{\circ}$, the maximum of the analyzing power. This configuration
allows a simultaneous measurement of the two components of polarization
transverse to the beam momentum direction. A Wien filter upstream of the
polarimeter is used to rotate the electron's spin 
from longitudinal to transverse
polarization for the Mott measurement. 
Multiple scattering in the foil target
leads to substantial uncertainty in the analyzing
power which is evaluated by measurements for
a range of target foil thicknesses and an extrapolation
to zero thickness.  It is believed \cite{Sinclair1}
that the theoretically
calculated single-atom analyzing power 
(Sherman function) is the correct number
to use for zero target thickness extrapolation.
The primary systematic errors of the device
were the extrapolation to 
zero target foil thickness (5\% relative) 
and background subtraction (3\%) \cite{Price},
see section \ref{sec:pol_1998}.

\subsubsection{M{\o}ller Polarimeter}
\label{sec:moller_method}

A M{\o}ller polarimeter measures the beam polarization
via measuring the asymmetry in
$\vec e, \vec e$ scattering, which depends on the
beam and target polarizations $P^{\rm beam}$ and $P^{\rm target}$, 
as well as on the
analyzing power $\ath_m$ of M{\o}ller scattering:

\begin{equation}
\label{moller_asy}
         \aexp_m = 
            \sum_{i=X,Y,Z} (\ath_{mi}\cdot{}{P}^{\rm targ}_{i}\cdot{}{P}^{\rm beam}_{i}) ,
\end{equation}
where $i = X,Y,Z$ defines the projections of the polarizations
($Z$ is parallel to the beam, while $X-Z$ is the scattering plane).
The analyzing powers $\ath_{mi}$ depend on 
the scattering angle $\theta_{\rm CM}$ in the
center-of-mass (CM) frame 
and are calculable in QED.
The longitudinal analyzing power is
\begin{equation}
\label{eq:moller_apower}
\ath_{mZ} = - \frac{ \sin^2 \theta_{\rm CM} 
( 7 + \cos^2 \theta_{\rm CM}) }
{ {(3 + \cos^2 \theta_{\rm CM})}^2 }.
\end{equation}

The absolute values of $\ath_{mZ}$ reach the maximum of 7/9 
at $\theta_{\rm CM}=90^{\circ}$. 
At this angle the transverse analyzing powers are $\ath_{mX}=-\ath_{mY}=\ath_{mZ}/7$.

The polarimeter target is a ferromagnetic foil
magnetized in a magnetic field of 24 mT along its plane.
The target foil can be oriented at various angles
in the horizontal plane providing both
longitudinal and transverse polarization
measurements.  The asymmetry is measured
at two target angles ($\pm 20^{\circ}$) 
and the average taken, which cancels
transverse contributions and reduces
the uncertainties of target angle measurements.
At a given target angle two sets of measurements
with oppositely signed target polarization
are made which cancels some false asymmetries
such as beam current asymmetries.  The target
polarization was (7.95 $\pm$ 0.24)\%.

\begin{figure}[bht]
    \begin{center}
        \includegraphics*[scale=0.36,angle=-90]{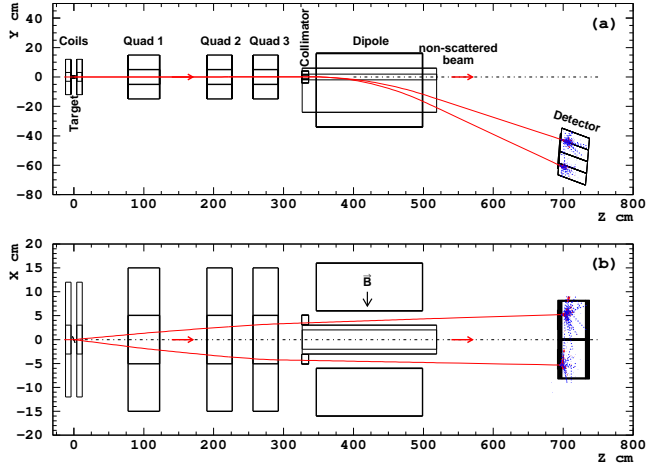}
    \end{center}
    \caption{Layout of the Hall A M{\o}ller polarimeter.
            }
    \label{fig10:moller_layout}
 \end{figure}

The M{\o}ller-scattered electrons were
detected in a magnetic spectrometer (see Fig. ~\ref{fig10:moller_layout})
consisting
of three quadrupoles and a dipole \cite{A-NIM}.

The spectrometer selects electrons in a
bite of $75^{\circ} \le \theta_{\rm CM} \le
105^{\circ}$ and $-5^{\circ} \le \phi_{\rm CM}
\le 5^{\circ}$ where $\phi_{\rm CM}$ is
the azimuthal angle.  The detector consists
of lead-glass calorimeter modules in two 
arms to detect the electrons in coincidence.
More details about the M{\o}ller polarimeter
are published in \cite{A-NIM}.  The total
systematic error that can be achieved is
3.2\% which is dominated by uncertainty in
the foil polarization.

\subsubsection{Compton Polarimeter}
\label{sec:cpt_exp_meth}
\begin{figure*}
\includegraphics[width=6.5in]{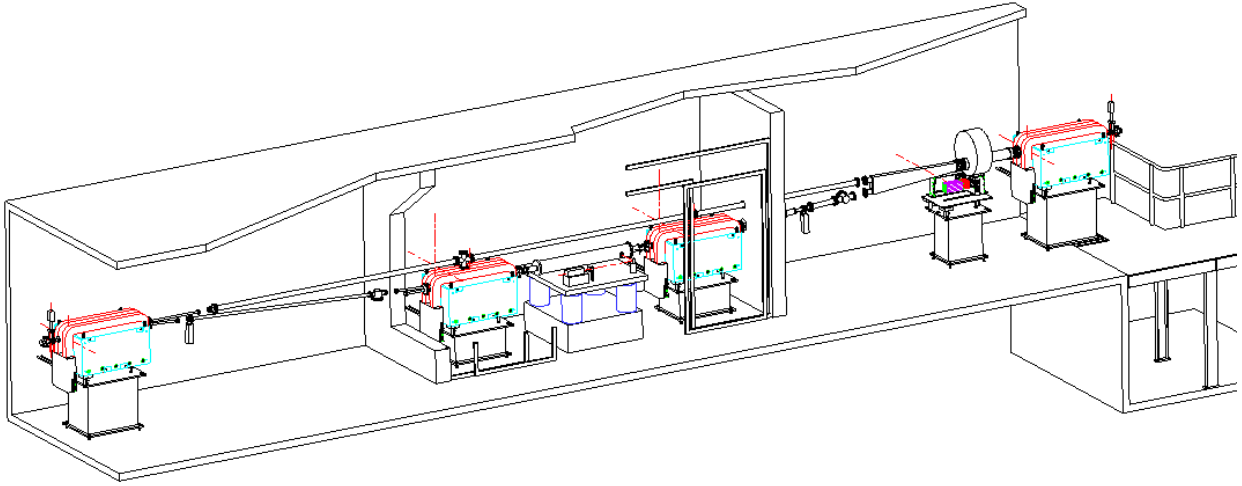}
\caption{Oblique view of the Compton polarimeter.  
The beam enters from the left and is bent 
down into a chicane where it intersects the laser
cavity.  The cavity is on the bench in the middle
of the chicane. The photon detector for backscattered
photons is on the bench just upstream of 
the last chicane magnet.}
\label{fig11:setup}
\end{figure*}

The Compton polarimeter performed its first measurements during the
second HAPPEX run in July 1999~\cite{Baylac}. 
It is installed on the beam line of
Hall  A (see Fig.\ref{fig11:setup}). The electron beam interacts with
a polarized ``photon target" in the center of a vertical magnetic
chicane that aims at separating the scattered electrons and photons
from the primary beam. The backscattered photons are detected in a
matrix of 25 $PbWO_4$ crystals \cite{Neyret}. 

The experimental
asymmetry $\aexp_c=(N^+-N^-)/(N^++N^-)$ is measured, where
$N^+\, (N^-)$ refers to Compton counting rates for right (left)
electron helicity, normalized to the beam intensity. This asymmetry is
related to the electron beam polarization via
\begin{equation}
P_e=\frac{\aexp_c}{P_\gamma \ath_c}
\label{eq:a_expc}
\end{equation}
where $P_\gamma$ is the photon polarization and $\ath_c$ the analyzing
power.  At typical JLab energies (a few GeV), the Compton cross-section
asymmetry is only a few percent. An original way to
compensate this drawback is the implementation of a Fabry-Perot cavity
\cite{Jorda} which amplifies the photon density of a standard
low-power laser at the integration point. An average power of 1200 W
is accumulated inside the cavity with a photon beam waist of the
order of 150 $\mu$m and a photon polarization above 99\%, monitored
online at the exit of the cavity \cite{Falletto}.

Since less than $10^{-9}$ of the beam undergoes Compton scattering,
and thanks to the zero total field integral of the magnetic
chicane, the primary beam is delivered unchanged to the experimental 
target. These features make Compton polarimetry an attractive 
alternative to other techniques, as it provides a non-invasive measurement simultaneous
with the running experiment.

\begin{figure}
\epsfig{figure=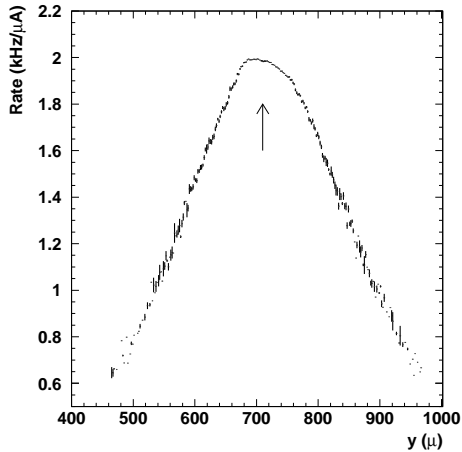,width=0.4\textwidth}
\caption{
Counting rate normalized to beam current versus
vertical position of the electron beam for the 
Compton polarimeter. The
sensitivity to beam position differences 
is proportional to the
derivative of this curve.  The arrow points
to where we run.}
\label{fig12:raty}
\end{figure}

The quality of the polarization measurement is driven by the tuning of
the electron beam in the center of the magnetic chicane. In the early
tests a large background rate was generated in the photon detector by
the halo of the electron beam scraping on the narrow apertures of the
ports in the mirrors of the cavity. Extra focusing in the horizontal plane, induced
by an upstream quadrupole dramatically reduces this background. Then a
fine adjustment of the electron beam vertical position optimizes the
luminosity at the Compton interaction point. Figure~\ref{fig12:raty}
illustrates that beyond maximizing the luminosity, standing near the
optimum position also reduces our sensitivity to electron beam
position differences correlated with the helicity. 

In the data-taking procedure, periods of cavity ON (resonant) and
cavity OFF (unlocked) are alternated in order to monitor the background
level and asymmetry. A typical signal over background ratio of 5 is
achieved and the associated errors are small.

The photon polarization is reversed for each ON period, reducing the
systematic errors due to electron helicity correlations. These
correlations are already minimized by our controls at the source
(see Sec.~\ref{sec:laser}). By summing the Compton asymmetries of the right and left
photon polarization states with the proper statistical weights we
expect the effects of helicity correlations to cancel out to first
order and the residual effects to be small. Nevertheless, extra slow drifts in
time of the beam parameters can occur and increase the sensitivity to
helicity correlations. In order to select stable running conditions we
apply cuts of $\pm 3~\mu A$ on the beam current and reject all the
coil-modulation periods in the analysis. This leads to the loss of 1/3 of
the events. In the end the residual helicity correlated 
luminosity asymmetry $A_F$ still contributed 1.2\%
to the experimental Compton asymmetry and remained 
its main source of systematic
error (cf. Table \ref{tab:error_bud}). 

An optical setup allows us to monitor the photon polarization at the
exit of the cavity. The connection with the ``true" polarization
$P_\gamma$ at the Compton interaction point is given by a transfer
function measured once during a maintenance period. Polarizations for
right and left handed photons are found to be stable in time and
given by $\pg^{R,L} = \pm 99.3^{+0.7}_{-1.1}\%$.

\begin{figure}
\epsfig{figure=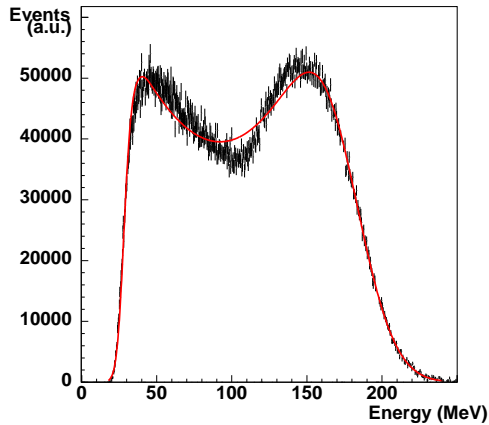,width=0.4\textwidth}
\caption{Compton spectrum as measured by the photon 
calorimeter. The curve is a fit of the Compton 
cross-section convoluted with a Gaussian
resolution of the calorimeter (see Eq. ~\ref{eq:cptres}).}
\label{fig13:fitcpt}
\end{figure}

The last ingredient of Eq.~\ref{eq:a_expc} is the analyzing power
$\ath_c$. The response function of the photon detector 
(see Fig. ~\ref{fig13:fitcpt}) is
parametrized by a Gaussian resolution $g(k')$ of width 
\begin{equation} \sigma_{\rm res}(k') = 
\sqrt{a +  \frac{b}{k'} + \frac{c}{(k')^2}} , 
\label{eq:cptres}
\end{equation}
where
$k'$ is the backscattered photon energy. 
A Gaussian was used because the complete study 
of the calorimeter 
response wasn't available at the time of this analysis; 
the corresponding errors in the calibration, efficiency,
and resolution are shown in Table ~\ref{tab:error_bud}
and explained here.
The coefficients $(a,b,c)$ are fitted to
the data (Fig.~\ref{fig13:fitcpt}).  
A ``smeared'' cross section is then obtained
\begin{equation}
\frac{d\sigma ^{\pm }_{\rm smeared}}{dk_{r}'}=\int ^{\infty
}_{o}\frac{d\sigma ^{\pm }_{c}}{dk'}\, g(k'-k'_{r})\, dk'
\end{equation}
where $k'_{r}$ is the energy deposited in the calorimeter and
$d\sigma ^{\pm }_{c}/dk'$ the helicity-dependent Compton cross
section. Experimentally, the energy spectrum has a finite width at the
threshold (see Fig.~\ref{fig13:fitcpt}) which is modeled by an error function
$p(k'_{s},k'_{r})={\rm erf}((k'_{r}-k'_{s})/\sigma _{s})$ where
$\sigma _{s}$ is fitted to the data as well. This width can be due
either to the fact that the threshold level itself is unstable, or to
the fact that  a given $k'_{r}$ can correspond to different voltages
at the discriminator level.

\begin{table}
\caption{Average relative error budget for 
the beam polarization measured using the Compton polarimeter,
based on 40 measurements in the 1999 run.  $S$ and $B$ refer
to signal and background, $A_B$ is the asymmetry in
the background, and $\af$ is the helicity correlated 
luminosity asymmetry.}
\label{tab:error_bud}
\begin{tabular}{l|ccc|c}
Source & \multicolumn{2}{c}{} & Systematic&Statistical \\
\tableline
$\pg$ & & & 1.1\% & \\ 
$\aexp_c$ & Statistical    & &  & 1.4\% \\
          & $B/S$          & 0.5\%  & & \\
          & $A_B$          & 0.5\%  & 1.4\% & \\
	  & $\af$          & 1.2\%  & & \\  
$\ath_c$     & Non-linearities & 1\% & & \\
          & Calibration & 1\% &  & \\
          & Efficiency/Resolution & 1.9\% & 2.4\% & \\
\tableline
{\bf Total}  &  &  &  \multicolumn{2}{c}{\bf \qquad 3.3\%} \\
\end{tabular}
\end{table}

Finally, the observed counting rates can be expressed as
\begin{equation}
N^{\pm }(k'_{s})=L\times \int ^{\infty }_{0}p(k'_{s},k'_{r})\, 
\frac{d\sigma ^{\pm }_{\rm smeared}}{dk_{r}'}\, dk'_{r}
\end{equation}
where $L$ stands for the interaction luminosity and the analyzing
power of the polarimeter can be calculated as
\begin{equation}
\ath_c=\frac{N^{+}(k'_{s})-N^{-}(k'_{s})}{N^{+}(k'_{s})+N^{-}(k'_{s})}
\end{equation}

The analyzing power is of the order of 1.7\%. To estimate the 
systematic error in the modeling of the calorimeter response, we varied
the parameters $ a,\, b,\, c,\, k'_{s}$, and $\sigma _{s} $ around their
fitted values. The sizes of those variations were chosen to reproduce 
the dispersion of the experimental data. The analyzing
power was then computed for each of the possible combinations of the
cross variations of the five parameters and the maximum deviation from
the nominal analyzing power was assigned as the systematic error. This
contributed a systematic error of 1.9 \%  \cite{Maud_these}. 

Other systematic errors
related to non-linearities in the electronics and
uncertainty in the energy calibration, which is performed by fitting the
Compton edge, make only a small contribution to the final error (cf. table
\ref{tab:error_bud}). Further information on the Compton polarimeter
is available in \cite{Baylac}. 

\section{Systematic Control}
\label{sec:systematics}

\subsection{Control of the Laser Light}
\label{sec:laser}

\newcommand{\be}{\begin{enumerate}}
\newcommand{\ee}{\end{enumerate}}
\def\hap{HAPPEX}

Section~\ref{source_optics} describes the optics of the polarized electron source.  Here, we discuss how those optics were used to control the laser beam's polarization and to suppress helicity-correlated beam asymmetries.

\subsubsection{Laser Polarization and the PITA Effect}
\label{sec:pita}
\begin{figure}
\includegraphics[width=4.0in]{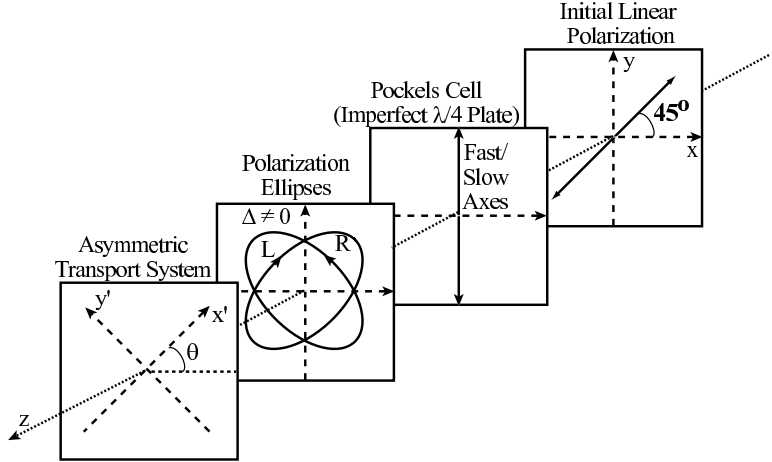}
\caption[]{ \label{fig14:axes} 
Incident linear polarization is nearly circularly polarized by the Pockels cell.  The error phase $\Delta$ causes the polarization ellipses for the two helicities to have their major and minor axes rotated by 90$^o$ from each other, causing helicity-correlated transmission through an optical element with an analyzing power. }
\end{figure}

  The Pockels cell that is used to circularly polarize the laser beam  acts as a voltage-controlled quarter-wave plate.  Depending on the sign of the voltage applied to it, it can produce light of either helicity.  The Pockels cell is an imperfect quarter-wave plate, however, and a convenient way to parameterize the phase shift it induces on the laser beam is
\begin{equation}
\label{eq:phaseshap}
\delta_{R} = -(\frac{\pi}{2}+\alpha)-\Delta,  \qquad \qquad \delta_{L} = +(\frac{\pi}{2}+\alpha)-\Delta,
\end{equation}
where $\delta_R$ ($\delta_L$) is the phase shift induced by the Pockels cell to produce right- (left) helicity light. The imperfections in the phase shift are given by $\alpha$ (``symmetric'' offset) and $\Delta$ (``antisymmetric'' offset), and perfect circular polarization is given by the condition $\alpha = \Delta = 0$.  When an imperfectly circularly polarized laser beam is incident on an optical element that possesses an analyzing power (as in Fig. ~\ref{fig14:axes}), an intensity asymmetry results that depends on the antisymmetric phase, $\Delta$.  To first order, this intensity asymmetry can be expressed as
\begin{equation}
A = -\frac{\epsilon}{T} \cos 2\theta\cdot(\Delta - \Delta^0),
\label{eq:pitafdbk}
\end{equation}
where the ratio $\epsilon/T<<1$ is the ``analyzing power'' of the
optical element defined in terms of the difference in optical
transmission fractions between two orthogonal 
axes ($x^\prime$ and $y^\prime$ in fig ~\ref{fig14:axes}), 
$\epsilon = T_{x^\prime} - T_{y^\prime}$, 
divided by the summed transmission 
fractions $T = T_{x^\prime} + T_{y^\prime}$, and $\theta$ is 
the angle between the Pockels cell's fast axis 
and the $x^\prime$ transmission axis of the analyzer, and $\Delta^0$ is an offset phase shift introduced by residual birefringence in the Pockels cell and the optics downstream of it.  This effect is referred to as  the Polarization-Induced Transport Asymmetry (PITA) effect~\cite{tbh_Cates,tbh_Humensky} and was one of the dominant sources of helicity-correlated beam asymmetries.  The intensity asymmetry is proportional to $\Delta$, and the constant of proportionality $(\epsilon/T)\cos 2\theta$ is referred to as the ``PITA slope''.  Any optical element downstream of the Pockels cell possesses a small analyzing power.  For the 1998 run, a glass slide was introduced into the laser beam to provide a small 
controlled analyzing power.  
For the 1999 run, the QE anisotropy of the strained GaAs cathode (which behaves in this case in a manner formally equivalent to an optical analyzing power) acted as the dominant source of analyzing power in the system.

By controlling the phase $\Delta$ we can control the size of the intensity asymmetry.  In particular, $\Delta$ can be chosen such that the intensity asymmetry is zero.  $\Delta$ can be adjusted by changing the voltage applied to the Pockels cell according to
$V_{\Delta} = \Delta \cdot V_{\lambda/2}/\pi$,
where $V_{\Delta}$ is the change in Pockels cell voltage  required to induce a phase shift $\Delta$ and $V_{\lambda/2}$ is the voltage required for the Pockels cell to provide a half wave of retardation ($\sim$ 5.5 kV).

The magnitude of the PITA slope 
is a key parameter in the source configuration.  For the 1998 run, the PITA slope was set by selecting the angle of incidence of the glass slide.  A value of $\sim 3$ ppm/V was used for production running.  This value was large enough to make the slide the dominant analyzing power in the system, while remaining small enough to suppress higher-order effects that can arise from residual linear polarization.  For the 1999 run, the strained cathode's QE anisotropy provided a PITA slope of as large as $\sim 30$ ppm/V; the value of the PITA slope could be set by choosing the orientation of the rotatable half-wave plate downstream of the Pockels cell as discussed below.  This much larger analyzing power made the glass slide unnecessary, but also enhanced higher-order helicity-correlated differences in beam properties, such as position differences.

In the remainder of section, we discuss the suppression of helicity-correlated beam asymmetries.  The primary techniques, described in more detail below, were to
\be
\item Suppress the intensity asymmetry via an active feedback, the ``PITA feedback.''
\item For the 1999 run, suppress position differences at the source by rotating an additional half-wave plate located downstream of the helicity-flipping Pockels cell (Fig. ~\ref{Fig2_KK_laser}) to an orientation at which position differences appeared to be intrinsically small.
\item Gain additional suppression of position differences by properly tuning the accelerator to take advantage of ``adiabatic damping''
(section ~\ref{sec:adiabatic_damp}).
\item For the 1999 run, suppress the intensity asymmetry of the Hall C beam by use of a second intensity-asymmetry feedback system.
\item Gain some additional cancellation of beam asymmetries by using the insertable half-wave plate (located just upstream of the Pockels cell in Fig. ~\ref{Fig2_KK_laser}) as a means of slow helicity reversal.
\ee

\subsubsection{PITA Feedback}
\label{sec:pita_feedback}
The linear relationship between the intensity asymmetry and the phase $\Delta$ allowed us to establish a feedback loop.  The intensity asymmetry was measured by a BCM located near the target and the phase $\Delta$ was corrected to zero the asymmetry by adjusting the high voltage applied to the Pockels cell by small amounts.  This feedback loop was called the ``PITA Feedback.''  The algorithm worked as follows.  The initial Pockels cell voltages for right- and left-helicity ($V_R^0$ and $V_L^0$, respectively, with $V_R^0 \approx -V_L^0$) were determined while aligning the Pockels cell.  We measured the PITA slope $M$ approximately every 24 hours, a time scale
on which it was reasonably stable.
During physics running, the DAQ monitored the intensity asymmetry in real time and, every 2500 window pairs (approximately every three minutes), adjusted the Pockels cell voltages to null the intensity asymmetry measured on the preceding 2500 pairs.  We referred to each set of 2500 pairs as a ``minirun.''  The feedback is initialized with the offset voltage set to zero and the voltages for right and left helicity set to their default values: 
\begin{eqnarray} \label{eq:pita1}
V_{\Delta}^1 &= 0, \nonumber \\
V_R^1 &= V_R^0, \\
V_L^1 &= V_L^0. \nonumber
\end{eqnarray}
Using the measured value of $M$, we
apply a correction for the $n^{th}$ minirun according to
the following algorithm.
For minirun $n$, the Pockels cell voltages were
\begin{eqnarray} \label{eq:pita2}
V_{\Delta}^n &= V_{\Delta}^{n-1} - \big( 
{\asy_I^{n-1}}/{M} \big), \nonumber \\
V_{R}^n &= V_{R}^{0} + V_{\Delta}^n, \\
V_{L}^n &= V_{L}^{0} + V_{\Delta}^n. \nonumber
\end{eqnarray}

The \hap\ DAQ was responsible for calculating the intensity asymmetry and the required correction to the Pockels cell voltages for each minirun.  The correction voltage $V_{\Delta}^n$ was transmitted back to the Injector over a fiber-optic line as indicated in Fig. ~\ref{Fig2_KK_laser}.  
This algorithm worked effectively; the intensity asymmetry averaged over the entire 1999 run was below one ppm, an order of magnitude smaller than the physics asymmetry.

The virtue of the PITA feedback lies in the fact that the dominant 
cause of intensity asymmetry is the
residual linear polarization in the laser beam.  By adjusting the phase $\Delta$ to suppress the intensity asymmetry, we are either minimizing the residual linear polarization or at least arranging the Stokes-1 and Stokes-2 components such that their effects cancel out.

\subsubsection{The Rotatable Half-Wave Plate}
\label{sec:rwhp}
The rotatable half-wave plate gives us control over the orientation of the laser beam's polarization ellipse with respect to the cathode's strain axes.  To describe its utility, we extend Eq. ~\ref{eq:pitafdbk} to include effects due to the half-wave plate and the vacuum window at the entrance to the polarized gun.  We assume that the half-wave plate is imperfect and induces a retardation of $\pi + \gamma$, where $\gamma \ll 1$.  In addition, we assume that the vacuum window possesses a small amount of stress-induced birefringence $\beta \ll 1$.  The result, to first order, is

\begin{eqnarray}
\label{eq:rwhp}
\asy_I = -\frac{\epsilon}{T} [(\Delta - \Delta^0) 
\cos (2\theta - 4\psi) - 
\\ \nonumber
\gamma \sin (2\theta - 2\psi) - \beta \sin (2\theta - 2\rho) ]
\end{eqnarray}
where $\psi$ and $\rho$ are orientation angles for the half-wave plate and the vacuum window fast axes, respectively, as measured from the horizontal axis.  In Eq. ~\ref{eq:pitafdbk}, the contributions from the half-wave plate and the vacuum window were included in the term $\Delta^0$.  This new expression has three terms:
\be
\item The first term, proportional to $\Delta$, is now modulated by the orientation of the half-wave plate with a $90^o$ period.  
\item The second term, proportional to $\gamma$, arises from using an imperfect half-wave plate and also depends on the half-wave plate's orientation but with a $180^o$ period.
\item The third term, proportional to $\beta$, arises from the vacuum window and is independent of the half-wave plate's orientation because the vacuum window is downstream of the half-wave plate.  This term generates a constant offset to the intensity asymmetry.
\ee

\begin{figure}
\includegraphics[width=3.4in]{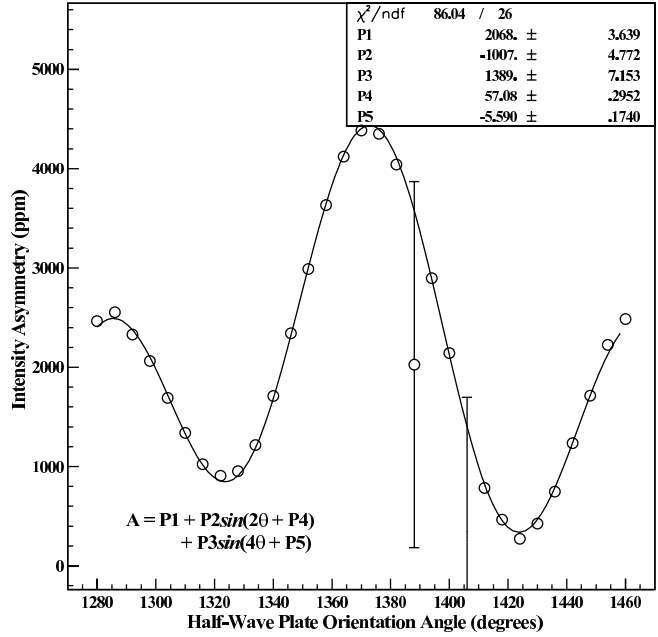}
\caption[Intensity asymmetry as a function of rotatable half-wave plate orientation.]{ \label{fig15:sweet} 
Intensity asymmetry as a function of rotatable 
half-wave plate orientation.  The error bars on
some points are smaller than the symbols.}
\end{figure}

Figure~\ref{fig15:sweet} shows a  measurement of intensity asymmetry as a function of half-wave plate orientation angle from the 1999 run.  The function fit to the data allowed us to extract the relative contributions of the half-wave plate error, the vacuum window, and the Pockels cell.  The three terms contributed at roughly the same magnitude, though the offset was large enough that the curve did not pass through zero intensity asymmetry.  In addition, we found, as discussed more below, that the PITA slope was usually maximized at the extrema of this curve.  These facts motivated us to choose to operate at an extremum (in this case, at 1425$^o$) in order to minimize the voltage offset required to null the intensity asymmetry.

\begin{figure}
\includegraphics[width=3.5in]{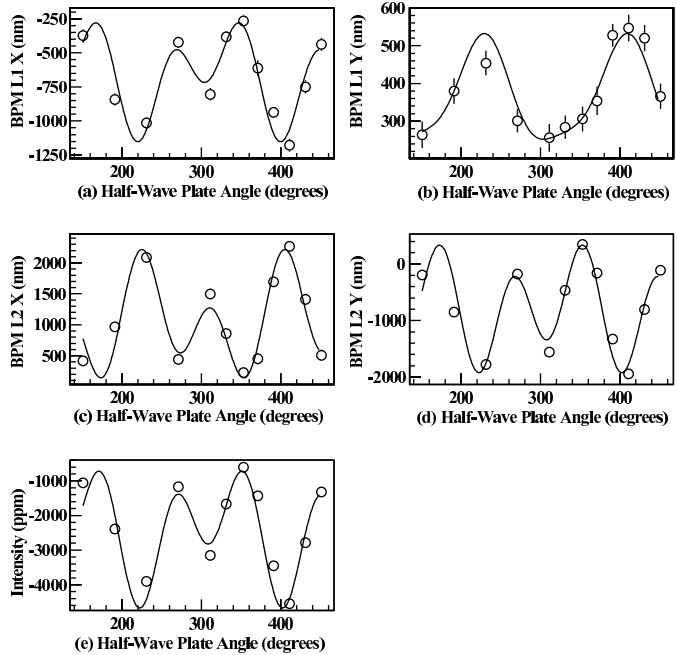}
\caption[Injector position differences as a function of rotatable half-wave plate orientation.]{ \label{fig16:poscorrel} 
Dependence of position differences measured by two BPMs at the 5 MeV point in the Injector (a-d) on the orientation of the rotatable half-wave plate.  The position differences show a strong correlation with the intensity asymmetry (e).  The error bars on some data points are smaller than the symbols.}
\end{figure}

Figure~\ref{fig16:poscorrel} shows the results of a study conducted prior to the start of the 1999 run in which the position differences were also measured using BPMs located at the 5 MeV point in the injector.  We observed a fairly strong correlation between the intensity asymmetry and the position differences.  It was not clear what the underlying cause of this correlation was, but it was certainly clear that by minimizing the intensity asymmetry we simultaneously suppressed position differences.  For this reason, during the 1999 run our strategy was to measure the intensity asymmetry as a function of half-wave plate orientation using a Hall A BCM and to choose an orientation angle which minimized the intensity asymmetry; this orientation angle would also minimize the position differences.  It would have been preferable to measure the position difference in the Injector and choose a half-wave plate orientation that minimized them directly, but such a study would have required interrupting beam delivery to Hall C for several hours, and that level of interference with an experiment running in another Hall was unacceptable.  Using this strategy, we achieved position differences below 500 - 1000 nm at the 5-MeV BPMs.  The position differences were further suppressed in the accelerator via adiabatic 
damping (section \ref{sec:adiabatic_damp})
and some additional cancellation was achieved via the insertable half-wave plate used for slow helicity reversal.

\subsubsection{Adiabatic Damping}
\label{sec:adiabatic_damp}

If the sections of the accelerator are well matched and free of XY
coupling, the helicity-correlated position differences become damped as
$\sqrt{(A/P)}$ where $A$ is a constant and $P$ is the momentum.  
This is due to the well-known
adiabatic damping of phase space area for a beam
undergoing acceleration~\cite{edwards_syphers}. 
The beam emittance, defined as the invariant
phase space area based on the beam density matrix, varies inversely as
the beam momentum.  The projected beam size and divergence, and thus the
difference orbit amplitude (defined as the size of the excursion from
the nominally correct orbit), are proportional to the square root of the
emittance multiplied by the beta function at the point of interest.
Ideally therefore the position differences become reduced by a factor of
$\sqrt{(3.3 \ {\rm GeV} / 5 \ {\rm MeV)}} 
\sim 25$ between the 5 MeV region and the target.  This
also implies that the 5 MeV region is a sensitive location to measure and
apply feedback on these position differences, if signals from the beams
of the different halls could be measured separately.

Deviations from this ideal reduction factor can however occur mainly due
to two effects. The presence of XY coupling can potentially lead to
growth in the emittance in both X and Y planes, while a mismatched beam
line often results in growth in the beta function.  Both effects, as can
be seen from the previous paragraph, can translate into growth in
difference orbit amplitude and a reduction in adiabatic damping actually
derived.  The Courant-Snyder parameters~\cite{courant_snyder}
calculated at different
sections of the accelerator based on such difference orbits are an
effective measure of the quality of betatron matching, with a constant
value at all sections for all orbits indicating perfect betatron
matching.

Imperfections or deviations from design in the magnetic elements at the
$10^{-3}$ level distributed across the magnet lattice, or $10^{-2}$ at
one point in the lattice, 
can lead to large coupling between position and angle, or
growth in one or more dimensions of phase space, and consequent
amplification of the position differences.  Matching the sections of the
accelerator is an empirical procedure in which the Courant-Snyder
parameters (or equivalently the transfer matrices) are measured by making
kicks in the beam orbit, and the quadrupoles are adjusted to 
fine-tune the matrix elements.  
This adjustment procedure is
being automated \cite{ychao} for future experiments.

\subsubsection{Suppressing the Hall C Intensity Asymmetry}
\label{sec:hallc_asy}
During the 1999 run, experiments were running in Hall C that required a high beam current (50 - 100 $\mu$A).  While the PITA feedback suppressed the intensity asymmetry in Hall A, it was  possible for a large intensity asymmetry to develop on the Hall C beam.   Cross talk between the beams in the accelerator allowed the intensity asymmetry in the Hall C beam to induce intensity, energy, and position asymmetries in the Hall A beam.  

A second feedback system on the laser power was used to control the Hall C intensity asymmetry.  This feedback was based on helicity-correlated modulation of Hall C's laser intensity rather than its polarization.  The modulation was introduced by adding an offset to the current driving its seed laser.  We found that by manually adjusting the offset once per hour to null the Hall C intensity asymmetry, we could maintain the asymmetry at the 10 ppm level, small enough to make its effects on the Hall A beam negligible.  

While adequate for a non-parity experiment, the laser-power feedback suffered
from two flaws that prevented it from replacing the PITA feedback.  
First, the laser beam's pointing was correlated with its drive current.  
Thus, changing the current in a helicity-correlated way induced position
differences.  Second, the laser-power feedback removed the intensity 
asymmetry directly without correcting the underlying problem of
residual linear polarization in the circularly polarized light.

\subsection{Beam modulation}
\label{bmod}

Modulation of beam parameters calibrated the response of
the detectors to the beam and permitted us to measure online
the helicity-correlated beam parameter differences.
The beam modulation system intentionally varied 
beam parameters concurrently with data taking.  
The relevant parameters were the
beam position in $x$ and $y$ at the
target, angle in $x$ and $y$ at the target, and energy.  
We measured position differences in $x$ and $y$ at two points 
1.3 and 7.5 m upstream of the target in a field free region,
and at a point of high dispersion in the magnetic arc
leading into Hall A, as well as several
other locations for redundancy.
False asymmetries due to these differences were found to be negligible.

The energy of the beam is varied by applying a control voltage to a
vernier input on a cavity in the accelerator's South Linac.  To vary
beam positions and angles, we installed seven air-core corrector coils
in the Hall A beam line upstream of the
dispersive arc.  These coils are interspersed with quadrupoles
in the beam line; their positions are chosen based on beam transport
simulations intended to verify that we could span the space of two
positions and two angles at the target using four of the seven coils.  The
additional coils are for redundancy, since a change in beam tune
could change our ability to span the required space.  The coils are
driven by power supply cards with a control voltage input to govern
their excitation.  Control voltages for the seven coils and energy
vernier are supplied by a VME DAC module in response to
requests sent from the HAPPEX DAQ.

The coils and vernier are modulated in sequence.  A modulation cycle
consists of three steps up, six down, and three up, forming a stepped
sawtooth pattern.  Each step is 200 ms in duration.  Typically the
total peak-to-peak amplitude of the coil modulation is 800 mA
corresponding to a beam deflection at the BPMs in the hall on the
order of $\pm100~\mu$m; for the vernier the typical amplitude is 900
keV, resulting in a deflection of similar size at the dispersion point
BPM.  After stepping through all seven coils and the vernier the
modulation system is inactive for 38 sec, resulting in a duty
factor of $\sim$33\%.

\begin{figure*}
\includegraphics[width=6.0in]{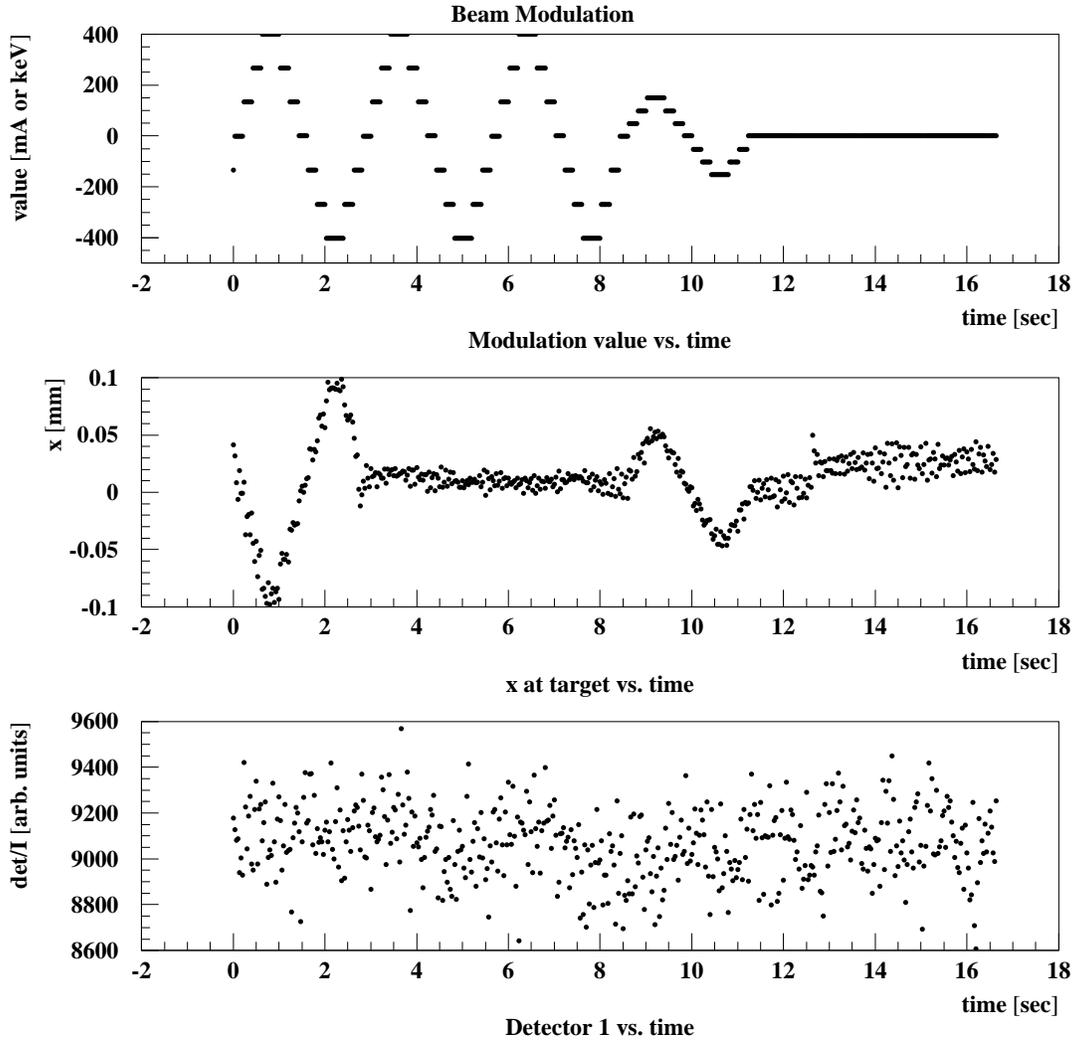}
\caption{Beam modulation to calibrate sensitivity.  
(top) Typical coil and energy vernier modulation
values as a function of time.  Four modulation pulses each about three
seconds long are seen: the first is a horizontal correction coil, the
next two are vertical coils, and the fourth is the energy vernier. 
(middle) Horizontal position at target versus time for the same data. 
The position responds to modulation of horizontal coil and energy
vernier but not to modulation of vertical coils.  (bottom) \v{C}erenkov
detector response versus time for the same data.  Sensitivity to
position and energy modulation is small compared to counting statistics.
}
\label{Fig17_RSH_Bmod}
\end{figure*}

Individual modulation cycles are evident in the BPM data
(Fig.~\ref{Fig17_RSH_Bmod}).  It should  be emphasized
that these data are integrated at a subharmonic of the 60 Hz line
frequency, which eliminates any 60 Hz noise in the beam position.
Typically the 60 Hz noise is significantly larger than the modulations
we impose.  Figure \ref{Fig17_RSH_Bmod} also shows that the response of
our detectors to the beam modulation is small compared to the
window-to-window noise, which is dominated by counting statistics.
Only by averaging over many modulation cycles can the effects of
modulation be seen in the detectors; therefore the modulation system
does not add significantly to our experimental error.
Section~\ref{sec:slopes} details how the sensitivities to beam differences
are extracted from the modulation data. 

\section{Asymmetries}
\label{sec:rsh_asy}

In this section we describe how data are selected for analysis, how
raw asymmetries are extracted from the data, and how these raw
asymmetries are corrected for systematic effects due to helicity-correlated 
differences in beam parameters and to pedestals and nonlinearities in
the measured signals.

\subsection{Data selection}
\label{sec:data_select}

The 1998 production quality data were generated by 
78 Coulombs of electrons striking the target; 
in 1999, 92 C struck the target.  
These totals
exclude runs taken for diagnostic purposes and a small number of runs
in which equipment malfunctions serious enough to compromise the
quality of the entire run occurred; a typical run was about one hour.

We define a `data set' as a group of consecutive runs taken with the same
state (in or out) of the insertable half-wave plate; the state of the 
half-wave plate was changed typically after 24--48 hours of data-taking.

In our analysis of the production data, we impose a minimal set of
cuts to reject unusable or compromised data.  Our philosophy was never
to cut on asymmetries (or helicity-correlated differences), rather only to cut 
on absolute quantities. We reject any data in which:

\begin{itemize}

\item
The integrated current monitor signal falls below a value
corresponding to 2\% of the maximum current.
In practice the threshold value was not critical since the beam was
almost always either close to fully on or off.

\item
Any of several redundant checks for synchronization between ADC data
and helicity information fails.  Since the helicity state arrives in
the data stream eight windows after the window it applied to,
incorrect helicity assignment could result if one or more windows are
missing from the data stream due to DAQ deadtime.  We
therefore check that the second window of each pair has helicity
opposite the first; that the sequence of helicity values read in
hardware matches the prediction of a software implementation of the
same pseudorandom bit generator; and that the scaler used to count
windows increments by one at each window.  

\end{itemize}

Whenever one or more consecutive windows fail one of these cuts, we
also reject some windows before and after the ones that failed.  For
example, when the current monitor threshold cut is imposed, we also
reject 10 windows before the BCM drops below threshold and 50 windows
after it comes back above threshold.  
This procedure eliminates not only beam-off data but also
conditions where the beam was ramping or the
gains of our devices were recovering from a beam trip.

Additional cuts are applied depending on what is being calculated.
In effect there are five different measurements being made using the
same data: raw
asymmetries in each of the two detectors, helicity-correlated
differences in beam parameters, and sensitivities of each of the two
detectors to changes in beam parameters.  The additional cuts
appropriate to each measurement are discussed in the following
subsections. 

Integrated signals for each event include: $D_1$ and $D_2$, the
\v{C}erenkov detectors in the two arms; $I_1$, $I_2$, $I_U$, 
three beam current monitors (the two cavity monitors and the Unser
monitor); 
$X_1$, $Y_1$, $X_2$, and $Y_2$, two pairs
of beam position monitors (BPMs) measuring horizontal and vertical
positions 7.5 and 1.3 m, respectively, upstream of the target;
and $X_E$, a horizontal BPM located in a region of high dispersion
72.6 m upstream of the target.  (These five BPMs are also
denoted $B_i$, where $i = 1..5$.)  The analysis uses detector signals
normalized to the beam current, $d_{1(2)} \equiv D_{1(2)} / I_1$. 

\subsection{Calculation of raw asymmetries}
\label{sec:rsh_rawasy}

For each window pair of each run we compute asymmetries for various
signals $S$,

\begin{equation}
A(S) = \frac{S^+-S^-}{S^++S^-}
\end{equation}

Superscripts $+$ and $-$ refer to the two states of the {\it Helicity}
signal originating at the polarized electron source; a change in this
signal corresponds to a helicity reversal of the source laser beam.
The relationship of this signal to the sign of the polarization of the
electron beam in the experimental hall depends on a number of factors:
whether the half-wave plate is present or not in the laser table
optics, the beam energy (due to precession in the accelerator arcs and
the Hall A line), and the setup of the
helicity Pockels cell electronics.  We use the Hall A polarimeters to
determine the actual polarization sign relative to the {\it Helicity}
signal.  
For our 1998 and July 1999
data, with the half-wave plate in (out), the $+$ {\it Helicity} state corresponds to
left (right) polarized electrons while the $-$ state corresponds to
right (left) polarized electrons; for the April-May 1999 data the
correspondence is opposite. A change in the Pockels cell
configuration between May and July accounts for the latter difference, the
small energy change having been compensated by adjustment of the
Wien filter at the source.

For example, we compute asymmetries for each \v{C}erenkov detector
normalized by the beam current, $\asy_{1(2)} \equiv A(d_{1(2)})$; the
summed normalized detectors, $\asy_s \equiv A(d_1+d_2)$; the average
value from the two detectors $\asy_a \equiv (A(d_1)+A(d_2))/2$;
and the beam
current, $\asy_I \equiv A(I_1)$.  We also compute asymmetries for
various non helicity-correlated voltage and current sources as a check
for electronic crosstalk.

In addition to the cuts on beam current and data acquisition dead
time, cuts are applied to reject data taken during a malfunction of
the beam current monitor.  For calculation of $\asy_{1(2)}$ and
$\asy_s$ we also reject data taken during a malfunction of the magnets
or detector in that arm, or during times when there was significant
boiling in the target.

For each run, we then compute averages of these asymmetries weighted by beam
currents,
\begin{equation}
\langle A(S)\rangle = \frac{\sum_k w_kA(S_k)}{\sum_k w_k}
\end{equation}
where the index $k$ denotes pulse pair in the run and $w_k =
I^+_{1k}+I^-_{1k}$.  Errors on these averages, denoted $\delta\langle
A(S)\rangle$, are estimated from widths of the distributions of
$A(S)$.

Finally, we compute average asymmetries over all runs in the data set
\begin{equation}
\langle\langle A(S)\rangle\rangle = \frac{\sum_j \epsilon_j W_j(S)
\langle A(S)\rangle_j}{\sum_j W_j(S)}
\end{equation}
where the index $j$ denotes the run,
$\epsilon_j = \pm 1$ depending on the sign of the measured beam
polarization, and $W_j(S) = 1/\delta^2\langle A(S)\rangle_j$.

\begin{figure}[phtb]
\includegraphics[width=2.6in,angle=90]{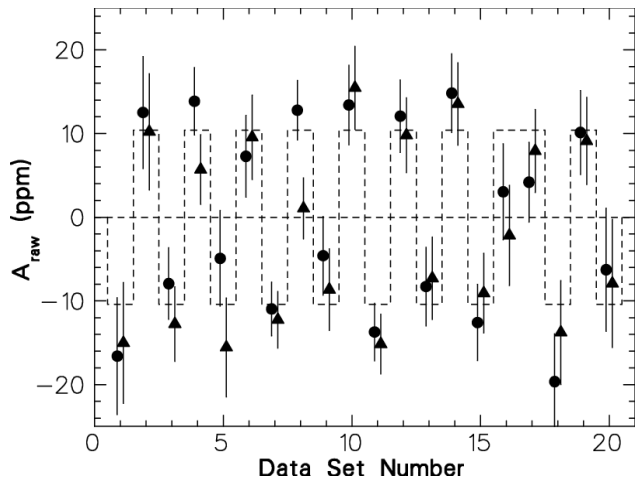}
\caption{Raw asymmetries for 1999 running period, in ppm, broken down
by data set.  The circles are for the left spectrometer, triangles for
the right spectrometer. The step pattern represents the effect of
insertion/removal of the half-wave plate between data sets combined
with a Pockels cell reconfiguration between data sets 16 and 17; see
text.  The amplitude of the step is the average value of the asymmetry
over the entire run.  }
\label{Fig18_RSH_Asym}
\end{figure}

Figure \ref{Fig18_RSH_Asym} shows the asymmetries for the 1999 running
periods broken down into data sets.  As expected, the asymmetry
changed sign when the half-wave plate was inserted, but the magnitude
of the asymmetry is statistically compatible for all data sets. Similar
behavior is seen for the 1998 data~\cite{aniol1}. 

\begin{figure}[pbth]
\includegraphics[width=3.5in]{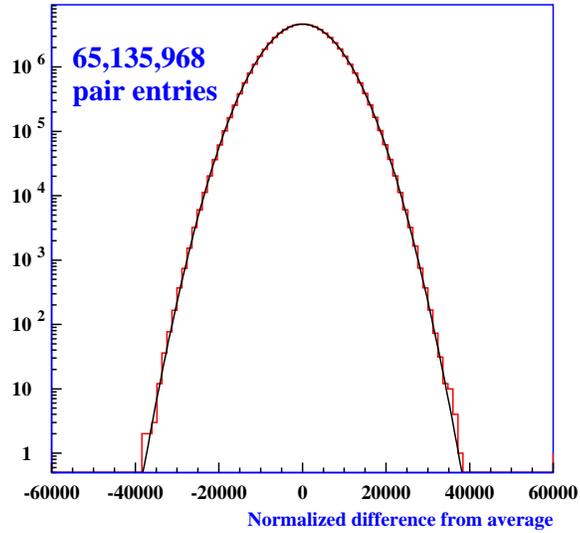}
\caption{Window pair asymmetries for 1999 running period, normalized
  by square root of beam intensity,
with mean value subtracted off, in ppm.}
\label{Fig19_RSH_GauPair}
\end{figure}

\begin{figure}[pbth]
\includegraphics[width=3.5in]{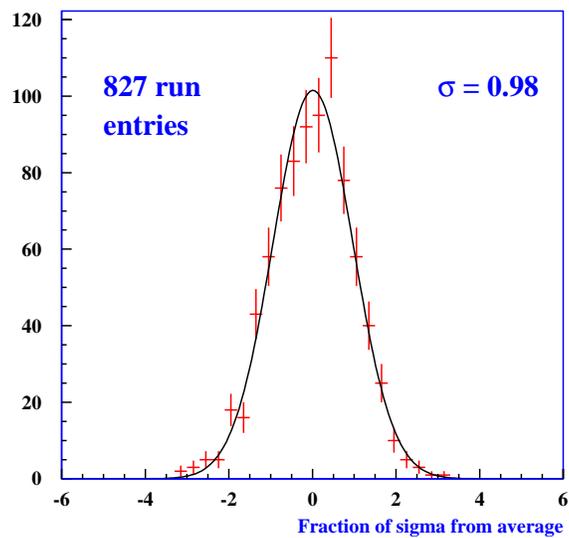}
\caption{Run asymmetries for 1999 running period, with mean subtracted
off and normalized by statistical error.}
\label{Fig20_RSH_GauRun}
\end{figure}

Our analysis assumes the asymmetry distributions are Gaussian with
widths dominated by counting statistics.  To check this, in Fig.~\ref{Fig19_RSH_GauPair} 
we plot the distribution of the quantity
$((\asy_s)_{jk} - \langle\asy_s\rangle) / \sqrt {2(I_1)_{jk}}$ for the 1999 running
periods.  If counting statistics dominate, then the distribution of
this quantity should be Gaussian.  We see that this is indeed the
case, over seven orders of magnitude with no tails.  Likewise, the run
averages behave statistically as can be seen in Fig.~\ref{Fig20_RSH_GauRun} 
where we plot the distribution of the quantity
$((\asy_s)_j - \langle\asy_s\rangle) / \delta(\asy_s)_j$ for the 1999 running periods; the
distribution is Gaussian with unit width. The 1998 data show
similar behavior. 

\subsection{Calculation of helicity-correlated beam differences}
\label{sec:rsh_beamdiff}

For calculation of helicity-correlated beam position and energy
differences, cuts are applied to reject data taken during a
malfunction of the position monitors and data taken while a beam
modulation device was ramping.  The difference in the $i$th BPM is
denoted $\Delta B_i = B^+_i - B^-_i$.

Averages over each run $\langle\Delta B_i\rangle$ and over all runs
in the data set $\langle\langle\Delta B_i\rangle\rangle$ are
computed similarly to the asymmetry averages.  For the latter,
differences are weighted in the average by $W_j =
1/\delta^2\langle \asy_s\rangle_j$, not by $1/\delta^2\langle\Delta
B_i\rangle_j$.  The reason is that in a computation of an average
corrected asymmetry $\langle\langle \asy_s\rangle^{\rm corr}\rangle = \langle
\langle \asy_s\rangle - \sum_j a_j\langle\Delta B_j\rangle \rangle$
(sec \ref{sec:slopes})
the dominant error is $\delta\langle \asy_s\rangle$
and the average over multiple runs of $\langle \Delta B_j\rangle$ weighted by
$1/\delta^2\langle \asy_s\rangle$ is the relevant quantity.

The BPM differences for the three running periods and two
half-wave plate settings are given in Table
\ref{Tab_RSH_Diff}. Note that the differences for the two different half-wave plate
states tend to have opposite sign and thus partially cancel, reducing
the size of their effect on the experimental asymmetry.

\begin{table*}
\caption{Beam position differences in nm, corrected for sign of beam polarization.\vspace*{2ex}}
\label{Tab_RSH_Diff}
\begin{tabular}{lrrrrr}
 & \multicolumn{1}{c}{$\Delta X_1$} & \multicolumn{1}{c}{$\Delta Y_1$} & \multicolumn{1}{c}{$\Delta X_2$} & \multicolumn{1}{c}{$\Delta Y_2$} & \multicolumn{1}{c}{$\Delta X_E$} \\
1998 half-wave out &  $ -2.7 \pm 2.9 $ &  $  1.9 \pm 1.9 $ &  $ -1.9 \pm 3.2 $ &  $  1.3 \pm 2.8 $ &  $ 20.9 \pm 8.5 $ \\
1998 half-wave in  &  $ -2.3 \pm 2.9 $ &  $ -1.1 \pm 1.9 $ &  $ -2.9 \pm 3.2 $ &  $ -0.1 \pm 3.0 $ &  $ -0.8 \pm 8.5 $ \\
All 1998 data      &  $ -2.5 \pm 2.0 $ &  $  0.4 \pm 1.4 $ &  $ -2.4 \pm 2.3 $ &  $  0.7 \pm 2.0 $ &  $ 10.0 \pm 6.0 $ \\
\hline
Apr/May 1999 
half-wave out      &  $-20.9 \pm 3.1 $ &  $-12.6 \pm 1.5 $ &  $-15.3 \pm 5.2 $ &  $ 12.7 \pm 0.7 $ &  $-47.3 \pm 4.6 $ \\
Apr/May 1999 
half-wave in       &  $ -1.0 \pm 3.4 $ &  $ -5.9 \pm 1.8 $ &  $ -5.8 \pm 5.7 $ &  $ -3.7 \pm 0.8 $ &  $ 18.6 \pm 5.1 $ \\
All Apr/May
1999 data          &  $-11.9 \pm 2.3 $ &  $ -9.8 \pm 1.2 $ &  $-11.0 \pm 3.8 $ &  $  5.2 \pm 0.5 $ &  $-17.5 \pm 3.4 $ \\
\hline
Jul 1999 
half-wave out      &  $ -9.5 \pm 5.5 $ &  $-44.8 \pm 10.4 $&  $ 87.2 \pm 12.4 $&  $  0.3 \pm 3.2 $ &  $-77.0 \pm 10.6 $\\
Jul 1999 
half-wave in       &  $ 13.9 \pm 4.6 $ &  $ 11.4 \pm 8.4 $ &  $-53.8 \pm 11.3 $&  $ -5.8 \pm 2.4 $ &  $ 60.2 \pm 9.6 $ \\
All Jul
1999 data          &  $  4.3 \pm 3.5 $ &  $-10.8 \pm 6.6 $ &  $ 10.2 \pm 8.4 $ &  $ -3.6 \pm 1.9 $ &  $ -1.2 \pm 7.1 $ \\
\end{tabular}
\end{table*}

\subsection{Calculation of sensitivities to beam parameters}
\label{sec:slopes}

The helicity-correlated differences in beam parameters originate in
the polarized source and can give rise to differences in rates in the
detectors and therefore false contributions to the asymmetries.  We
compute normalized detector asymmetries corrected for beam differences
using

\begin{equation}
\langle A\rangle^{\rm corr} = \langle A\rangle - \langle \Delta A\rangle.
\end{equation}

The asymmetry correction $\Delta A$ is
calculated by
\begin{equation}
\langle\Delta A\rangle = \frac {1}{2 \langle d\rangle} \left ( \sum_{j=1}^5 \left
( \frac {\partial d} {\partial B_j} \right ) \langle \Delta
B_j\rangle \right )
\end{equation}
where $\langle d\rangle$ is the average normalized signal for the
detector.  We assume that the cross section is a linear function of $(x, y,
\theta_x, \theta_y, E)$.  Then $\partial d/\partial B_j$ is a quantity
which describes the sensitivity of the detector signal to changes in a
combination of beam parameters measured by the BPM.  We obtain these
partial derivatives by starting with the system of linear equations

\begin{equation}
\frac{\partial d}{\partial C_i} =
\sum_{j=1}^5 \frac{\partial d}{\partial B_j}
	\frac{\partial B_j}{\partial C_i}
\end{equation}

and solving by matrix inversion

\begin{equation}
\frac{\partial d}{\partial B_j} =
\sum_{i=1}^5 \frac{\partial d}{\partial C_i}
	\left (\frac{\partial B_j}{\partial C_i}\right )^{-1}
\end{equation}

The slopes $\partial d/\partial C_i$ and $\partial B_j/\partial C_i$
describe the sensitivities of the normalized detectors and the BPMs to
changes in the beam modulation devices; the index $i$ refers to the
five devices (four coils and one vernier). These slopes are calculated
in offline analysis using the beam modulation data.  For each
modulation cycle the BPM and detector data versus coil or vernier
offset value, $C_i$, are fit to straight lines, and the resulting
slopes are averaged over each run.  
Values for these slopes $\partial d / \partial B_j$
averaged over each run period are given in 
Table~\ref{Tab_RSH_slopes}.

\begin{figure*}[phtb]
\includegraphics[width=6.5in]{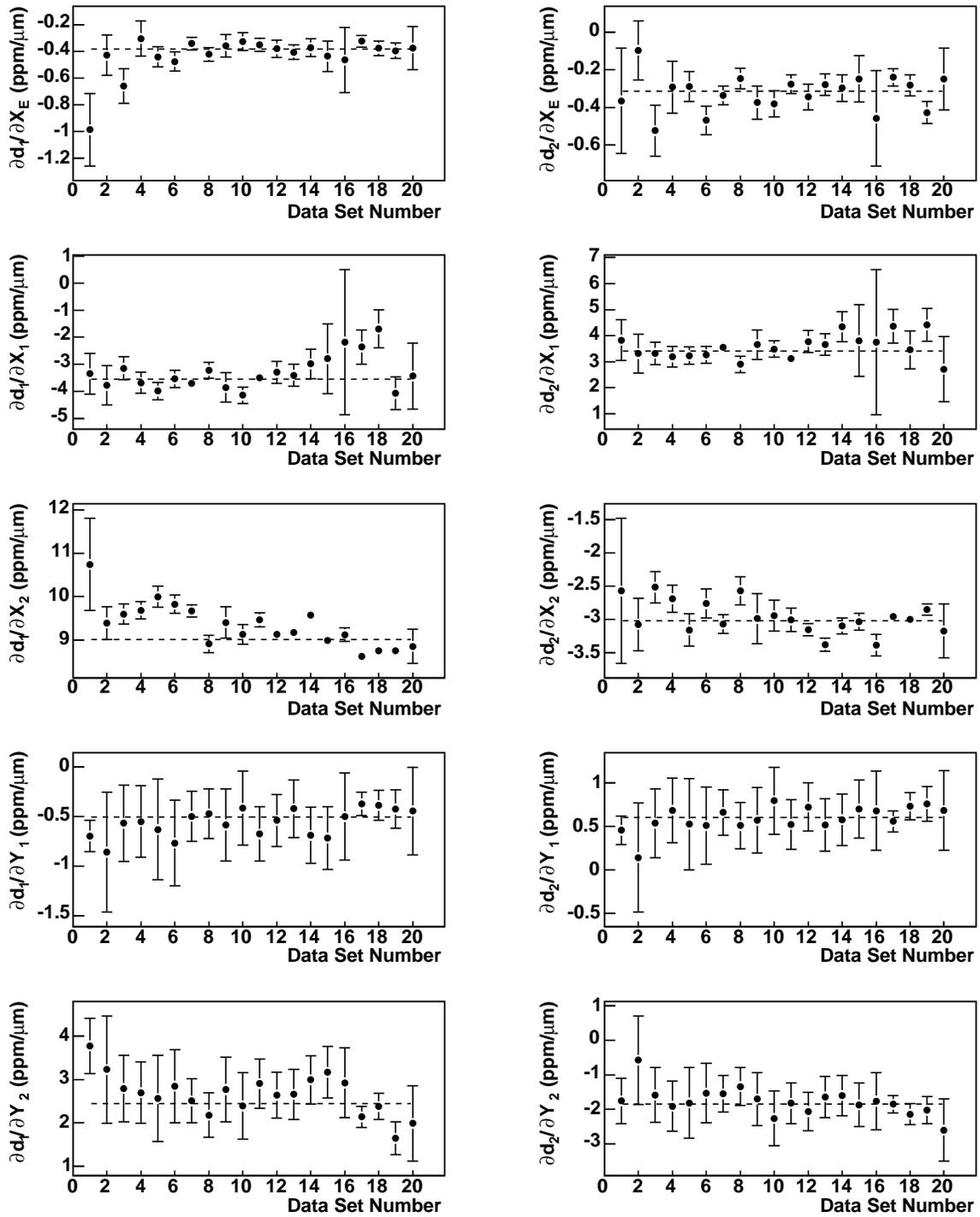}
\caption{Representative sensitivity coefficients $a_j = ( \partial d /
\partial B_j ) / 2 \langle d\rangle$ {\em vs.\/} data set for 1999
run, for energy-sensitive position (top row), horizontal positions
at locations on the beamline 7.5 m and 1.3 m upstream of the target
(second and third rows), and
vertical positions at 7.5 m and 1.3 m (fourth and
fifth rows).  Left and right columns correspond to the two detectors.
Units in all cases are ${\rm ppm}/\mu{\rm m}$.  Coefficients are seen
to be stable at the level of estimated errors.}
\label{Fig21_RSH_Corrcoef}
\end{figure*}

\begin{figure*}[phtb]
\includegraphics[width=6.5in]{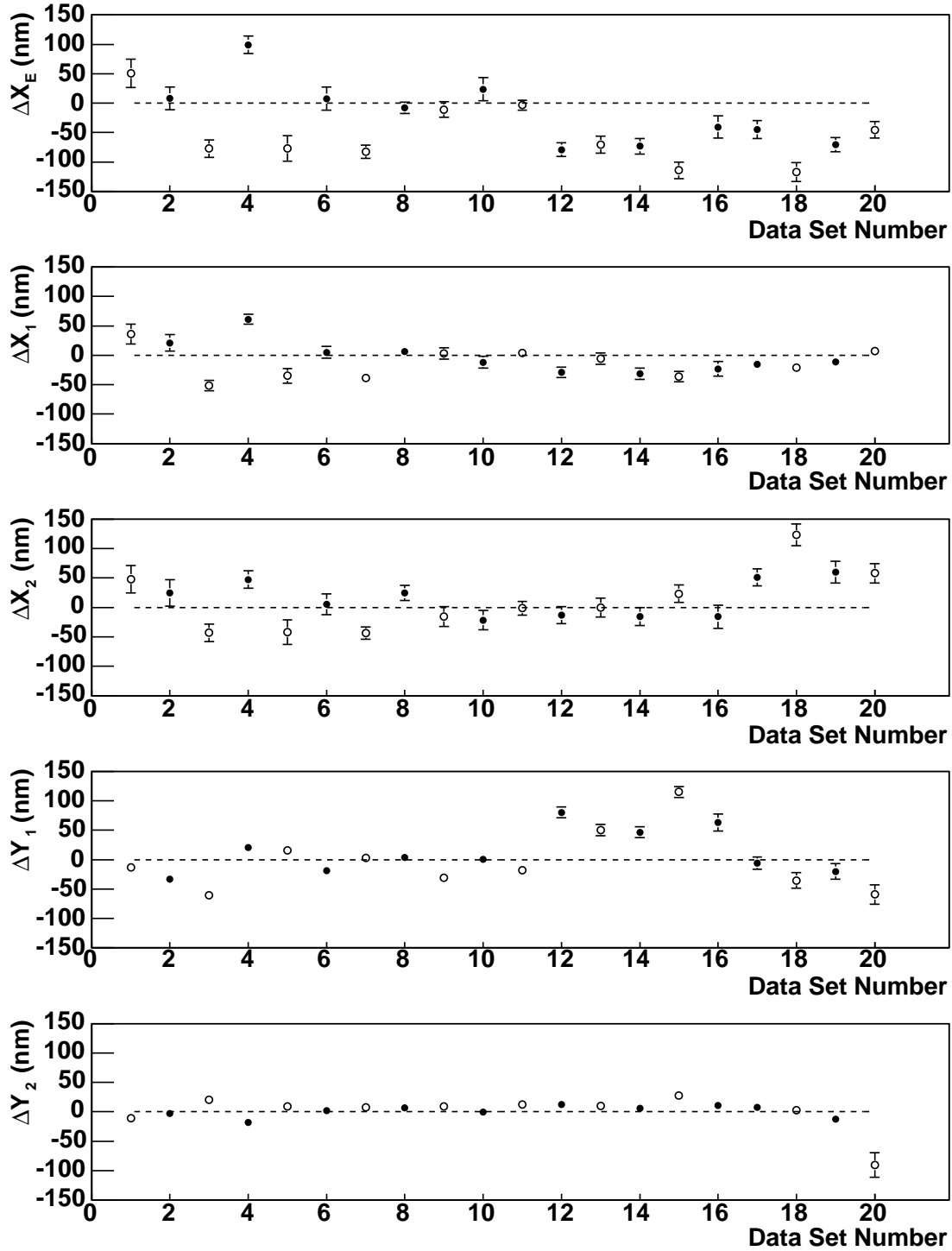}
\caption{Helicity-correlated position differences 
for 1999 run {\em vs.\/} data set, for energy-sensitive position (top plot), horizontal positions at locations on the beamline
7.5 m and 1.3 m upstream of the target
(second and third plots), and
vertical positions at 7.5 m and 1.3 m (fourth and
fifth plots). 
The closed (open) circles correspond to positive (negative)
polarization of the electron beam in the experimental hall.
The data are plotted without correction for sign 
of the electron beam polarization.}
\label{Fig22_RSH_differences}
\end{figure*}

We can write $\langle\Delta A\rangle = \sum_{j=1}^5 a_j \langle\Delta
B_j\rangle$ where $a_j = ( \partial d / \partial B_j ) / 2 \langle
d\rangle$.  The coefficients $a_j$ are stable against changes in the
gains of the detectors and BCMs, as shown in Fig.~\ref{Fig21_RSH_Corrcoef}.  
The helicity-correlated position differences in the beam
monitors are shown in Fig.~\ref{Fig22_RSH_differences}.
Assuming negligible correlations between
these coefficients and the BPM differences, we may compute corrections
to asymmetries averaged over multiple runs using

\begin{table}[h]
\caption[Summary of the detector asymmetry dependence on BPMs.]{Summary
of the detector asymmetry dependence on BPMs for the 1998 and 1999 
runs.  All values are given in units of ppm/$\mu$m.}
\label{Tab_RSH_slopes}
\begin{center}
\begin{tabular}{lrrrr}

BPM   &  \multicolumn{2}{c}{Detector 1} &  \multicolumn{2}{c}{Detector 2}
\\
      &      1998~~~     &        1999~~~~    &       1998~~~~    &         1999~~~~ \\
\hline
$X_E$   & $ -0.2 \pm 0.05 $ & $  -0.39 \pm 0.02 $ & $  0.5 \pm 0.05 $ & $
-0.32 \pm 0.02$ \\
$X_1$   & $ -5.0 \pm 0.7~  $ & $  -3.59 \pm 0.09 $ & $  3.5 \pm 0.6~  $ & $
3.43 \pm 0.09$ \\
$X_2$   & $ 10.4 \pm 0.9~  $ & $   9.07 \pm 0.03 $ & $ -4.8 \pm 0.8~  $ & $
-3.04 \pm 0.03$ \\
$Y_1$   & $ -0.50 \pm 0.06 $ & $  -0.51 \pm 0.06 $ & $  0.10 \pm 0.05 $ & $
~~0.61 \pm 0.06$ \\
$Y_2$   & $  4.7 \pm 0.03 $ & $   ~~~2.48 \pm 0.12 $ & $  ~~0.90 \pm 0.03 $ & $
-1.88 \pm 0.12$ \\
\end{tabular}
\end{center}
\end{table}

\begin{equation}
\langle\langle\Delta A\rangle\rangle = \sum_{j=1}^5 \langle
a_j\rangle \langle\langle\Delta B_j\rangle\rangle\quad.
\end{equation}

\begin{figure*}[phtb]
\includegraphics[width=6.5in]{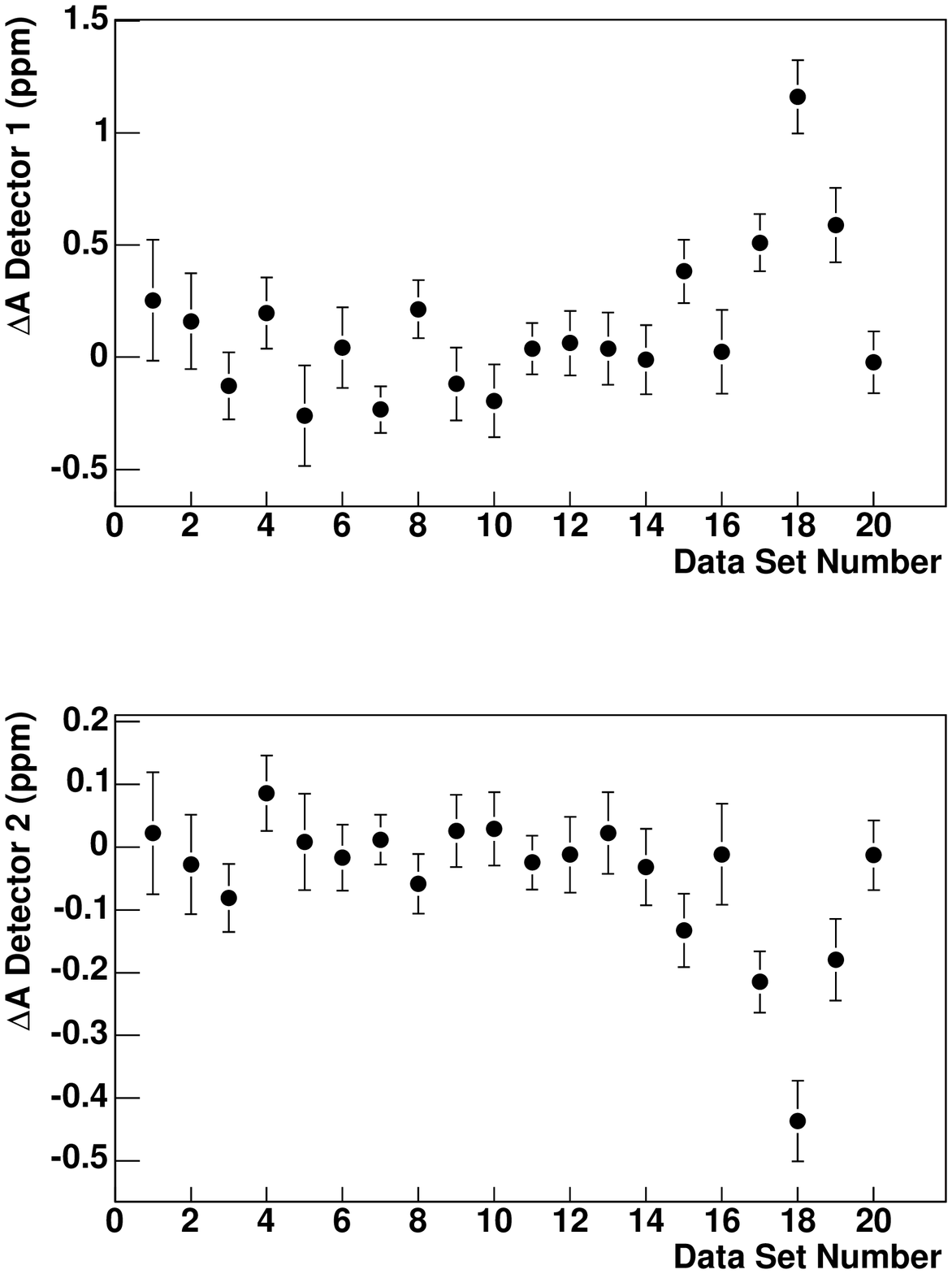}
\caption{Detector correction coefficients for 
1999 run {\em vs.\/} data set.  Note that corrections are generally
consistent with zero at the level of the estimated errors.
The data are plotted
without correction for polarization sign.}
\label{Fig23_RSH_Corrslug}
\end{figure*}

The corrections for each detector as a function of the data set
 are shown in Fig.~\ref{Fig23_RSH_Corrslug}.  
The overall averages of the corrections are
shown in Table \ref{Tab_RSH_Corr}.  The corrections are negligibly
small, as are their contribution to our systematic error.

\begin{table}[tbhp]
\caption{Asymmetry corrections in parts per billion (ppb), 1999 data.
\vspace*{2ex}}
\label{Tab_RSH_Corr}
\begin{tabular}{cccc}
half-wave & Detector 1  & Detector 2 & Average\\
plate state  & (ppb) & (ppb) & correction (ppb) \\
\hline
 & & & \\
out       &   $69 \pm 49$    &  $-45 \pm 21$  &  $14 \pm 27$  \\
in        &   $151 \pm 51$   &  $-39 \pm 21$  &  $60 \pm 28$  \\
 & & & \\
combined  &  $-36 \pm 35$    &  $-3 \pm 15$   &  $-20 \pm 20$ \\
\end{tabular}
\end{table}

\subsection{Pedestals and linearity}
\label{sec:pedlin}

The signals produced by the beam monitors and \v{C}erenkov detectors
ideally are proportional to the actual rates in those devices.  In
reality, however, these signals can deviate from linearity over the
full dynamic range and in general do not extrapolate to a zero
pedestal.  For illustrative purposes, suppose a measured signal,
$S_{\rm meas}$, is a quadratic function of the true rate, $S$:
\begin{equation}
S_{\rm meas} = s_0 + s_1 S + s_2 S^2 .
\end{equation}
Then in the approximation where $|s_0|
\ll |s_1S|$ and $|s_2S^2| \ll |s_1S|$, the {\it measured\/} asymmetry is
\begin{equation}
A(S_{\rm meas}) \approx A(S) \left( 1 + \frac{s_2 S^2}{s_1 S} - \frac{s_0}{s_1 S} \right),
\end{equation}
{\em i.e.} the measured asymmetry is the true asymmetry, $A(S)$, increased by the
size of the quadratic piece relative to the linear piece, and
decreased by the size of the pedestal relative to the linear piece (in
the case where all the coefficients are positive).  

For the normalized detector asymmetries we have $A(D_i/I) \approx
A(D_i) - A(I)$.  Since the average of $A(D_i)$ is an order of magnitude
larger than $A(I)$, we are an order of magnitude more sensitive to
detector pedestals and nonlinearities than we are to beam cavity
monitor pedestals and nonlinearities.

To study the linearity of the detectors and cavity monitors, 
we compared them to an Unser monitor~\cite{Unser},
a parametric current transformer
which can be used as an absolute reference of current.  
For our purposes the Unser monitor's advantage
is its excellent linearity at low currents
which allows us to obtain the cavity monitor pedestals.
However, the fluctuations in the Unser monitor's pedestals, 
which drift significantly on a time scale of 
several minutes, and the ordinarily small 
range of beam currents limited 
the precision of such comparisons during production data taking.
Instead, we use calibration data in which the beam
current is ramped up and down from zero to more 
than 50 $\mu$A.  One cycle takes about a minute.  
The result is that for any given beam
current we have about sixty samples spread over 
a half hour run.  This breaks any random correlation 
between Unser pedestal fluctuations and
beam current and converts the Unser pedestal systematic 
to a random error.

Calibration data exist only for the 1999 run, but studies of the 1998
production data indicate nonlinearities and pedestals during that run
were small in comparison to the 1998 statistics and polarimetry
uncertainties.

\subsubsection{Linearity}
\label{sec:linear}

In order to study linearity, we make scatterplots of one signal versus
another and  fit each scatterplot to a straight line, using only
events where $24~\mu{\rm A} < I_1 < 34~\mu{\rm A}$, a range in which
exploratory fits suggested everything was fairly linear.  We then
examine the residuals between the scatterplots and the fits, relative
to the signal size corresponding to about 32 $\mu$A, over the full
range of beam current.

\begin{figure}
\begin{center}
\includegraphics[width=4.2in]{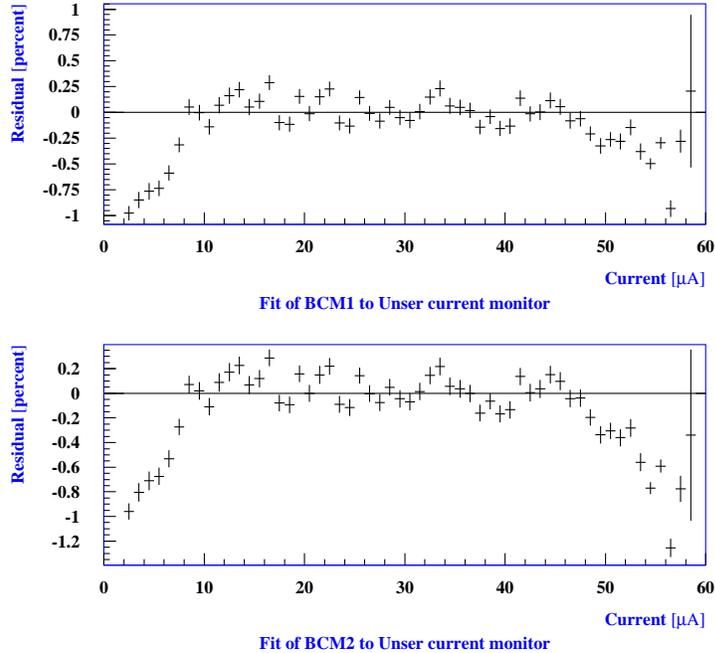}
\caption{(top) Residuals from fit of BCM1 to Unser data, as a fraction
of the BCM1 pulse height at 32 $\mu$A, versus beam current.  (bottom)
Same for fit of BCM2 to Unser.}
% Figure created 20 Oct 1999 on uhep2 using macro 5515
\label{Fig24_RSH_Linearbu} 
\end{center}
\end{figure}

\begin{figure}
\begin{center}
\includegraphics[width=4.2in]{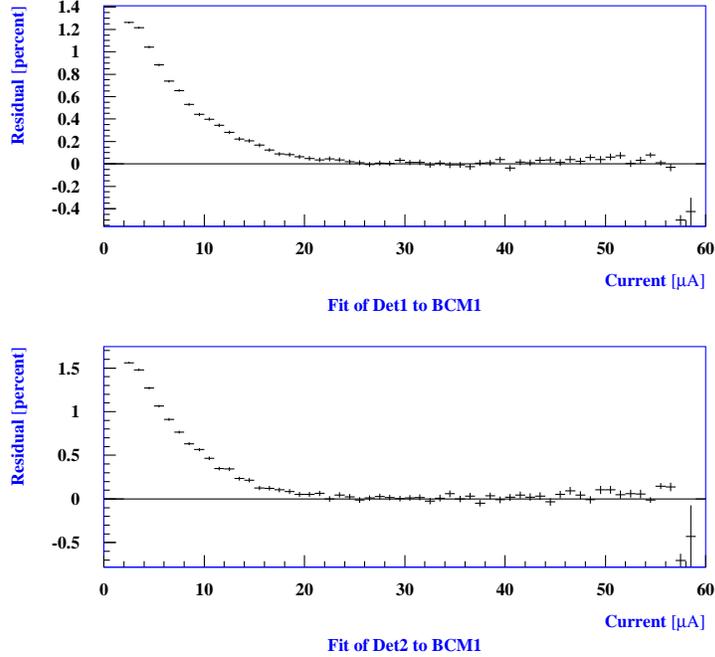}
\caption{(top) Residuals from fit of detector 1 to BCM1 data, as a fraction
of the detector 1 pulse height at 32 $\mu$A, versus beam current.  (bottom)
Same for fit of detector 2 to BCM1.}
% Figure created 20 Oct 1999 on uhep2 using macro 5515
\label{Fig25_RSH_Lineardb1} 
\end{center}
\end{figure}

Figures \ref{Fig24_RSH_Linearbu} to \ref{Fig25_RSH_Lineardb1} show the 
results as a
function of $I_1$.  
In Fig.~\ref{Fig24_RSH_Linearbu}
we see the behavior of the two cavity monitors relative to the
Unser monitor.  Both show deviations from linearity below about 14
$\mu$A and above about 47 $\mu$A, though the high-current problem for
$I_1$ is not as clear-cut as for $I_2$ and the nonlinearities are at
worst about 1\% of the signal.

In Fig.~\ref{Fig25_RSH_Lineardb1} we see residuals for fits of the
two detector signals versus $I_1$.  The nonlinear behavior at low
current is due mainly to the cavity monitors.  From 32 $\mu$A to over
50 $\mu$A the detectors are linear to well under 0.2\%.

We may conclude that the detectors and cavity monitors are linear to
well within the required tolerances.

\subsubsection{Pedestals}
\label{sec:peds}

Detector pedestals were measured easily, by averaging the detector
signals during times when the beam is off.  The resulting pedestals
were always less than 0.3\% of the signal corresponding to the lowest
stable beam current in the production data set, and typically less
than 0.06\%; these pedestals are negligible.

The cavity monitor pedestals cannot be measured this way, since the
cavity signals are meaningless when the beam is off.  Instead, we fit
$I_{1(2)}$ to $I_U$ in the calibration data and extrapolate to zero
current.  Such an extrapolation requires knowledge of the average
Unser pedestal, which is obtained from the beam-off data in the same
run.  The resulting pedestals are less than 2\% of the signal
corresponding to the lowest stable beam current in the production data
set.

Are the cavity monitor pedestals obtained in the calibration data
typical of the 1999 data?  In order to answer this, we must make the
reasonable assumption that the cavity monitor linearities are stable
at the negligible level seen in the calibration data.  If that is the
case, then with negligible pedestals and nonlinearities for the
detectors, a straight line fit to a scatterplot of $A(D_{\rm meas})$
{\it vs.} $A(I_{\rm meas})$ should give a 
slope equal to 1.0 if $A(I_{\rm meas})$ is
computed with a {\it corrected} BCM signal in which the pedestal
measured in the calibration data is subtracted off.  Any residual
pedestals would give a deviation from unity equal to the size of the
pedestal relative to the size of the signal.  We find that such
deviations are negligible.

\subsubsection{Pedestal and linearity conclusions}
\label{sec:pedlinconcl}

No corrections for pedestals or nonlinearities need to be applied.
The nonlinearities of the detectors and cavity monitors
were negligible over the dynamic range of the beam current we ran.
The pedestals for detectors and cavity monitors were negligible.

\section{Normalization}
\label{sec:phys_asy}

To extract physics results from the raw 
measured asymmetry, one needs to 
correct the beam polarization,
estimate and correct
for any contributions from background processes, 
and determine the
average $Q^2$ of the elastically-scattered electrons, 
weighted by the response of the
detectors. In addition one must apply radiative
corrections and correct for the finite acceptance.
This section describes each of these 
steps of the data analysis. 

\subsection{Beam polarization}
\label{sec:beam_pol}

Transverse components of the beam polarization are a negligible source
of systematic error; the maximum analyzing power for a point
nucleus is $< 10^{-8}~$\cite{Mott} and the transverse
component bounded by M{\o}ller polarimetry results
was $\le P_Z \sin(10^{\circ})$ where $P_Z$ is
the longitudinal polarization.
Explicit calculations of the vector
analyzing power arising from two-photon exchange diagrams, including
proton structure effects, yield an analyzing power of less than 0.1
ppm~\cite{Afanasev} for our kinematics.  At different kinematics, a
larger analyzing power, ($-15.4 \pm 5.4$)ppm, was measured in the
SAMPLE experiment~\cite{Wells}, in reasonable agreement with the
predicted value~\cite{Afanasev}; the much smaller value expected for
our kinematics is a consequence of the higher beam energy and small
scattering angle. The left-right symmetry of the apparatus further
suppresses our sensitivity to transverse components. 
The determination of the magnitude of the polarization proceeded
differently in the two running periods, and is described below. 

\subsubsection{1998 Run}
\label{sec:pol_1998}

For the 1998 running period, we used the Mott and M{\o}ller
measurements to determine the absolute beam polarization, averaged over
the entire running period.  This average was used to correct the
asymmetry averaged over the running period.
The Compton polarimeter was not yet available. 
The average of 16 Mott measurements
yielded a polarization of $(40.5 \pm 2.8)$\%. The quoted error is
dominated by the systematic error due to extrapolation to zero target
foil thickness (5\% relative error), background subtraction (3\%), and
observed variations in the measured $P_e$ with beam current (3\%).

The average of several M{\o}ller measurements yielded $\langle
P_e\rangle = (36.1 \pm 2.5)$\%, in reasonable agreement with the Mott results
(note that the M{\o}ller results are 3\% lower than those reported in 
\cite{aniol1}, due to a subsequent recalibration of the polarization 
of the target foil). The uncertainty was dominated by knowledge of the 
foil polarization (5\% relative error). 

Averaging the Mott and M{\o}ller results we obtain the final result for the 
1998 run of $\langle P_e\rangle = (38.2 \pm 2.7)$\%. Note that we conservatively 
choose not to reduce the error by $\sqrt{2}$ when averaging the results. 

\subsubsection{1999 Run}
\label{sec:pol_1999}

For the 1999 running periods, we used the M{\o}ller measurements to
determine the absolute beam polarization for each of the 20 data sets.  
These averages were used to correct the asymmetries averaged over each
data set.
Typically there were between one and three
M{\o}ller measurements during each data set; these measurements were
averaged to determine $\langle P_e\rangle$ for that data set.  For two
data sets there were no M{\o}ller measurements and $\langle P_e\rangle$
was set to the average of $\langle P_e\rangle$ for the preceding and
following data sets. The polarization average over all the data sets
was $(68.8 \pm 2.2)\%$.

At the time of this run, the M{\o}ller was fully commissioned, and
the systematic errors were reduced by more than a factor of two. 
Thus we did not make regular Mott measurements, however those that 
were done were in reasonable agreement with the  M{\o}ller results.

The M{\o}ller measurement is invasive, as it involves significantly reducing the 
beam current and inserting the M{\o}ller target in the beam, and so
these measurements were only made at intervals. A possible concern
is that the polarization may be varying between M{\o}ller measurements,
and thus a non-invasive, continuous measurement of the beam polarization
was desirable. This was provided in the 1999 run by the Compton polarimeter. 

\subsubsection{Compton Polarimeter: 1999 Run Results}
\label{sec:cpt_exp_res}

Under the conditions of the 1999 run (electron beam energy of 3.3 GeV and current
of 40 $\mu$A) the measured Compton rate was 58 kHz and the
experimental asymmetry was 1.3\%. Due to the high gain of the
Fabry-Perot cavity coupled to a standard 300 mW laser, a relative
statistical accuracy of 1.4\% was achieved within an hour, inside the
analysis cuts. All the systematic errors of the measurement discussed
above in section \ref{sec:cpt_exp_meth} are listed in Table~\ref{tab:error_bud} 
and lead to a total uncertainty of 3.3\%.

\begin{figure*}
\epsfig{figure=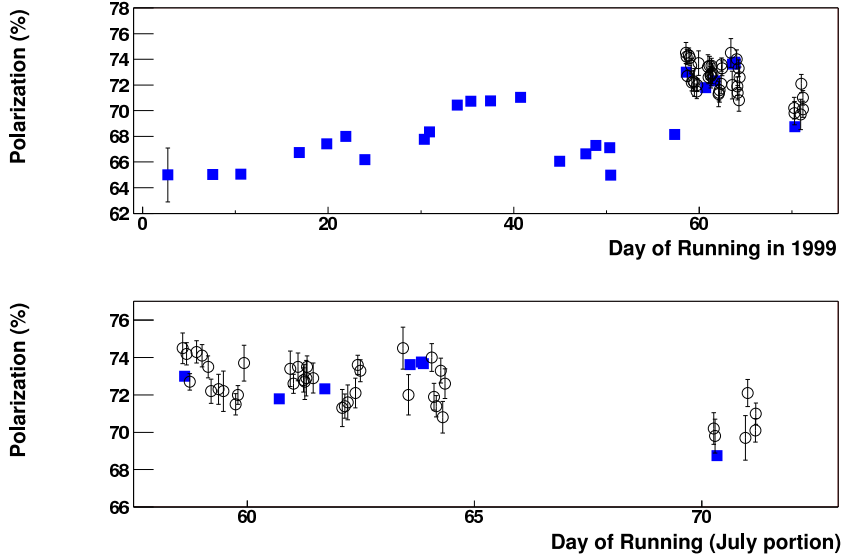,angle=-90,width=0.8\textwidth}
\caption{Polarization of the JLab electron beam measured by the
M\o ller (solid squares) and the Compton (open circles) polarimeters 
during the entire 1999 run (upper plot) and
July portion (lower plot) where the Compton polarimeter
was available.
The error bar on the left-most M\o ller point in
the upper plot is its total error (dominated by systematic
error 3.2\% relative) while all other points show only
the statistical error, which for M\o ller data
is smaller than the symbol (0.2\% relative).}
\label{fig26:polar_time}
\end{figure*}

Forty polarization measurements were 
performed by the Compton polarimeter
in July 1999 in good agreement with
measurements from the M\o ller polarimeter (see Fig.~\ref{fig26:polar_time}). 
They provide, for
the first time, an essentially continuous monitoring of the electron beam polarization with a
total relative error from run-to-run of less than 2\% (due to the correlations of the systematics
on $\ath_c$ between consecutive runs). Large variations of the beam
polarization between two M\o ller measurements are excluded by the 
Compton data. More details on the Compton results are available in 
a separate publication~\cite{Baylac}.

Several hardware improvements have been added to the setup since then,
including new front-end electronic cards and electron beam position
feed-back. An electron detector made of 4 planes of 48 micro-strips is
now operational and reduces the systematic errors related to the
detector response.

\subsubsection{Experimental asymmetries}

The experimental asymmetries for the three running periods and two
half-wave plate settings, corrected for the signs and magnitudes of
the measured beam polarizations, are given in Table
\ref{Tab_RSH_Asym}.  For each running period, all the asymmetries are
statistically compatible.  The Apr/May 1999 and July 1999 results
would be negligibly different if we used asymmetries and polarizations
averaged over all data sets.

\begin{table*}[tbhp]
\caption{Asymmetry results (ppm).
$\aexp_1$ and $\aexp_2$ are the asymmetries
of our two detectors normalized to beam current and corrected
for sign and magnitude of beam polarization.
$\aexp_s$ is the asymmetry of the summed detectors,
$\aexp_a$ is the average of the asymmetries of the
detectors, see
section \ref{sec:rsh_rawasy}.  
$\asy_I$ is the beam current asymmetry corrected for
sign of beam polarization.}

\label{Tab_RSH_Asym}
\begin{tabular}{lrrrrr}
 & \multicolumn{1}{c}{$\aexp_1$} & \multicolumn{1}{c}{$\aexp_2$} & \multicolumn{1}{c}{$\aexp_s$} & \multicolumn{1}{c}{$\aexp_a$} & \multicolumn{1}{c}{$\asy_I$} \\

1998 half-wave out & $ 13.1 \pm 3.7$ & $ 16.0 \pm 3.8$ & $ 14.4 \pm 2.7$ & $ 14.5 \pm 2.6$ & $  0.50 \pm 0.21$ \\
1998 half-wave in  & $  8.5 \pm 4.0$ & $ 20.8 \pm 4.1$ & $ 14.2 \pm 2.9$ & $ 14.6 \pm 2.9$ & $  0.18 \pm 0.26$ \\
All 1998 data      & $ 11.0 \pm 2.7$ & $ 18.2 \pm 2.8$ & $ 14.3 \pm 2.0$ & $ 14.5 \pm 1.9$ & $  0.37 \pm 0.09$ \\
\hline
Apr/May 1999 
half-wave out      & $ 14.8 \pm 2.2$ & $ 17.1 \pm 2.3$ & $ 16.0 \pm 1.6$ & $ 15.9 \pm 1.6$ & $ -0.79 \pm 0.11$ \\
Apr/May 1999 
half-wave in       & $ 17.1 \pm 2.3$ & $ 10.9 \pm 2.4$ & $ 13.9 \pm 1.7$ & $ 14.1 \pm 1.7$ & $ -0.76 \pm 0.14$ \\
All Apr/May
1999 data          & $ 15.9 \pm 1.6$ & $ 14.2 \pm 1.6$ & $ 15.0 \pm 1.2$ & $ 15.1 \pm 1.1$ & $ -0.78 \pm 0.09$ \\
\hline
Jul 1999 
half-wave out      & $  9.2 \pm 4.5$ & $ 11.7 \pm 4.7$ & $ 10.7 \pm 3.3$ & $ 10.4 \pm 3.3$ & $ -0.10 \pm 0.81$ \\
Jul 1999 
half-wave in       & $ 20.6 \pm 6.2$ & $ 15.8 \pm 6.6$ & $ 18.1 \pm 4.5$ & $ 18.4 \pm 4.5$ & $  0.56 \pm 0.61$ \\
All Jul
1999 data          & $ 13.2 \pm 3.7$ & $ 13.1 \pm 3.8$ & $ 13.3 \pm 2.7$ & $ 13.2 \pm 2.6$ & $  0.32 \pm 0.49$ \\
% 1998 data from c2_pn.ssums divided by 0.388.
% 1999 data from spreadsheet and apr_may.sum, jul.sum.
\end{tabular}
\end{table*}

Note that, for all the groups of data, $\aexp_s$ (asymmetry of the summed
signal from the two detectors) and $\aexp_a$ (average asymmetry from the
two detectors) are essentially identical, with identical widths. This
indicates that the two detectors are statistically independent,
demonstrating that both false asymmetries and target density
fluctuations are negligibly small.

\subsection{Backgrounds}
\label{sec:backgrounds}

The two backgrounds that we observed were: 
1)~electrons that scattered inelastically
and then rebounded into the detector;
and 2)~electrons from the target aluminum walls.
In addition, we put an upper limit on the contribution
from magnetized iron in the spectrometer, based on
measurements using a ``proton tagging'' technique, which
was confirmed by simulation.   
In this section we describe the corrections and
systematic errors due to these backgrounds.

\subsubsection {Electrons from Inelastic Scattering}
\label{sec:inel_bgr}

The main background to proton elastic scattering
in the Hall A spectrometers near the HAPPEX
kinematics comes from electrons that scattered
inelastically and then re-scattered
inside the spectrometer after the dipole.
Much of this re-scattered debris is
in the form of low energy charged or neutral 
particles which contribute little to the 
integrated signal in our calorimeter detector. 
The validity of this ``re-scattering model''
was studied with simulation of the optics, 
as well as
with a data set of e-P elastic scattering runs
with energies and angles nearby the HAPPEX
running conditions.  The energies varied from 3.2 to 4.0 GeV
and angles from 12.5$^\circ$ to $35^\circ$.
Several observables of background were
studied from this data set, to verify that they tracked with our model.
The model was applied to the HAPPEX kinematics to obtain
the correction and systematic error for re-scattering from the
$\pi$ threshold through the $\Delta$ resonance region.

The re-scattering model is based on the assumption 
that the background, as a fraction of the elastic scattering
signal, is given by the following integral over
the energy of the scattered electron:
\begin{equation}
\label{eq:scattsig}
B = \int_{E_{\rm thr}}^{E_{\rm max}}
dE \ P_{\rm rs}(E) \times R(E) 
\end{equation} 
where $P_{\rm rs}$ is the product of 
the probability to re-scatter in the spectrometer
and the energy deposited by the scattered electron
\begin{equation}
\label{eq:prescatt}
P_{\rm rs} = {\rm (energy \  deposited)} \times
{\rm (re-scatter \ probability) }
\nonumber
\end{equation}
and $R(E)$ is the ratio of inelastic to elastic cross section, 
\begin{equation}
\label{eq:ratioinel}
R(E) = {\left(\frac{d\sigma}{d\Omega dE}\right)_{\rm{inel}}}
\ /\  
\left(\frac{d\sigma}{d\Omega}\right)_{\rm{elastic}} 
\nonumber
\end{equation}
and the integral extends from the inelastic 
threshold $E_{\rm thr}$ to the maximum
energy loss $E_{\rm max}$ that could contribute, 
about 20\% below the beam energy.  
 
\begin{figure}
\includegraphics[scale=0.50]{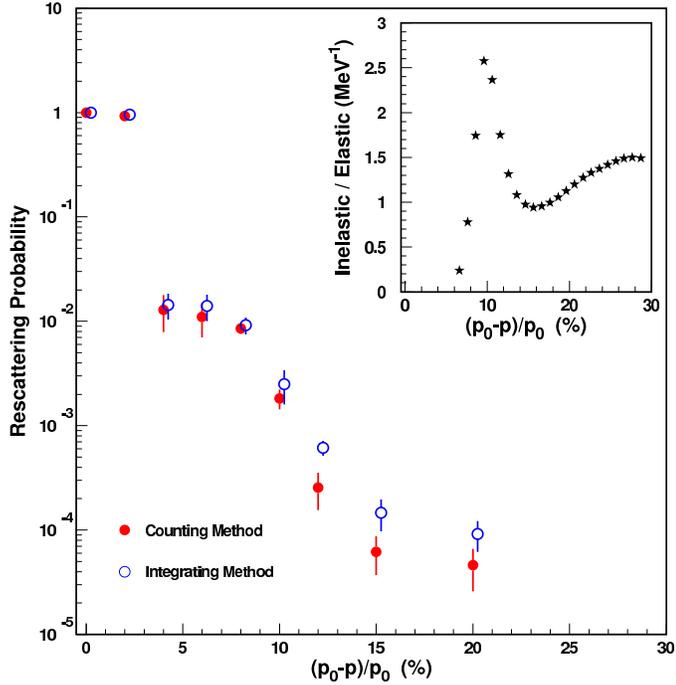}
\caption{\label{fig27_field_scan}
Results of scan of spectrometer magnetic fields to measure the
probability to re-scatter into the detector {\em vs.} the fractional difference
from the nominal momentum setting. Inset: the ratio of inelastic to elastic cross sections
at the HAPPEX kinematics, 
$(d\sigma/d{\Omega}dE)_{\rm inel} / (d\sigma/d{\Omega})_{\rm elast}$.}
\end{figure}

Measurements of the re-scattering function
$P_{\rm rs}$ are shown in Fig.~\ref{fig27_field_scan}.
The measurement was performed by scanning the magnetic
fields in the spectrometer to
force the elastically scattered electrons to follow
trajectories that simulate inelastically scattered 
electrons; we measured the signal in the 
detectors as a function of the field increase.
The measurements were done both with the counting technique, using
the standard spectrometer DAQ, and with the integrated technique, using
the integrated HAPPEX detector signal.
For the individual counting technique, one measures
a rate above a threshold used to trigger the DAQ,
and one multiplies this rate by the
amplitude in the detector; the integrating technique
measures this product directly.
The $\Delta$ resonance contribution is suppressed
by two orders of magnitude by the spectrometers.
The inelastic and elastic e-P cross
sections were taken 
from a parameterization of SLAC data~\cite{ep_whitlow}.
As an example, we show in Fig.~\ref{fig27_field_scan} the ratio
$R(E)$ for the HAPPEX kinematics ($Q^2 = $ 0.48 (GeV/c)$^2$).

\begin{figure}
\includegraphics[width=3.4in]{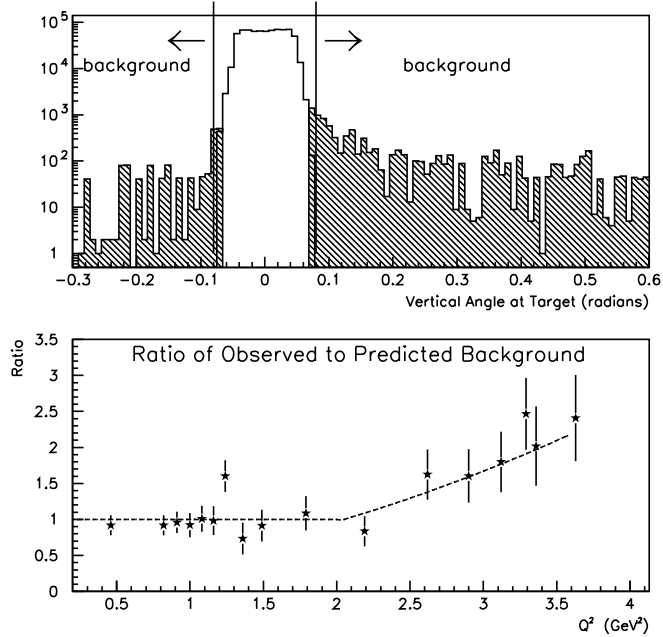}
\caption{\label{fig28_thtgt}
Top: Reconstructed vertical angle at the target, from triggered data; background from re-scattering
of inelastic electrons indicated by hatched area.  Bottom: The ratio of observed to
predicted re-scattering background {\em vs.} $Q^2$; the ratio is 1 in the region of 
our  kinematics ($Q^2 = 0.48$ (GeV/c)$^2$). The line is a guide to the eye. }
\end{figure}

In the spectrometer event-trigger data, 
backgrounds are identified using the following observables:
1) energy in lead glass too low; 2) momentum
of electron too high; and 3) target variables outside the normal
region.  The target variables used were 
the position in the scattering plane perpendicular to the 
central trajectory, as well as the vertical and horizontal angles
reconstructed at the collimator.  
The observable best correlated to the re-scattering background is
the vertical angle at the target, because inelastically scattered
electrons which strike near the focal plane create secondaries
which have an angle that extrapolates to a position above
the collimator.  In Fig.~\ref{fig28_thtgt} we show the definition of
this background observable and its agreement with the model.
The validity of the re-scattering model is demonstrated by the
ratio of observed to predicted background, 
which is close to 1.0 at the HAPPEX 
kinematics for most observables.
For some of the other observables, the ratio was less than one 
since the observables measure only part of the background.
Note that for this comparison, instead of using the
energy-weighted re-scattering function, we use the probability
to re-scatter into the focal plane which is measured by
the magnet scan using the individual counting technique.

Above $Q^2 = 2$ (GeV/c)$^2$ the model under-predicts
the observed backgrounds and there was a growing rate 
of pions seen with particle identification cuts that use the
\v{C}erenkov and lead glass detectors.
However, the model works fairly well within the range 
$Q^2 = 0.5$ to 1.0 (GeV/c)$^2$ where there are no pions.
We conclude that re-scattering in the spectrometer
is the main source of background to e-P elastic scattering
and is $B = (0.20 \pm$ 0.05)\% of our detected signal
(Eq. \ref{eq:scattsig}).

The background is mainly due to 
the $\Delta$ resonance (see Fig.~\ref{fig27_field_scan}).
To compute the correction to our data, we
use the predicted parity-violating asymmetry from the $\Delta$ 
resonance \cite{bob_mukhopadhyay}
\begin{equation}
\label{eq:deltaasy}
\apv_{\Delta} \approx 
{\frac{-G_F | Q^2 |}{2 \sqrt{2} \pi \alpha}} 
( 1 - 2{\rm sin}^2 \theta_W)
\end{equation}
The asymmetry is $(-47 \pm 10)$ ppm at our $Q^2$ which is 3 times
as large as the asymmetry for elastic scattering.
In Ref.~\cite{bob_mukhopadhyay}, 
various small additional terms and 
theoretical uncertainties are discussed in detail,
including non-resonant hadronic vector current background,
axial vector coupling, and hadronic contributions to 
electroweak radiative corrections.
The extra terms are typically 4\%
and have opposite signs that tend to cancel.
We therefore ascribe a conservative error of 20\%
to the asymmetry and arrive at a correction to
our experimental asymmetry of (0.06 $\pm$ 0.02)~ppm,
where the error includes the estimated systematic error
of the re-scattering model.

\subsubsection {Quasielastic Scattering from the Target Walls}
\label{sec:qe_bgr}

Scattering from the target aluminum windows contributed
(1.4 $\pm$ 0.1)\% to our detected signal.
This background can be observed in the reconstructed
target position in the region
of momentum above the elastic peak, where one sees an
enhancement in the target window regions which is
due to quasielastic scattering.  
A more direct
measure of this background was performed by
inserting into the beam an empty
aluminum target cell, similar to the one used to
contain liquid hydrogen, and measuring the 
signal in our detector. 
The thickness of the empty target cell walls is about 10 times
that of the walls used in the hydrogen cell, in
order to compensate for the radiative
losses in the hydrogen cell.  

The correction to our data arises from the neutrons
in the aluminum target. 
The kinematic setup of the spectrometer selects electrons
which have scattered quasielastically from protons and
neutrons in the aluminum.  
For quasielastic scattering from a nucleus with $Z$ protons
and $N$ neutrons, the expected parity-violating asymmetry is 
\cite{bob_donmus}
\begin{equation}
\label{eq:asyqe}
\apv_{\rm QE} = 
\frac{-G_F | Q^2 |}{4 \sqrt{2} \pi \alpha}
\frac{ W^{\rm PV} }{W^{\rm EM}} 
\end{equation}
where, following the notation of \cite{bob_donmus},
\begin{equation}
\label{eq:wem}
{W^{\rm EM}}  =   
\epsilon \lbrack Z {(G_E^p)}^2 + N {(G_E^n)}^2 \rbrack 
 + 
\tau \lbrack Z {(G_M^p)}^2 + N {(G_M^n)}^2 \rbrack 
\nonumber
\end{equation}
and 
\begin{equation}
\label{eq:wpv}
{W^{\rm PV}}  =
\epsilon \lbrack Z G_E^p \tilde G_E^p 
+ N G_E^n \tilde G_E^n \rbrack 
 + 
\tau \lbrack Z G_M^p \tilde G_M^p 
+ N G_M^n \tilde G_M^n \rbrack 
\nonumber
\end{equation}

where the $G$'s are nucleon electromagnetic form factors, the
$\tilde G$'s are the weak nucleon form factors, 
$\epsilon, \tau$ are 
the usual 
kinematic quantities (see definitions after
Eq. ~\ref{eq:pvasy})
and we have neglected small
axial vector and radiative correction terms.
The predicted asymmetry for quasielastic aluminum
scattering is -24 ppm at our $Q^2$.
We obtain a correction (0.12 $\pm$ 0.04) ppm,
where we have assumed that the asymmetry from this
process is known with a relative accuracy of 30\%.

\subsubsection {Magnetized Iron in the Spectrometer}
\label{sec:mag_iron}

Scattering from the magnetized iron in the spectrometer
is a potential source of systematic error because 
of the polarization dependent asymmetry in $\vec e, \vec e$ 
scattering (M{\o}ller scattering).
In this section we describe the
analysis which led to an upper bound for this effect.

Using the two HRS spectrometers we performed
``proton tagging'' measurements
in which we used protons from elastic e-P scattering to
tag the trajectories of electrons. 
We set up the two spectrometers
slightly mispointed, so that for electrons that come 
close to the edge
of the acceptance, the corresponding protons are well within
the proton arm acceptance.
Thus, the protons can tag electrons which might hit the
magnetized iron of the pole tips.

To measure the backgrounds in the electron spectrometer
we use the lead glass detector, which is read out
in a bias--free way for every proton trigger or 
other triggers.
In the low-energy tail
of the energy spectrum, which contains backgrounds,
we measure the excess energy 
for events in which the electrons come closest
to the pole tips.  The excess is measured relative to
the energy spectra for electrons in the middle of
the acceptance.  No enhancement was seen for the
``poletip scattering'' candidate events, and we
placed an upper bound that $\ll 10^{-4}$ of
the energy in our detector arises from poletip scattering.

Simulations of the magnetic optics confirmed
these observations.
The acceptance of the spectrometer is defined
primarily by the collimators, and secondarily by the
first two quadrupoles in the QQDQ design.
Practically no high-energy rays strike magnetized iron.
In addition, secondaries from
reactions in which particles which have
struck the first elements of the spectrometer
tend to be low energy and get swept away
before hitting the detector.

The correction to our data from poletip scattering is
\begin{equation}
\label{eq:pol_da}
dA = 
f \ P_{e1} \ P_{e2} \ A 
\end{equation}
where $f$ is the fraction of our signal ($f \ll 10^{-4}$),
\hskip 0.05in $P_{e1}, P_{e2}$ are the polarizations of
the scattered electron and the electron
in the iron ($P_{e1} \sim 0.8$ and $P_{e2} \sim 0.03$),
and $A$ is the analyzing power $A \le 0.11$.
The result is conservatively $dA \ll 0.26$ ppm
and we make no correction for this effect.

\subsubsection {Backgrounds in HAPPEX Triggered Data}
\label{sec:happex_trig}

Backgrounds could be studied under
the conditions of the experiment by using the HAPPEX
detector to define the trigger.
A signal above a discriminator threshold was
used to trigger the spectrometer DAQ and read out the drift
chambers and other detectors.
  
One small source of backgrounds was electron
scattering from the aluminum
frame of the HAPPEX detector, observed in a correlation
between the amplitude in the detector and the track
position.  At the location of the detector frame
a small enhancement $\sim 10^{-3}$ in low 
energy background was seen which 
in addition should have the same asymmetry
and is therefore a negligible systematic.
The neutral particle component of background from the HRS
was measured as the energy-weighted sample of events
which had no track activity, and was a $\le 0.2$\% background.
For the charged particle component, the method of 
analyzing the background was similar to what
was described above for the e-P runs.
We reconstructed tracks and traced them back through 
the spectrometer to the collimator.   
The percentage of tracks that miss an aperture 
is a measure of the background as well as other problems
including mis-reconstruction.
One complication of placing the HAPPEX
detector near the drift chambers was that secondaries
from showers splashed back into the chambers, causing
confusion in the reconstruction.  
In event displays such events were often ambiguous with 
other background candidate events and could not be
easily subtracted by a pattern recognition algorithm.
Other chamber problems included
inefficiency, scattering inside a chamber, two-track
confusion due to overlap of two events, 
and events in which an abnormal 
array of hits with bad fit $\chi^2$
existed in only one of
the four chambers.
This latter category was easily eliminated.  
We eliminated many of the
two-track events by rejecting events in which one
of the tracks had a good fit and was within
0.2 GeV of the elastic peak.  From the remaining
sample, we obtained an upper bound $\le$ 0.5\% background
which is a weaker upper bound than that obtained from
the re-scattering model.
Because of the limitations in reconstructing events
at the $10^{-3}$ level we consider the re-scattering
model to be a more accurate assessment of our background.

\subsubsection{Summary on Backgrounds}
\label{sec:sum_bgr}

Table ~\ref{bob_tablebg} lists the backgrounds,
the correction to our data, and the systematic
error. The total correction was $+(0.18 \pm 0.04)$ ppm,
which represents a $(1.2 \pm 0.3)$\% correction to the 
experimental asymmetry.

\begin{table}\centering
\caption[] {\bf Backgrounds and Corrections.\vspace*{2ex}}
\begin{tabular}{l|c|c|c|c }  
Source &  Fraction Events &  A (ppm) & Correction (ppm) \\
\hline\hline
Inelastic ${\rm e}^-$  &  0.2 \%   &  -47  &  0.06 $\pm$ 0.02  \\
Al walls & (1.4 $\pm$ 0.1)\% & -24  & 0.12 $\pm$ 0.04  \\
Magn. Iron & $\ll 10^{-4}$ & $\le$ 2700 & none  \\
\end{tabular}
\label{bob_tablebg}
\end{table}

\subsection{Measurement of $Q^2$}
\label{sec:qsq}

The square of the four-momentum transfer is 
$Q^2 = 2 E E^{\prime} (1- \cos(\theta))$
where the three ingredients needed are the incident energy $E$, 
final energy of the electron $E^{\prime}$, and
the scattering angle $\theta$.  For elastic scattering 
one may eliminate one of the three variables,
which provides a consistency check.
The kinematics
were $E \sim$ 3.3 GeV, $\theta = 12.5^\circ$ 
(see table Table~\ref{bob_tableq2}).
  
\par The beam energy is measured by two methods to an accuracy
of about 1 MeV.  One apparatus, called 
the arc method \cite{bob_arcmeas}, measures
the deflection of the beam in the arc of magnets that lead into
the experimental hall, for which the integral of the field is precisely
known.  A second apparatus, called 
the e-P method \cite{bob_epmeas}, measures the
kinematics in e-P coincidences on hydrogen.
When we assumed that beam energy was correctly measured 
in the 1999 run, we found that an $-8$ MeV
($-0.2$\%) adjustment was needed for the $Q^2$
in the 1998 run to be consistent with elastic scattering
after known corrections for angle and momentum 
calibration of the scattered electron.
Based on this, and based on the history of comparisons of the
two energy apparatus, we have assigned a very conservative
10 MeV error to our energy measurement.

\par A second ingredient required for the $Q^2$ determination
is the momentum of the scattered electron.  
We adjusted the momentum scale by a few tenths of a percent 
in order to satisfy the missing mass constraint for elastic scattering.
Subsequently, the magnet constants were measured by an independent
group and found to agree within 0.1\% of our values.

\par The largest error in $Q^2$ comes from the scattering angle.
There are two ingredients here:  \hskip 0.05in 1) surveys 
measure the angle of the spectrometer's optic axis relative to
the incident beam direction; and \hskip 0.05in
2) the spectrometer reconstruction code
reconstructs the horizontal and vertical angles at the target
relative to the optic axis
using tracking detectors in the focal plane.
Calibration of the optical transfer
matrix for the spectrometers is performed
by sieve slit runs in which the
optical transfer matrix of the spectrometers
is calibrated in the following way.
A 0.5 cm thick tungsten plate with a
rectangular pattern of holes covering the
acceptance (sieve slit) is placed at the 
entrance of the spectrometers,
and tracks in the focal plane are used to
reproduce the hole pattern through
a $\chi^2$ minimization procedure.
Location of this sieve slit requires additional
survey information.  The combined error in these
ingredients gives a 1 mrad error in the scattering angle.

\begin{figure}
\includegraphics[width=3.5in]{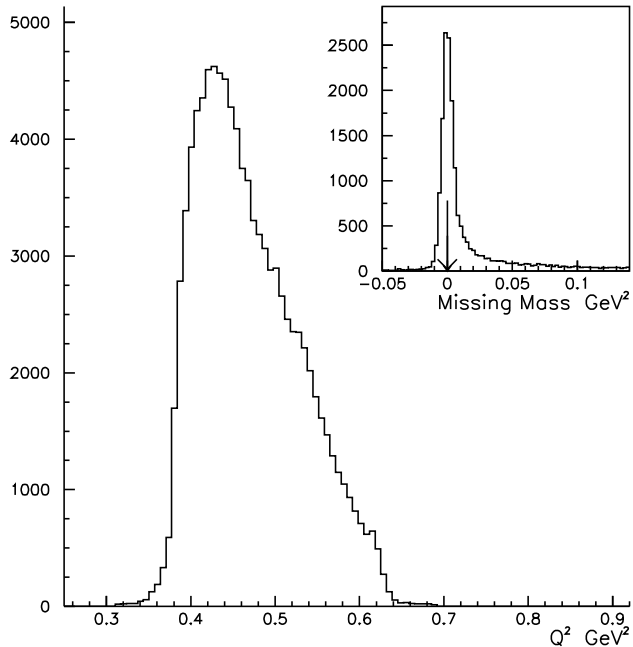}
\caption{\label{fig29_qsq_typical}
Typical $Q^2$ spectrum measured during HAPPEX.
In the inset is a missing mass spectrum from the same data.}
\end{figure}

The measurements of $Q^2$ from the 1998 and 1999 runs
are given in Table~\ref{bob_tableq2}. These take into account
the average energy loss in the target and
a weighting by amplitudes in the HAPPEX
detector according to
$Q^2 = (\Sigma Q_i^2 A_i) / (\Sigma A_i)$ where
$A_i$ are ADC amplitudes in bin $i$ and 
$Q_i^2$ is the corresponding measurement. 
This weighting shifted $Q^2$ by 
$(-0.38 \hskip 0.02in \pm \hskip 0.02in 0.05)$\%.
A typical $Q^2$ distribution and missing mass
spectrum is shown in Fig.~\ref{fig29_qsq_typical}.

\par In Table~\ref{bob_tableeq2} we summarize the errors which add in
quadrature to 1.2\% or $\pm 0.006$ (GeV/c)$^2$
for each spectrometer.  
The matrix element error is an estimate
of the instability in the fitting procedure for
the sieve slit calibration.  The estimate of time drifts
was based on the observed variation with time of $Q^2$
and the observed time variation in the results from
sieve slit runs and surveys.

\begin{table}
\caption{\label{bob_tableq2}
$Q^2$ for 1998 and 1999 HAPPEX Runs}
\begin{ruledtabular}
\begin{tabular}{cccc}
 &  1998 Run  &  1999 Run (I) & 1999 Run (II) \\
 &  &   &  \\
\hline
Beam Energy &   3.345  &   3.353  & 3.316  \\
(GeV) &  &   &  \\
L-arm Angle &  12.528${}^\circ$ & 12.527${}^\circ$ & 12.527${}^\circ$ \\
R-arm Angle &  12.558${}^\circ$ & 12.562${}^\circ$ & 12.562${}^\circ$ \\
L-arm $Q^2$  & 0.473 & 0.477 & 0.466 \\
R-arm $Q^2$  & 0.475 & 0.477 & 0.466 \\
(GeV/c)$^2$ &  &   &  \\
$Q^2$ Error  & $\pm$0.006 & $\pm$0.006 & $\pm$0.006 \\
\end{tabular}
\end{ruledtabular}
\end{table}

\begin{table}
\caption{\label{bob_tableeq2} Summary of Errors in $Q^2$}
\begin{ruledtabular}
\begin{tabular}{cccc}  
Error Source  & Error  & Error in $Q^2$   \\
Timing Calibration         &  $\le$ 5 nsec  &   $ \le $ 0.1\% \\
Beam Position              &   0.5 mm       &     0.5\%       \\
Survey of Spectr. Angle    &   0.3 mrad     &     0.3\%       \\
Survey of Mispointing      &   0.5 mm       &     0.5\%       \\
Survey of Collimator       &   0.5 mm       &     0.5\%       \\
Target Z position          &   2 mm         &     0.3\%  \\
Momentum Scale             &    3 MeV       &     0.1 \%      \\
Beam Energy                &   10 MeV       &     0.3 \%      \\
Matrix Elements            &                &     0.4 \%      \\
Drifts in Time             &                &     0.5 \%      \\
{\bf Total Systematic Error} &              &     1.2 \%      \\
\end{tabular}
\end{ruledtabular}
\end{table}

The asymmetries presented in Table~\ref{Tab_RSH_Asym} were 
obtained at slightly different 
values of $Q^2$ (see Table~\ref{bob_tableq2}). 
We used $\aexp_a$, the average of the asymmetries of the
detectors.
To combine these, 
the asymmetries were first corrected for background as
described in the previous section, and then extrapolated to a common
$Q^2 = 0.477$ (GeV/c)$^2$ using the leading $Q^2$ dependence from
Eq.~\ref{eq:pvasy}. The resulting weighted average asymmetry was
\begin{eqnarray}\label{a_exp}
\aexp = -15.05 \pm 0.98 \pm 0.56 \; {\rm ppm}, 
\end{eqnarray}
where the first error is statistical and the second error is
systematic. This latter includes the errors in the beam polarization,
background subtraction, helicity-correlated beam properties, and
$Q^2$.

\subsection{Finite Acceptance}
\label{sec:acceptance}

To interpret the experimental asymmetry given in Eq.~\ref{a_exp}, one must correct for 
the effect of averaging over the finite acceptance of the detectors and the 
effect of radiation on the effective kinematics of the measurement.
A Monte Carlo simulation was developed for this purpose, and is described below.

%%in order to investigate
%the effects of finite acceptance, radiative corrections, and
%uncertainties in the known electromagnetic form factors on the
%measured asymmetry. 

\subsubsection{Monte Carlo}
\label{sec:monte}
   As the acceptance of the HRS spectrometers is dictated
by their entrance collimators, the simulation involved generating
elastically scattering electrons along the length of the target, with
realistic account of the materials in the target region, and tracking
the events to the collimators. First-order magnetic optics of the
spectrometers were then used to determine the location and momentum of
the electrons at the focal plane detectors. The measured analog
response of the focal plane detectors, as a function of the position
of the hit along the detector, was taken as a weighting factor on the
asymmetry (this weighting had a $\sim 1\%$ effect compared to pure
counting statistics).  Account was taken of ionization energy loss in the 
target, both before and after the scattering. 

%\subsection{Radiative Corrections}
%\label{sec:rad_corr}

Bremsstrahlung was included in the simulation in both the initial and
final state. In the extreme relativistic limit, hard photon radiation
is strongly peaked in the forward angle, and so the angle peaking
approximation \cite{Stein} was adopted.
 
The radiated cross section $\sigma^{\rm rad}$was calculated as a convolution of 
integrals along the incident and scattered electron directions
\cite{MoTsai}.
With $E_s$ the incident electron energy, $E_p$ the final electron
energy, $t$ the location of the scattering along the target of 
length $T$, $t_1(t)$ and $t_2(t)$ the material thickness in radiation lengths 
before and after the scattering respectively, we have
\begin{eqnarray}\label{radiativecross}
\sigma^{\rm rad}(E_s,E_p)  =  (1 + \delta_f) \int^{T}_0 
\frac{dT}{T} \int^{1}_{0}dy_1 I_1(y_1,t_1) \nonumber \\
\int^{1}_{0}dy_2 I_2(y_2,t_2)  \sigma(E_s',E^{\rm max}_p) 
\Theta(E_p-E_{\rm cut})
\end{eqnarray}
where $y_1$ and $y_2$ are the fractional radiative energy losses before and after the 
scattering, $\sigma(E_s',E^{\rm max}_p)$ is the unradiated cross section for elastic scattering
of electrons of energy $E_s' = E_S(1-y_1)$ into energy  $E^{\rm max}_p$, where 
\begin{equation}
E^{\rm max}_p = \frac{E_s'}{1 + 2 (E_s'/M) \sin^2(\theta/2)}
\end{equation}
with  $M$ the proton mass and $\theta$ the scattering angle; 
the final electron energy is therefore $E_p = E^{\rm max}_p(1-y_2$).
The lower-energy cutoff in the spectrometer acceptance is $E_{\rm cut}$. The intensity
factors $I_1(y_1,t_1)$ and $I_2(y_2,t_2)$ are given by 
\begin{equation}
I(y,t) = \frac{\Phi(y,t)}{y} \; \exp\left(\int_1^y\negthickspace{dy' \frac{\Phi(y',t)}{y'}}\right)
\end{equation}
with 
\begin{equation}\label{Phi}
\Phi(y,t) = t_v(1-y) + \frac{4}{3}t\left(1 - y + \frac{3}{4}y^2\right) \;\;\;.
\end{equation}
The first term represents the effect of 
internal bremsstrahlung, which was dealt with using an equivalent virtual radiator \cite{MoTsai} 
of thickness 
\begin{equation}
t_v = \frac{\alpha}{\pi} \left[ \ln \left(\frac{Q^2}{m^2}\right) -1 \right] \;\;\; .
\end{equation}
The second term in Eq.~\ref{Phi} represents the 
`complete screening approximation'~\cite{Tsai} calculation of external bremsstrahlung. 

Finally, the factor  $(1 + \delta_f)$ in Eq.~\ref{radiativecross} is the 
lowest order correction to the running coupling constant $\alpha^2(Q^2)$, 
\begin{equation}
 \delta_f(Q^2) \approx \frac{2\alpha}{\pi} \left[ \frac{13}{12} \ln\left(\frac{Q^2}{m^2}\right) -\frac{28}{18} \right] \;\;\; .
\end{equation}

The primary effect of bremsstrahlung was to radiate about 20\% of the 
elastic events out of the detector acceptance, and to lower the 
effective $Q^2$ by about 0.1\%, a negligible amount.  

\subsubsection{Effective Kinematics}
\label{sec:kine}

Due to both the finite acceptance of the spectrometer and radiative
energy losses, the measured asymmetry represents a convolution over a
range of $Q^2$. To account for this, and to present a value of the
asymmetry for a single $Q^2$, we calculated an average incident
electron energy and effective scattering angle for the experiment, and
then used the simulation to calculate the factor needed to correct the
acceptance-averaged asymmetry to that from point scattering at the
effective kinematics.

The effective kinematics were calculated from the most probable value of 
the incident beam energy $E_s$, including energy loss in the
target, as
\begin{equation}
E_s = \left\langle E_{\rm beam} -\frac{dE}{dx}t \right\rangle \;\;\; .
\end{equation}
Using the measured average $Q^2$, the effective scattering angle $\theta_{\rm eff}$ was 
found from 
\begin{equation}
\cos(\theta_{\rm eff}) = \frac{1- (Q^2/(2E_s^2))(1 + E_s/M)}
{1-(Q^2/(2E_s^2))(E_s/M)} \;\;\; .
\end{equation}

To obtain the correction factor, the simulation was
run using a theoretical point asymmetry $A(E_s, \theta_{\rm eff})$ 
at the effective 
kinematics. The ratio of this to the averaged asymmetry $A_{\rm MC}$ extracted from the 
simulated data was then  used to extract the correction factor
\begin{equation}
C_{\rm finite} = \frac{A(E_s,\theta_{\rm eff})}{A_{\rm MC}} = 0.993 \pm 0.010
\end{equation}

This correction factor was then applied to the measured asymmetry $\aexp$
(Eq.~\ref{a_exp}) to yield
a physics asymmetry $\aphys$ at the effective kinematics:
\begin{equation}\label{a_phys} 
\aphys = C_{\rm finite} \, \aexp = -14.92 \pm 0.98 \; ({\rm stat}) \pm 0.56 \; ({\rm syst})
\end{equation}
for the average kinematics $Q^2 = 0.477$ (GeV/c)$^2$ 
and $\theta = 12.3^\circ$.

In the calculation of $C_{\rm finite}$, the default values for the
electromagnetic form factors discussed below in Section~\ref{sec:formfact} were used. The strange
quark form factors were assumed to be zero for the baseline value of
$C_{\rm finite}$.  Various available models for the $Q^2$ evolution of
non-zero strange form factors were also simulated, and the most
extreme case was used to estimate a model-dependent error on $C_{\rm
finite}$ of 0.9\%.  As mentioned, $C_{\rm finite}$ includes the effect
of bremsstrahlung, and the weighting by the detector's analog
response. Uncertainties due to these effects, including errors in the
beam energy and direction contributed 0.15\% to the error in $C_{\rm
finite}$. Note that overall correction due to finite acceptance {\em etc.}
to the measured asymmetry is much smaller than the statistical error on
our measurement. 

\begin{table}[tbhp]
\caption{Asymmetry corrections and systematic errors.\vspace*{2ex}}
\label{Tab_RSH_Syst}
\begin{tabular}{lccc} 
Source		&Correction	&  $\delta A/A $(\%)	& $\delta A/A $(\%)\cr
   & &  (1998)	& (1999)\cr
\hline
Statistics	& $- $	& 13.3	&7.2\cr
 $P_{e} $	& $- $	& 7.0	&3.2\cr
 $Q^2 $		& $- $	&1.8	&1.8\cr
Backgrounds	&1.2	& 0.6	&0.6\cr
Radiative corrections  & -0.1   & 0.1    & 0.1 \cr
Finite acceptance      & 0.7    & 0.9  & 0.9 \cr
\end{tabular} 
\end{table}

Table \ref{Tab_RSH_Syst} summarizes all corrections and systematic
errors applied to the measured asymmetries in Table
\ref{Tab_RSH_Asym} to
obtain the physics asymmetry of Eq.~\ref{a_phys}.  

\section{Results and Interpretation}
\label{sec:interpretation}

\subsection{Electromagnetic and Axial Form Factors}
\label{sec:formfact}

The extraction of the strange quark form factors $G^s_E$ and 
$G^s_M$ from the measured
asymmetry (Eq.~\ref{a_phys}) requires knowledge of the other form factors entering
into Eq.~\ref{eq:pvasy}: the purely electromagnetic form factors 
$G^{\gamma p}_E$,
$G^{\gamma p}_M$,
$G^{\gamma n}_E$, 
and 
$G^{\gamma n}_M$, 
as well as the neutral weak axial form factor $G^{Z p}_A$. 
Uncertainties in these form factors contribute significantly to the 
total uncertainty in the extracted strange form factors.

In the time since our initial publications \cite{aniol1,aniol2}, 
there has been considerable progress made on precision measurements 
of these form factors (see \cite{gaoreview} for a review). In the 
following we describe how
values for the form factors, interpolated to our $Q^2$, 
 were extracted from world data, and we reassess our extraction of
the strange form factors in light of the recent data.

\subsubsection{$G^{Z p}_A$}

 As mentioned earlier, the contribution of the neutral weak axial form
factor $G^{Z p}_A$ to the measured asymmetry is suppressed
for our kinematics (forward-angle scattering, where $\epsilon'$ is
small). This form factor can be decomposed into terms involving the
well-known charged-current axial form factor and $\Delta s$, the first
moment of the strange quark momentum distributions.  The latter, as
measured in deep inelastic scattering, while not precisely measured,
is small for our purposes \cite{abe}. The former at $Q^2 = 0$ is the
axial vector coupling constant $g_A$, which is well measured
\cite{barnet}, and the $Q^2$ evolution of the form factor is well reproduced
with a dipole form. However, $G^{Z p}_A$ suffers from
large electroweak radiative corrections, which include hadronic 
uncertainties, and which are problematic to calculate. 
These corrections have been calculated by Zhu {\em et al.} \cite{zhu},
and lead to a predicted effect on our measured asymmetry of 
$0.56 \pm 0.23$ ppm (the hadronic uncertainties in the axial
radiative correction dominate the error on this prediction). 

This prediction was cast into some doubt with the results from the
SAMPLE collaboration on backward-angle parity-violating quasielastic
scattering from a deuterium target \cite{hasty}. When combined with
their measurement on a hydrogen target \cite{spayde,mueller}, they extracted a
value for $G^{Z p}_A$ in significant disagreement with the
calculation of Zhu {\em et al.}, leading to speculation of large
`anapole moment' contributions. However, more recent data from SAMPLE,
along with a reanalysis of the earlier data \cite{ito} now yields 
excellent agreement with the Zhu {\em et al.} calculation, and so there is
no longer reason to doubt that the axial contribution is under adequate
control.

\subsubsection{$G^{\gamma p}_M$}
 
The proton's magnetic form factor is quite precisely known at our
kinematics, and it deviates only slightly from the dipole form factor
$G_D = \left[1 + Q^2/(0.71\ ({\rm GeV/c})^2)\right]^{-2}$.  
We adopt the value $G^{\gamma p}_M/\mu_p G_D = 0.9934$ at $Q^2 = 0.477$ (GeV/c)$^2$ 
using the
recent fit of Brash {\em et al.}  \cite{brash}. This fit is a
reanalysis of the magnetic form factor obtained from Rosenbluth
separation data, using as an additional constraint the 
results on $G^{\gamma p}_E/G^{\gamma p}_M$ obtained with polarization transfer techniques.
An almost identical value $G^{\gamma p}_M/\mu_p G_D = 0.9940$ 
at $Q^2 = 0.477$ (GeV/c)$^2$ was found from
the empirical fits of Friedrich and Walcher \cite{friedrich}. The
value also agrees within 0.3\% with the one we adopted in our earlier
publication \cite{aniol2}.

As the other electromagnetic form factors are often measured
relative to $G^{\gamma p}_M$, we will express them 
relative to this value, and subsume its small uncertainty
in the errors assigned to the other form factors. 

\subsubsection{$G^{\gamma p}_E$}

The situation with regard to the proton's electric form factor is
unsettled at present. The recent high-precision measurements from Jefferson
Lab of the
ratio $G^{\gamma p}_E/G^{\gamma p}_M$ using
recoil polarization techniques \cite{jones, gayou}
differ significantly from older results that used Rosenbluth separation
techniques (see \cite{arrington} for a review of the situation). There
are recent suggestions that this discrepancy could be the result of
contributions from two-photon exchange \cite{vanderhaeghen, blunden}
which may have a large effect on the Rosenbluth separation data at
large $Q^2$. We note, however, that at our lower $Q^2$ the difference
between the values of $G^{\gamma p}_E$ extracted from the
recoil polarization data and those from the Rosenbluth data is small.  Adopting
the empirical fit of Friedrich and Walcher \cite{friedrich}, which is
based on both polarization data and Rosenbluth data at lower $Q^2$,
yields $G^{\gamma p}_E/(G^{\gamma p}_M/\mu_p) = 0.98$
 at $Q^2 = 0.477$ (GeV/c)$^2$; a similar value  of $G^{\gamma p}_E/(G^{\gamma p}_M/\mu_p) = 0.97$ is obtained from the 
empirical fit of Arrington \cite{arrington}. We adopt the former value
with a 2\% uncertainty, which is essentially the same as we used
previously \cite{aniol2}.

\subsubsection{$G^{\gamma n}_E$}

In our previous publications \cite{aniol1,aniol2} the largest
uncertainty arising from an electromagnetic form factor was that due
to the electric form factor of the neutron, $G^{\gamma n}_E$.  Since those publications appeared, the situation for $G^{\gamma n}_E$ has improved dramatically, due to new precise
results using polarization techniques now available from Jefferson Lab
\cite{zhu,madey,warren} and Mainz \cite{bermuth}, as well as a new
analysis that obtained $G^{\gamma n}_E$ from data on the quadrupole form factor in elastic electron-deuteron
scattering \cite{schiavilla}. Individual measurements now have
uncertainties at roughly the 10\% level, and the recent results,
which conveniently bracket our $Q^2$, are satisfactorily consistent. 

To extract the value of $G^{\gamma n}_E$ at our $Q^2$, we
use the fit to a Galster form \cite {galster} 
provided by Madey {\em et al.}~\cite{madey}, 
which gives $\mu_p G^{\gamma n}_E/G^p_M = 0.161 \pm 0.006$ at $Q^2= 0.477$ (GeV/c)$^2$. 
This fit was based on the world
data from polarization measurements as well as the analysis of the
deuteron quadrupole form factor. It did not include the very recently
reported result of Warren {\em et al.} \cite{warren}, however the fit
agrees with the Warren {\em et al.} datum at $Q^2= 0.5$ within $1
\sigma$.  A similar fit was presented by Warren {\em at al.}, which
did not include the Madey {\em et al.} datum, but nevertheless agreed
with the Madey {\em et al.} result at their $Q^2 = 0.447$ (GeV/c)$^2$. 
That fit also
gave a value consistent within 3.6\% with that from the Madey {\em et
al.} fit at our $Q^2$. 
To be conservative, we enlarge the error
from the Madey fit to 5\% and thus adopt the value 
$\mu_p G^{\gamma n}_E/G^p_M = 0.161 \pm 0.008$
at $Q^2 = 0.477$ (GeV/c)$^2$. The central value is essentially unchanged 
from that used previously
\cite{aniol2}, however the error bar has been reduced by almost a factor 
of 4. The contribution to the error in $\apv$ due to the 
uncertainty in $G^{\gamma n}_E$ is now less than those due to
other form factors ({$G^{\gamma n}_M$ and 
$G^{\gamma p}_E$); see Table~\ref{Tab_formfactors}.

\subsubsection{$G^{\gamma n}_M$}

The largest contribution to our error due to electromagnetic form
factors is that due to the neutron's magnetic form factor,  $G^{\gamma n}_M$. Results from two new experiments have appeared
since our earlier publications \cite{aniol1, aniol2}. These are the
measurements from Mainz of Kubon {\em et al.} \cite{kubon} and from
JLab of Xu {\em et al.} \cite{xu}. The former span a range of $Q^2$
from 0.071 to 0.894 GeV$^2$, and the later, while somewhat less
precise, report data for $Q^2$ ranging from 0.3 to 0.6 GeV$^2$,
including points ($Q^2 = 0.4$, 0.5 (GeV/c)$^2$) close to our own
kinematics.

 Kubon {\em et al.} \cite{kubon} provide an empirical fit to 
their data along with other recent data on $G^{\gamma n}_M$
\cite{anklin, gao, xu00}. While the recent results of Xu {\em et al.} 
\cite{xu} were not included in the fit, the fit does an excellent job of 
reproducing them, with agreement to better than 2\%. We note that
this agreement exhibits the compatibility of results obtained from
very different experimental techniques, with different model
dependences, and thus builds confidence in the values of 
 $G^{\gamma n}_M$ extracted. Thus we adopt
the Kubon {\em et al.} fit to interpolate to $Q^2 = 0.477$ (GeV/c)$^2$ 
and extract the value  $(G^{\gamma n}_M)/\mu_n)
/(G^{\gamma p}_M/\mu_p) = 1.004 \pm 0.040$ (in order to be conservative,
we have inflated the uncertainty in the fit from Kubon {\em et al.} by a factor
of 3). This new value is somewhat lower than
the value of $1.05 \pm 0.02$ adopted previously by us \cite{aniol2}. 

A different fit for $G^{\gamma n}_M$, using a somewhat
different database of results, and a very different functional form,
was obtained by Friedrich and Walter \cite{friedrich}, and it yields
the value  $(G^{\gamma n}_M)/\mu_n)
/(G^{\gamma p}_M/\mu_p)
 = 1.039$ at $Q^2 = 0.477$ (GeV/c)$^2$,
in reasonable agreement with the fit of  Kubon {\em et al.}.

 We note, however, that both fits discard the results of Bruins {\em et
al.} \cite{bruins} and Markowitz {\em et al.} \cite{markowitz}. The former
has been criticized \cite{jourdan} due to potential difficulties with
the extraction of their neutron detection efficiency, however a direct
measurement of that efficiency is planned \cite{bruins2}. If the results
from Bruins {\em et al.} are adopted at face value, this would have a 
very significant effect on our extracted strange form factors. Finally,
we note that there are new data from the CLAS at JLab presently under 
analysis, which should help clarify the situation \cite{brooks}. 

In summary, the two significant changes that recent data have made to
the information on the electromagnetic form factors, compared to that
of our previous result \cite{aniol2}, are the significantly improved
precision on $G^{\gamma n}_E$ (without a change in the central
value) and a change in the best estimate of $G^{\gamma n}_M$. The latter causes a shift in the extracted strange quark
contribution compared to that presented in Ref. \cite{aniol2} (the
shift is small compared to the statistical error). The effect of the 
form factors on the predicted asymmetry is summarized in Table~\ref{Tab_formfactors}.

\begin{table}[tbhp]
\caption{Electromagnetic form factors at the present $Q^2$, normalized
to $(G^{\gamma p}_M/\mu_p)$, and their contribution to the error
in ppm on the theoretical asymmetry $\apv$. \vspace*{2ex}}
\label{Tab_formfactors}
\begin{tabular}{lcc} 
Form Factor         & Value  &  $\delta A$ (ppm) \\
\hline
 & & \\
$G^{\gamma p}_E/(G^{\gamma p}_M/\mu_p)$ & $0.98 \pm 0.02$ &   0.33 \\
$G^{\gamma n}_E/(G^{\gamma p}_M/\mu_p)$ & $ 0.161 \pm 0.008$ &  0.15  \\
$(G^{\gamma n}_M)/\mu_n) /(G^{\gamma p}_M/\mu_p)$ & $1.004 \pm 0.040$ & 0.48   \\

\end{tabular} 
\end{table}

\subsection{Strange Quark Form Factors}
\label{sec:strange}

Using Eq.~\ref{eq:pvasy} and the result in
Eq.~\ref{a_phys}, along with the calculated $G^{Z p}_A$ and the known values of the proton and
neutron form factors in Table~\ref{Tab_formfactors}, we may solve for the linear combination of strange
form factors $G_E^s + \beta G_M^s$ where $\beta = \tau G_M^{\gamma p}
/ \epsilon G_E^{\gamma p} = 0.392$ at our kinematics.  We obtain
\begin{equation} 
\label{eq:strff_expt}
G_E^s + \beta G_M^s = 0.014 \pm 0.020 \pm 0.010  
\end{equation}
where the first error is the total experimental error
(statistical and systematic errors added
in quadrature) and the second error
is the error due to the 
``ordinary'' electromagnetic form factors
and is dominated by $G_M^n$.
Since \cite{aniol2} the central value 
has reduced slightly, though less than the error bar,
and the error due to electromagnetic form factors
has reduced.   
This result is consistent with zero 
strangeness contribution
to the vector matrix elements of the proton.
However, the result could also be zero due to a 
cancellation of $G_E^s$ and $G_M^s$ at our $Q^2$.
The SAMPLE experiment \cite{sample_prl, spayde, sample2},
which is sensitive to $G_M^s$ at $Q^2 = 0.1 {\rm (GeV/c)}^2$
as well as the axial form factor $G^{Z p}_A$, 
also found a very small strangeness contribution 
which is consistent with zero.

Numerous theoretical models have been formulated to
predict the strangeness form factors.
The problem is one of nonperturbative QCD
since $m_s \simeq \Lambda_{\rm QCD}$.
In some cases the models are considered to be 
only an order of magnitude estimate, and in
other cases only an upper bound to the strangeness effects.
The large variety of models with very
different physics assumptions is indicative
of the difficulty in making solid predictions.
See also the discussion in section \ref{sec:motivation} and 
refs \cite{jaffe1,hammer1,hammer2,forkel1,forkel2,
chiral_bag1,geiger_isgur,pquark,ma1,
riska1,riska2,skyrme_1,skyrme_2,
weigel_nlj,meissner1,koepf1,musolf1,ito1,
soliton,hemmert1,hemmert2,lattice_1,lattice_2}.

\begin{figure}
\includegraphics[angle=270,width=4.2in]{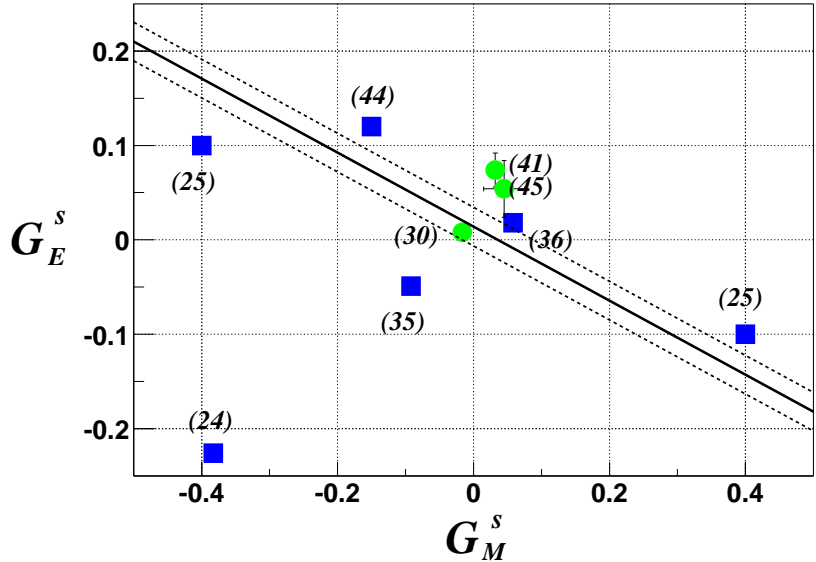}
\caption{\label{fig30_straff}
Plot of $G_E^s$ vs $G_M^s$ at $Q^2 = 0.477 {\rm (GeV/c)}^2$.
The band is the allowed region derived from our results.
The width of the band is the total error computed in
quadrature.  The points are estimates from various
models that make predictions at our $Q^2$.  The numbers
in ref. \cite{hammer2} are plotted twice due to an ambiguity
in the predicted sign.  This plot is similar to Fig. 4
in \cite{aniol2} except that the central value and 
error bars have both reduced slightly, and
three new models
shown in circles have since been published.}
\end{figure}

Most models focus on predictions of the static
moments $\rho_s$ and $\mu_s$ at $Q^2 = 0$.
A subset of the models attempt to predict the
form factors at our $Q^2$, shown as points in
Fig. ~\ref{fig30_straff},
together with our result for $G_M^s + \beta G_M^s$
displayed as a line with an error band representing
the two errors in Eq. \ref{eq:strff_expt} added in quadrature. 
The numbers near the points in Fig ~\ref{fig30_straff}
are the references for those models. 
The square points are the models displayed in
Fig. 4 of our previous publication \cite{aniol2}.
The circle points are from three models published
since \cite{aniol2} which predict 
relatively small strangeness form factors 
that are in good agreement with our data.
In two of the models \cite{soliton,lattice_2}
the authors predicted
a likely range (1$\sigma$) of form factors
which is indicated by the 
error bars in the figure for
those two points.  

Several of the models make predictions
which will be tested by future measurements,
including the HAPPEX-2 experiment \cite{happex2_prop}, 
He4 parity \cite{he4_prop}, G$^0$ \cite{g0_prop}, 
and the Mainz A4 parity experiments \cite{a4_prop}.
These measurements will be necessary to separate
$G_E^s$ and $G_M^s$ and determine their $Q^2$
dependence.

\section{Conclusions}
\label{sec:conclusions}

The HAPPEX results reported in this paper have 
provided a stringent test of strange $q \bar q$
contributions to the vector matrix elements 
of the proton. 
Our results still allow for strangeness
effects of a few percent or the possibility
of accidental cancellation at our kinematics.
It will be important to complete
the program of approved
parity experiments at Jefferson Lab 
\cite{happex2_prop,he4_prop,g0_prop}
and elsewhere \cite{a4_prop} to quantify
the strangeness effects over a range
of kinematics and over various
distance scales in the nucleon.
These experiments should yield a 
detailed mapping of the spatial
dependence of $s \bar s$ contributions
to nucleon structure.

In this paper we have reported 
details of the experimental technique 
and data analysis.
We have described methods for minimizing helicity
correlations of the polarized electron beam
from a strained GaAs crystal.
Because of the highly stable beam at 
Jefferson Lab we were able to acquire precise data 
that were nearly free of systematic error.
This bodes well for future applications of
parity violating electron scattering to various
physics topics including future
searches for strange sea effects 
\cite{happex2_prop,he4_prop,g0_prop,a4_prop},
precision studies of the standard model 
\cite{qweak,dis-parity}
and measurements of neutron densities in nuclei 
\cite{horowitz1,hpsm,prex}.

\section{Acknowledgments}
\label{sec:ackn}

We wish to thank the entire staff at 
Jefferson Lab for their exemplary  
effort in developing and operating 
the facility, and particularly 
C. K. Sinclair and M. Poelker for 
their essential work on the polarized 
source.  This work was supported 
by DOE contract DE-AC05-84ER40150
under which the Southeastern Universities 
Research Association
(SURA) operates the Thomas Jefferson 
National Accelerator Facility, 
and by the Department of 
Energy, the National Science Foundation, 
the Korean Science 
and Engineering Foundation (Korea), 
the Istituto Nazionale di Fisica Nucleare (Italy), 
the Natural Sciences and Engineering 
Research Council of Canada, 
the Commissariat \`a l'\'Energie Atomique (France), 
and the Centre National
de Recherche Scientifique (France).


\begin{thebibliography}{}\label{sec:lpaper_bib}

\bibitem{emc} J. Ashman {\em et al.}, Phys. Lett. B 
{\bf 206} (1988) 364; Nucl. Phys. B {\bf 328} (1988) 527.
Nucl. Phys. B {\bf 328} (1989) 1.

\bibitem{kaplan_manohar} D. B. Kaplan and A. Manohar,
Nucl. Phys. B {\bf 310} (1988) 527.

\bibitem{mckeown1} R. D. McKeown, Phys. Lett. B {\bf 219}
(1989) 140.

\bibitem{beise_mckeown} E. J. Beise and R. D. McKeown,
Comments Nucl. Part. Phys. {\bf 20} (1991) 105.

\bibitem{beck1} D. H. Beck, Phys. Rev. D {\bf 39} (1989) 3248.

\bibitem{aniol1} K. Aniol {\em et al.}, [HAPPEX Collaboration], Phys. Rev. Lett. {\bf 82}, 1096 (1999). 
\bibitem{aniol2} K. Aniol {\em et al.}, [HAPPEX Collaboration], Phys. Lett. {\bf B509}, 211 (2001). 

\bibitem{Gary} G.A. Rutledge, Ph.D. thesis, College of William and Mary, 2001.
\bibitem{Wilson} G.W. Miller, Ph.D. thesis, Princeton U., 2001.
\bibitem{Johan} J. Jardiller, Ph.D. thesis (in French), U. Blaise Pascal, 2001.
\bibitem{Bill} W.E. Kahl, Ph.D. thesis, Syracuse U., 2000. 
\bibitem{Brian} T.B. Humensky,  Ph.D. thesis, Princeton U., 2003.
\bibitem{Baris} B.T. Tonguc,  Ph.D. thesis, Syracuse U., 2003.

\bibitem{charm1} A. O. Bazarko {\em et al.}, Z. Phys. C {\bf 65} (1995) 189;
M. Gonchorov  {\em et al.}, Phys. Rev. D{\bf 64} (2001) 112006.

\bibitem{dis1} B. Adeva {\em et al.}, Phys. Lett. B {\bf 302},
(1993) 533; P.L. Anthony {\em et al.}, Phys. Rev. Lett.
{\bf 71} (1993) 959; D. Adams {\em et al.}, Phys. 
Lett. B {\bf 329} (1994) 399.

\bibitem{abe}K. Abe {\em et al.}, Phys. Rev. Lett {\bf 74} (1995) 346.

\bibitem{ellis_jaffe} J. Ellis and R.L. Jaffe, Phys. Rev. D
{\bf 9} (1974) 1444; {\bf 10} (1974) 1669.

\bibitem{badeva98} B. Adeva {\em et al.}, Phys. Rev. D
{\bf 58} (1998) 112002.

\bibitem{kumar_souder} K.S. Kumar and P.A. Souder,
Prog. Part. Nucl. Phys. 45 (2000) S333.

\bibitem{beck_mckeown} D.H. Beck and R.D. McKeown,
Ann. Rev. Nucl. Part. Sci. 51 (2001) 189.

\bibitem{beck_holstein} D.H. Beck and B.R. Holstein,
Int. J. Mod. Phys. E10 (2000) 1.

\bibitem{musolf_physrep94} M.J. Musolf {\em et al.},
Phys. Rep. 239 (1994) 1.

\bibitem{jaffe1} R.L. Jaffe, Phys. Lett. B {\bf 229}, 275 (1989).

\bibitem{hammer1} H.-W. Hammer, U.-G. Meissner, and D. Drechsel,
Phys. Lett. B {\bf 367}, 323 (1996).

\bibitem{hammer2} H.-W. Hammer and M.J. Ramsey-Musolf,
Phys. Rev. C {\bf 60},  045204 (1999); {\em ibid} {\bf 60},  045205 (1999);
{\em erratum} {\em ibid} {\bf 62}, 049902 (2000); 63, 049903 (2000).

\bibitem{forkel1} H. Forkel, Phys. Rev. C {\bf 56}, 510 (1997)..

\bibitem{forkel2} H. Forkel, M. Nielsen, X. Jin, and T. Cohen,
Phys. Rev. C {\bf 50}, 3108 (1994).

\bibitem{chiral_bag1} S.-T. Hong, B.-Y. Park, and D.-P. Min,
Phys. Lett. B {\bf 414}, 229 (1997).

\bibitem{geiger_isgur} P. Geiger and N. Isgur,
  Phys. Rev. D {\bf 55}. 299 (1997).

\bibitem{pquark} V. E. Lyubovitskij, P. Wang, 
Th. Gutsche, and A. Faessler,
Phys. Rev. C {\bf 66}, 055204 (2002).

\bibitem{ma1} B.-Q. Ma, Phys. Lett. B {\bf 408}, 387 (1997).

\bibitem{riska1} D.O. Riska, Nucl. Phys. A {\bf 678}, 79 (2000).

\bibitem{riska2} L. Hannelius, D.O. Riska, and L.Ya. Glozman,
Nucl. Phys. A {\bf 665}, 353 (2000).

\bibitem{skyrme_1} N.W. Park, J. Schechter, and H. Weigel,
Phys. Rev. D {\bf 43}, 869 (1991).

\bibitem{skyrme_2} N.W. Park and H. Weigel, 
Nucl. Phys. A {\bf 541}, 453 (1992).

\bibitem{weigel_nlj} H. Weigel {\em et al.}, 
Phys. Lett. B {\bf 353}, 20 (1995).

\bibitem{meissner1} U.-G. Meissner, V. Mull, J. Speth, 
J.W. Van Orden, Phys. Lett. B {\bf 408}, 381 (1997).

\bibitem{koepf1} W. Koepf, E.M. Henley, and J.S. Pollock, 
Phys. Lett. B {\bf 288}, 11 (1992).

\bibitem{musolf1} M.J. Musolf and M. Burkhardt, 
Z. Phys. C {\bf 61}, 433 (1994).

\bibitem{ito1} H. Ito, Phys. Rev. C {\bf 52}, R1750 (1995).

\bibitem{soliton} A. Silva, H.-C. Kim, and K. Goeke, 
Phys. Rev. D {\bf 65}, 014016 (2002),
Erratum-ibid. D {\bf 66}, 039902 (2002).

\bibitem{hemmert1} T. Hemmert, U.-G. Meissner, and S. Steininger,
Phys. Lett. B {\bf 437}, 184 (1998).

\bibitem{hemmert2} T. Hemmert, B. Kubis, and U.-G. Meissner,
Phys. Rev. C {\bf 60}, 045501 (1999).

\bibitem{lattice_1} S.J. Dong, K.F.Liu, and A.G. Williams, 
Phys. Rev. D {\bf 58}, 074504 (1998).

\bibitem{lattice_2} R. Lewis, W. Wilcox, and R.M. Woloshyn,
Phys. Rev. D {\bf 67}, 013003 (2003).

\bibitem{sample_prl} B. Mueller {\em et al.}, Phys. Rev. Lett
{\bf 78} (1997) 3824.

\bibitem{barnet}Particle Data Group, R.M. Barnet {\em et al.}, Phys. Rev. D {\bf 54} (1996) 1.

\bibitem{axial_corr1} M.J. Musolf and B.R. Holstein,
Phys. Rev. Lett B {\bf 242} (1990) 461.

\bibitem{stripline}T. Powers, L. Doolittle, R. Ursic, and J. Wagner, 
Proc. 7th Workshop on Beam Instrumentation, AIP Conf.Proc. {\bf 390},
Ed. A.. Lumpkin and C.E. Eyberger (1997); JLAB-TN-96-021.

\bibitem{A-NIM} J. Alcorn {\em et al.}, 
accepted by Nucl. Instrum. Methods (2003). 

\bibitem{Unser}K. Unser, IEEE Trans. Nucl. Sci. NS-28 (1981) 2344.
T. Powers, L. Doolittle, R. Ursic, and J. Wagner, 
Proc. 7th Workshop on Beam Instrumentation, AIP Conf.Proc. {\bf 390},
Ed. A. Lumpkin and C.E. Eyberger (1997); JLAB-TN-96-021.

\bibitem{gar_CERNOX}
CERNOX resistor, Lakeshore Cryogenics.

\bibitem{gar_ALLEN_BRADLEY}
Allen-Bradley,  http://www.ab.com. 

\bibitem{gar_EPICS}
Experimental physics and industrial control system (EPICS), 
http://www.aps.anl.gov/epics/.

\bibitem{CODA} W.A. Watson {\em et al.}, 
CODA: A Scalable, Distributed
Data Acquisition System, in Proc. of
the Real Time 1993 Conference, p. 296;
G. Heyes {\em et al.}, The CEBAF Online
Data Acquisition System, in Proc. of the
CHEP Conference, 1994, p. 122;
D.J. Abbott {\em et al.}, CODA Performance
in the Real World, 11th IEEE NPSS Real Time
199 Conference, JLab-TN-99-12 (1999).

\bibitem{Sinclair1} C.K. Sinclair, ``Electron
Beam Polarimetry'', Tech Note JLAB-ACC-98-04;
T.J. Gay and F.B. Dunning, Rev. Sci. Instrum.
{\bf 63} (1992) 1635; T.J. Gay {\em et al.},
Rev. Sci. Instrum. {\bf 63} (1992) 114.  
S. Mayer {\em et al.}, Rev. Sci. Instrum.
{\bf 64} (1993) 952. 

\bibitem{Price}J.S. Price {\em et al.}, Proc. 13th Symposium 
on High Energy Spin Physics (SPIN98) Protvino, Russia, 8-12 September 
1998, Ed. N.E. Tyurin, V.L. Solovianov, S.M. Troshin, and  A
. G. Ufimtsev, (World Scientific) 1999;
JLab Technical Note ACC-97-27, 1997 (unpublished). 

\bibitem{Baylac} M.~Baylac {\em et al.}, Phys. Lett. B {\bf  539}, 8 (2002).

\bibitem{Neyret}  D.~Neyret {\em et al.}, Nucl. Instrum. Methods {\bf A 443}, 231 (2000).

\bibitem{Jorda} J.P.~Jorda {\em et al.}, Nucl. Instrum. Methods {\bf A 412}, 1 (1998). 

\bibitem{Falletto} N.~Falletto {\em et al.}, Nucl. Instrum. Methods {\bf A 459}, 412 (2001). 

\bibitem{Maud_these} M. Baylac, Ph.D. thesis, U. Claude Bernard Lyon I, \# 212-2000 ; Report
CEA/DSM/DAPNIA/SPhN-00-05-T (unpublished). 

\bibitem{tbh_Cates} G. D. Cates {\em et al.}, Nucl. Instrum. Methods,
{\bf A 278} (1989) 293-317.
\bibitem{tbh_Humensky} T. B. Humensky {\em et al.}, SLAC-PUB-9381. Submitted to Nucl. Instrum. Methods {\bf A} (2002).

\bibitem{edwards_syphers} D.A. Edwards and M.J. Syphers, 
An Introduction to the Physics of High Energy Accelerators, 
Wiley Interscience (1993).

\bibitem{courant_snyder} E.D. Courant and H.S. Snyder, 
Annals of Physics 3(1) (1958) 1.

\bibitem{ychao} Y. Chao, ``Measuring and Matching Transport 
Optics at Jefferson Lab'',
Proceedings of the 2003 Particle Accelerator Conference,  Portland,
Oregon, 2003.

\bibitem{Mott}N.F. Mott, Proc. R. Soc. London, Ser A {\bf 135}, 429 (1932); 
{\it ibid.} {\bf 124}, 425 (1929). 

\bibitem{Afanasev}A. Afanasev, I. Akushevic, N.P. Merenkov, 
Workshop on Exclusive Process at High Momentum Transfer,
  Jefferson Lab, May 15-18, 2002; hep-ph/0208260. 

\bibitem{Wells}S.P. Wells {\em et al.}, 
Phys. Rev. C {\bf 63}, 064001 (2001). 

\bibitem{ep_whitlow} L.W. Whitlow {\em et al.}, Phys. Lett. B
{\bf 282} (1992) 475; Phys. Lett. B {\bf 250} (1990) 193.

\bibitem{bob_mukhopadhyay} N.C. Mukhopadhyay {\em et al.}, 
Nucl. Phys. {\bf A 633}, 481 (1998).

\bibitem{bob_donmus} M.J. Musolf and T.W. Donnelly, 
Nucl. Phys. {\bf A 546}, 509 (1992); \hskip 0.05in
Erratum-{\em ibid} {\bf A 550}, 564 (1992).

\bibitem{bob_arcmeas} D. Marchand, Ph.D. Thesis, U. Clermont-Ferrand, 1998.

\bibitem{bob_epmeas} O. Ravel {\em et al.}, Nucl. Instrum. Methods,
{\bf A 409}, 611 (1998).

\bibitem{Stein} S. Stein {\em et al.}, Phys. Rev. D {\bf 12}, 1884 (1975). 
\bibitem{MoTsai} L.W. Mo and Y.S. Tsai, Rev. Mod. Phys. {\bf 41}, 205 (1969). 
\bibitem{Tsai} Y.S. Tsai, Rev. Mod. Phys. {\bf 46}, 815 (1974). 

\bibitem{gaoreview} H. Gao, 
Int. J. Mod. Phys E {\bf 12}, 1 (2003). 

\bibitem{zhu_musolf} S.L. Zhu, S.J. Puglia, B.R. Holstein, and M.J. Ramsey-Musolf.
Phys. Rev. D {\bf 62}, 033008 (2000); M.J. Musolf and B.R. Holstein, 
Phys. Lett. B {\bf 242}, 461  (1990). 

\bibitem{hasty} R. Hasty {\em et al.}, Science, {\bf 290}, 2117 (2000).

\bibitem{mueller} B. Mueller {\em et al.}, Phys. Rev. Lett. {\bf 78}, 3824 (1997).

\bibitem{spayde} D.T. Spayde {\em et al.}, Phys. Rev. Lett. {\bf 84}, 1106  (2000).

\bibitem{ito} T.M. Ito {\em et al.}, nucl-ex/03100001; E. Beise, private communication (2003).

\bibitem{brash} E.J. Brash, A. Kozlov, Sh. Li, and G.M. Huber,
Phys. Rev. C {\bf 65}, 051001(R) (2002). 

\bibitem {friedrich} J. Friedrich and Th. Walcher, 
Eur. Phys. J A {\bf 17}, 607 (2003). 

\bibitem{zhu} H. Zhu {\em et al.}, Phys. Rev. Lett. {\bf 87}, 081801 (2001). 

\bibitem{madey} R. Madey {\em et al.}, Phys. Rev. Lett. {\bf 91}, 122002 
(2003). 

\bibitem{warren} G. Warren {\em et al.}, nucl-ex/0308021

\bibitem{bermuth} J. Bermuth {\em et al.}, Phys. Lett B {\bf 564}, 
199 (2003). 

\bibitem{galster} S. Galster, H. Klein, K.H. Scmidt, D. Wegener, and
J. Bleckwenn, Nucl. Phys. B {\bf 32}, 221 (1971). 

\bibitem{schiavilla} R. Schiavilla and I. Sick, Phys. Lett C {\bf 64}, 
041002(R) (2001). 

\bibitem{jones} M.K. Jones {\em at al.}, Phys. Rev. Lett. {\bf 84}, 1398 (2000).

\bibitem{gayou} O. Gayou {\em at al.}, Phys. Rev. Lett. {\bf 88}, 092301 (2002).

\bibitem{arrington} J. Arrington, Phys. Rev. C {\bf 68}, 034325 (2003). 

\bibitem{vanderhaeghen} P.A.M. Guichon and M. Vanderhaeghen
 Phys. Rev. Lett. {\bf 91}, 142303 (2003).

\bibitem{blunden} P.G. Blunden, W. Melnitchouk, J.A. Tjon, Phys. Rev. Lett. {\bf 91}, 142304
(2003).

\bibitem{kubon} G. Kubon {\em et al.}, Phys. Lett. B {\bf 524}, 26 (2002). 

\bibitem{xu} W. Xu {\em et al.}, Phys. Rev. C {\bf 67}, 012201(R) (2003). 

\bibitem{anklin} H. Anklin {\em et al.}, Phys. Lett. B {\bf 336}, 313 (1994); 
H. Anklin {\em et al.} Phys. Lett B {\bf 428}, 248 (1998). 

\bibitem{gao} H. Gao {\em et al.}, Phys. Rev. C {\bf 50}, R546 (1994). 

\bibitem{xu00} W. Xu {\em et al.}, Phys. Rev. Lett. {\bf 85}, 2900 (2000). 

\bibitem{bruins} E.E.W. Bruins {\em et al}, Phys. Rev. Lett. {\bf 75}, 21 
(1995).

\bibitem{markowitz} P. Markowitz {\em et al.}, Phys. Rev. C {\bf 48}, R5 
(1993). 

\bibitem{jourdan} J. Jourdan, I. Sick, J. Zhao, Phys. Rev. Lett. {\bf 79}, 5186
(1997).

\bibitem{bruins2} E.E.W. Bruins {\em et al.}, Phys. Rev. Lett. {\bf 79}, 5187
(1997).

\bibitem{brooks} Jefferson Lab Expt. 94-017, W. Brooks and M. Vineyard,
spokespersons; W. Brooks, private communication (2003). 

\bibitem{sample2} E.J. Beise nucl-ex/0309008; R. Hasty {\em et al.},
Science {\bf 290} (2000) 2117

\bibitem{happex2_prop} HAPPEX-II Experiment E99-115,
K.S. Kumar and D. Lhuillier, spokespersons.

\bibitem{he4_prop} He4 Parity Experiment E00-114,
D. Armstrong and R. Michaels, spokespersons.

\bibitem{g0_prop} G0 Experiment E00-006,
D.H. Beck, spokesperson.

\bibitem{a4_prop} Mainz A4 Parity Experiment,
D. von Harrach, spokesperson.

\bibitem{qweak} The $Q_{\rm weak}$ Experiment,
J. Bowman, R. Carlini, J.M. Finn, S. Kowalski, and S. Page,
spokespersons.

\bibitem{dis-parity} Letter of Intent  to Measure Parity Violating DIS 
from Deuterium, X. Zheng, spokesperson.

\bibitem{horowitz1} C. J. Horowitz, Phys. Rev. C {\bf 57}, 3430 (1998).

\bibitem{hpsm} C. J. Horowitz, S. Pollock, P. Souder, and R. Michaels, 
Phys. Rev. C {\bf 63}, 025501 (2001).

\bibitem{prex} The PREX Experiment, JLab proposal E03-011,
P. Souder, G. Urciuoli, R. Michaels, spokespersons 

\end{thebibliography}
\end{document}